\documentclass[preprint]{emulateapj}

\usepackage{amsfonts}
\usepackage{amsmath}
\usepackage{mathrsfs}
\usepackage{times}
\usepackage{hyperref}

\hypersetup{
  bookmarksnumbered = true,
  bookmarksopen=false,
  pdfborder=0 0 0,         
  pdffitwindow=true,      
  pdfnewwindow=true, 
  colorlinks=true,           
  linkcolor=blue,            
  citecolor=magenta,    
  filecolor=magenta,     
  urlcolor=cyan              
}

\newcommand{\alphat}{\tilde{\alpha}}

\newcommand{\betau}[1]{\beta^{#1}}
\newcommand{\betad}[1]{\beta_{#1}}

\newcommand{\boostubdh}[2]{\Lambda^{\bar{#1}}_{\mbox{ }\hat{#2}}}
\newcommand{\boostuhdb}[2]{\Lambda^{\hat{#1}}_{\mbox{ }\bar{#2}}}
\newcommand{\cuhdhdh}[3]{\Gamma^{\hat{#1}}_{\mbox{ }\hat{#2}\hat{#3}}}
\newcommand{\cubdbdb}[3]{\Gamma^{\bar{#1}}_{\mbox{ }\bar{#2}\bar{#3}}}
\newcommand{\cudd}[3]{\Gamma^{#1}_{\mbox{ }#2#3}}

\newcommand{\cbudd}[3]{\bar{\Gamma}^{#1}_{\mbox{ }#2#3}}

\newcommand{\covderiv}[1]{\nabla_{#1}}
\newcommand{\covderivb}[1]{\bar{\nabla}_{#1}}
\newcommand{\collision}[1]{\mathbb{C}\left[#1\right]}

\newcommand{\collisionh}[1]{\hat{\mathbb{C}}\left[#1\right]}
\newcommand{\collisionEnergyEquation}{\mathcal{C}^{\scriptscriptstyle\mathcal{E}}}
\newcommand{\collisionEnergyEquationt}{\tilde{\mathcal{C}}^{\scriptscriptstyle\mathcal{E}}}

\newcommand{\CollisionEnergyEquation}{C^{\scriptscriptstyle E}}
\newcommand{\CollisionEnergyEquationt}{\tilde{C}^{\scriptscriptstyle E}}
\newcommand{\CollisionEnergyEquations}{C^{\scriptscriptstyle E}_{\scriptscriptstyle s}}

\newcommand{\collisionMomentumEquation}[1]{\mathcal{C}^{{\scriptscriptstyle\mathcal{F}}{#1}}}
\newcommand{\collisionMomentumEquationd}[1]{\mathcal{C}_{~#1}^{{\scriptscriptstyle\mathcal{F}}}}
\newcommand{\collisionMomentumEquationOrderOned}[1]{\mathcal{C}_{~#1}^{{\scriptscriptstyle\mathcal{H}}}}

\newcommand{\CollisionMomentumEquation}[1]{C^{{\scriptscriptstyle F}{#1}}}

\newcommand{\CollisionMomentumEquationd}[1]{C^{{\scriptscriptstyle F}}_{~#1}}
\newcommand{\CollisionMomentumEquationOrderOned}[1]{C^{{\scriptscriptstyle H}}_{~#1}}

\newcommand{\CollisionMomentumEquationOrderOneds}[1]{C^{{\scriptscriptstyle H}}_{{\scriptscriptstyle s}{#1}}}

\newcommand{\collisionNumberEquationt}{\tilde{\mathcal{C}}^{\scriptscriptstyle\mathcal{N}}}

\newcommand{\CollisionNumberEquation}[1]{C^{\scriptscriptstyle N}_{#1}}
\newcommand{\CollisionNumberEquationt}[1]{\tilde{C}^{\scriptscriptstyle N}_{#1}}
\newcommand{\deltaud}[2]{\delta^{#1}_{\mbox{ }{#2}}}

\newcommand{\deltaubdh}[2]{\delta^{\bar{#1}}_{\mbox{ }{\hat{#2}}}}
\newcommand{\deltauhdb}[2]{\delta^{\hat{#1}}_{\mbox{ }{\bar{#2}}}}

\newcommand{\etadbdb}[2]{\eta_{\bar{#1}\bar{#2}}}

\newcommand{\Fourvectoru}[2]{{#1}^{#2}}
\newcommand{\Fourvectortu}[2]{\tilde{#1}^{#2}}
\newcommand{\Fourvectorua}[3]{{#1}^{~#2}_{\scriptscriptstyle{#3}}}
\newcommand{\Fourvectortua}[3]{\tilde{#1}^{~#2}_{\scriptscriptstyle{#3}}}

\newcommand{\Fourvectoruas}[4]{{#1}^{~#2}_{{\scriptscriptstyle{#3}},{#4}}}
\newcommand{\Fourvectord}[2]{{#1}_{#2}}
\newcommand{\Fourvectorda}[3]{{#1}_{{\scriptscriptstyle #3}#2}}

\newcommand{\fourvectoru}[2]{\mathcal{#1}^{#2}}
\newcommand{\fourvectortu}[2]{\tilde{\mathcal{#1}}^{#2}}
\newcommand{\fourvectorua}[3]{\mathcal{#1}^{~#2}_{\scriptscriptstyle\mathcal{#3}}}
\newcommand{\fourvectortua}[3]{\tilde{\mathcal{#1}}^{~#2}_{\scriptscriptstyle\mathcal{#3}}}

\newcommand{\fourvectord}[2]{\mathcal{#1}_{#2}}

\newcommand{\fourvectorub}[2]{\mathcal{#1}^{\bar{#2}}}

\newcommand{\fourvectoruh}[2]{\mathcal{#1}^{\hat{#2}}}
\newcommand{\fourvectoruha}[3]{\mathcal{#1}^{~\hat{#2}}_{\scriptscriptstyle\mathcal{#3}}}

\newcommand{\fourvelocityLd}[1]{u_{#1}}
\newcommand{\fourvelocityLu}[1]{u^{#1}}
\newcommand{\fourvelocityLdb}[1]{u_{\bar{#1}}}
\newcommand{\fourvelocityLub}[1]{u^{\bar{#1}}}
\newcommand{\fourvelocityLdh}[1]{u_{\hat{#1}}}
\newcommand{\fourvelocityLuh}[1]{u^{\hat{#1}}}
\newcommand{\fourvelocityEd}[1]{n_{#1}}
\newcommand{\fourvelocityEu}[1]{n^{#1}}
\newcommand{\genasis}{GenASiS}
\newcommand{\gmuu}[2]{\gamma^{#1#2}}
\newcommand{\gmdd}[2]{\gamma_{#1#2}}
\newcommand{\gmbuu}[2]{\bar{\gamma}^{#1#2}}
\newcommand{\gmbdd}[2]{\bar{\gamma}_{#1#2}}
\newcommand{\guu}[2]{g^{#1#2}}
\newcommand{\gdd}[2]{g_{#1#2}}

\newcommand{\mdet}{\sqrt{-g}}
\newcommand{\smdet}{\sqrt{\gamma}}
\newcommand{\smbdet}{\sqrt{\bar{\gamma}}}
\newcommand{\puh}[1]{p^{\hat{#1}}}
\newcommand{\pdh}[1]{p_{\hat{#1}}}
\newcommand{\pub}[1]{p^{\bar{#1}}}
\newcommand{\pu}[1]{p^{#1}}
\newcommand{\pAbs}{|\vect{p}|}
\newcommand{\uuh}[1]{u^{\hat{#1}}}
\newcommand{\nuh}[1]{n^{\hat{#1}}}
\newcommand{\vbub}[1]{\bar{v}^{\bar{#1}}}
\newcommand{\vbdb}[1]{\bar{v}_{\bar{#1}}}
\newcommand{\vbuh}[1]{\bar{v}^{\hat{#1}}}
\newcommand{\vbdh}[1]{\bar{v}_{\hat{#1}}}
\newcommand{\lorentzApp}{\tilde{W}}
\newcommand{\luhd}[2]{{L}^{\hat{#1}}_{\mbox{ }#2}}
\newcommand{\ludh}[2]{{L}^{#1}_{\mbox{ }\hat{#2}}}
\newcommand{\mathcalh}[1]{\hat{\mathcal{#1}}}

\newcommand{\mathcalt}[1]{\tilde{\mathcal{#1}}}
\newcommand{\mathcalhmcd}[2]{\hat{\mathcal{#1}}_{\scriptscriptstyle{\mathcal{#2}}}}
\newcommand{\mathcaltmcd}[2]{\tilde{\mathcal{#1}}_{\scriptscriptstyle{\mathcal{#2}}}}
\newcommand{\mathcalmcd}[2]{\mathcal{#1}_{\scriptscriptstyle{\mathcal{#2}}}}

\newcommand{\Tensordd}[3]{{#1}_{{#2}{#3}}}
\newcommand{\Tensoruu}[3]{{{#1}}^{{#2}{#3}}}
\newcommand{\Tensoruua}[4]{{#1}^{~{#2}{#3}}_{\scriptscriptstyle{#4}}}

\newcommand{\Tensorud}[3]{{#1}^{#2}_{~#3}}
\newcommand{\Tensoruda}[4]{{#1}_{{\scriptscriptstyle #4}~#3}^{~#2}}

\newcommand{\Tensoruuu}[4]{{{#1}}^{{#2}{#3}{#4}}}
\newcommand{\tensordd}[3]{\mathcal{#1}_{{#2}{#3}}}

\newcommand{\tensoruu}[3]{\mathcal{#1}^{{#2}{#3}}}

\newcommand{\tensorud}[3]{\mathcal{#1}^{{#2}}_{~{#3}}}

\newcommand{\tensordu}[3]{\mathcal{#1}_{{#2}}^{~{#3}}}

\newcommand{\tensorubub}[3]{\mathcal{#1}^{\bar{#2}\bar{#3}}}

\newcommand{\tensoruhuh}[3]{{\mathcal{#1}}^{\hat{#2}\hat{#3}}}

\newcommand{\tensoruuu}[4]{\mathcal{{#1}}^{{#2}{#3}{#4}}}

\newcommand{\tensorduu}[4]{\mathcal{{#1}}_{{#2}}^{~{#3}{#4}}}
\newcommand{\tensordud}[4]{\mathcal{{#1}}_{{#2}~{#4}}^{~{#3}}}

\newcommand{\tensorddu}[4]{\mathcal{{#1}}_{{#2}{#3}}^{~~{#4}}}
\newcommand{\tensorububub}[4]{\mathcal{#1}^{\bar{#2}\bar{#3}\bar{#4}}}
\newcommand{\tensoruhuhuh}[4]{{\mathcal{#1}}^{\hat{#2}\hat{#3}\hat{#4}}}

\newcommand{\threevelocityEd}[1]{v_{#1}}
\newcommand{\threevelocityEu}[1]{v^{#1}}
\newcommand{\tetudb}[2]{e^{#1}_{~\bar{#2}}}
\newcommand{\tetubd}[2]{e^{\bar{#1}}_{~#2}}

\newcommand{\eddingtontensoruhuh}[3]{{{#1}}^{\hat{#2}\hat{#3}}}
\newcommand{\eddingtontensoruhuhuh}[4]{{{#1}}^{\hat{#2}\hat{#3}\hat{#4}}}

\newcommand{\mathcaluh}[2]{\mathcal{#1}^{\hat{#2}}}

\newcommand{\mathcalu}[2]{\mathcal{#1}^{#2}}

\newcommand{\mathcaluhuh}[3]{\mathcal{#1}^{\hat{#2}\hat{#3}}}

\newcommand{\f}[2]{\frac{#1}{#2}}
\newcommand{\vect}[1]{\boldsymbol{#1}}
\newcommand{\pderiv}[2]{\frac{\partial #1}{\partial #2}}
\newcommand{\psit}{\tilde{\psi}}
\newcommand{\cderiv}[2]{\frac{D #1}{D #2}}
\newcommand{\tauderiv}[1]{\frac{\partial #1}{\partial\tau}}

\newcommand{\nue}{\nu_{e}}
\newcommand{\nueb}{\bar{\nu}_{e}}

\defcitealias{cardallMezzacappa_2003}{CM03}
\defcitealias{muller_etal_2010}{MJD10}

\shorttitle{CONSERVATIVE RADIATION MOMENT EQUATIONS}
\shortauthors{Endeve et al.}

\begin{document}

\title{Conservative Moment Equations for Neutrino Radiation Transport with Limited Relativity}

\author{Eirik Endeve\altaffilmark{1}, Christian Y. Cardall\altaffilmark{2,3}, and Anthony Mezzacappa\altaffilmark{1,2,3}}

\altaffiltext{1}{Computer Science and Mathematics Division, Oak Ridge National Laboratory, Oak Ridge, TN 37831-6354, USA; endevee@ornl.gov}
\altaffiltext{2}{Physics Division, Oak Ridge National Laboratory, Oak Ridge, TN 37831-6354, USA}
\altaffiltext{3}{Department of Physics and Astronomy, University of Tennessee, Knoxville, TN 37996-1200, USA}

\begin{abstract}
We derive conservative, multidimensional, energy-dependent moment equations for neutrino transport in core-collapse supernovae and related astrophysical systems, with particular attention to the consistency of conservative four-momentum and lepton number transport equations.  
After taking angular moments of conservative formulations of the general relativistic Boltzmann equation, we specialize to a conformally flat spacetime, which also serves as the basis for four further limits.  
Two of these---the multidimensional special relativistic case, and a conformally flat formulation of the spherically symmetric general relativistic case---are given in appendices for the sake of comparison with extant literature.  
The third limit is a weak-field, `pseudo-Newtonian' approach \citep{kim_etal_2009,kim_etal_2012} in which the source of the gravitational potential includes the trace of the stress-energy tensor (rather than just the mass density), and all orders in fluid velocity $v$ are retained.
Our primary interest here is in the fourth limit: `$\mathcal{O}(v)$' moment equations for use in conjunction with Newtonian self-gravitating hydrodynamics.  
We show that the concept of `$\mathcal{O}(v)$' transport requires care when dealing with both conservative four-momentum and conservative lepton number transport, and present two self-consistent options: `$\mathcal{O}(v)$-plus' transport, in which an $\mathcal{O}(v^2)$ energy equation combines with an $\mathcal{O}(v)$ momentum equation to give an $\mathcal{O}(v^2)$ number equation; and `$\mathcal{O}(v)$-minus' transport, in which an $\mathcal{O}(v)$ energy equation combines with an $\mathcal{O}(1)$ momentum equation to give an $\mathcal{O}(v)$ number equation.  
\end{abstract}


\section{INTRODUCTION}
\label{sec:introduction}

Detailed simulations of neutrino transport are central to understanding the core-collapse supernova explosion mechanism \citep[e.g.,][]{mezzacappa_2005,woosleyJanka_2005,kotake_etal_2006,janka_2012}.  
Neutrinos of all flavors (i.e., electron, mu, and tau neutrinos and antineutrinos) carry away $\sim99\%$ of the gravitational energy released ($\sim10^{53}$~erg) during core collapse of a massive star ($M\gtrsim 8~M_{\odot}$).  
As the dominant energy source (in the case of slowly-rotating progenitors), it is likely that the neutrinos streaming from the neutrinospheres of a proto-neutron star (PNS; the collapsed core of a massive star) deposit enough energy into the stellar fluid to power the supernova: about 5-10\% ($\sim10^{51}$~erg) of the neutrino energy emitted in the first second after core bounce ($\gtrsim10^{52}$~erg) must be transferred to give rise to a neutrino-driven supernova.  

Certainly, robust and accurate numerical methods must be deployed.  
Numerical simulations involving multi-physics codes (coupling neutrino radiation transport, magnetohydrodynamics, nuclear reaction kinetics and equations of state, and gravity, preferably---and ultimately---in general relativity) must be evolved for $\sim\mathcal{O}(10^{5}$-$10^{6})$ timesteps to study in detail the energy transfer from the neutrino radiation field to the supernova matter, and the onset and development of the explosion.  
In particular, conserved quantities (i.e., total energy and lepton number) must be preserved within tolerable margins, before firm conclusions can be drawn from the simulation outcomes.  

So far, only spherically symmetric simulations have reached the desired level of realism in neutrino transport, where general relativistic, angle- and energy-dependent neutrino radiation hydrodynamics simulations can be performed \citep[e.g.,][]{liebendorfer_etal_2001,liebendorfer_etal_2004,sumiyoshi_etal_2005}.  
However, spherically symmetric simulations do not result in neutrino-driven explosions \citep[][except in the case of the lightest progenitors with O-Ne-Mg cores; \citealt{kitaura_etal_2006}]{ramppJanka_2000,liebendorfer_etal_2001,thompson_etal_2003}, and, based on results from 2D (axisymmetric) simulations, there is now an emerging consensus that multidimensional effects due to hydrodynamics instabilities are important in shaping the core-collapse supernova explosion \citep[e.g.,][]{herant_etal_1994,burrows_etal_1995,jankaMuller_1996,fryerWarren_2002,burrows_etal_2006,burrows_etal_2007,bruenn_etal_2009,marekJanka_2009,suwa_etal_2010,muller_etal_2012a,muller_etal_2012b,bruenn_etal_2012}.  

Neutrino transport in multidimensional simulations is far less mature.  
Discrepant results from 2D simulations by different research groups, in part due to the physics included in the models and the approximations made, remain to be fully investigated.  
It appears clear, however, that energy-dependent neutrino transport (including ``observer corrections" due to fluid motions and inelastic scattering; e.g., \citealt{lentz_etal_2012a}) as well as a comprehensive set of weak-interactions are required \citep[e.g.,][]{lentz_etal_2012b}.  
3D simulations are still in their infancy, although recent progress has been made \citep[e.g.,][]{fryerYoung_2007,takiwaki_etal_2012}.  

Current approaches to neutrino transport in multidimensional simulations include the so-called ray-by-ray approximation with the multigroup (energy-dependent) flux-limited diffusion approximation \citep[e.g.,][]{bruenn_etal_2009,bruenn_etal_2012}, the two-moment model with approximate Boltzmann closure \citep[e.g.,][]{marekJanka_2009,muller_etal_2012a}, and the so-called isotropic diffusion source approximation \citep[e.g.,][]{liebendorfer_etal_2009,suwa_etal_2010}.  
Other recent approaches to neutrino radiation transport include multidimensional, multigroup flux-limited diffusion---valid to $\mathcal{O}(v)$ \citep[e.g.,][]{swestyMyra_2009,zhang_etal_2012} or to $\mathcal{O}(1)$ \citep[e.g.,][]{burrows_etal_2006}---or multidimensional, energy-integrated (grey) or multigroup two-moment models with analytic closure \citep[e.g.,][]{kuroda_etal_2012,obergaulingerJanka_2011,shibata_etal_2011}.  
\citet{ott_etal_2008} simulated the post-bounce phase of core-collapse supernovae in axial symmetry using $\mathcal{O}(1)$ multigroup and multiangle neutrino transport \cite[see also the recent developments by][]{sumiyoshiYamada_2012}.  

Lack of sufficient computational resources is the primary reason for the various approximations made in multidimensional simulations of neutrino transport in core-collapse supernovae.  
Gravitational collapse, a steep density cliff at the contracting surface of the PNS, and regions of post-shock turbulence recommend regions of high spatial resolution.  
Moreover, differences in the evolution of important hydrodynamic phenomena in two and three spatial dimensions---for example, the Standing Accretion Shock Instability \citep[e.g.,][]{blondin_etal_2003,blondinMezzacappa_2007}, or turbulence \citep[][]{ishihara_etal_2009,boffettaEcke_2012}---suggest that simulations should ideally (and eventually) be performed in 3D.  
Recent explorations using so-called neutrino light-bulb models emphasize differences between 2D and 3D simulations \citep[e.g.,][]{nordhaus_etal_2010,hanke_etal_2012,burrows_etal_2012,couch_etal_2012}.  
Models based on evolving angular moments of the neutrino distribution function are therefore attractive for 3D simulations, where the solution space must be reduced.  

With an eye toward future multidimensional simulations of angle- and energy-dependent neutrino transport, \citet[][hereafter \citetalias{cardallMezzacappa_2003}]{cardallMezzacappa_2003} derived \emph{conservative} general relativistic kinetic equations intended for implementation in numerical codes for simulations of neutrino transport in core-collapse supernovae.  
They adopted distinct position and momentum space coordinates, by employing spacetime position coordinates associated with a global coordinate basis and momentum space coordinates associated with an orthonormal comoving basis.  
The kinetic equations derived by \citetalias{cardallMezzacappa_2003} describe the evolution of the specific particle number density and four-momentum (six-dimensional phase space densities).  
Such conservative equations allow for convenient treatment of neutrino-matter interactions in the comoving-frame, and may be particularly useful when constructing numerical methods for the multidimensional Boltzmann equation with desirable conservation properties (i.e., conserves total lepton number and energy).  
However, Boltzmann neutrino transport in 3D core-collapse supernova simulations will not be possible for some time to come (exascale computational resources will be required, even with moderate phase space resolution), and approximate methods will continue to play a dominant role in the foreseeable future.  

In preparation for development of numerical methods for neutrino radiation transport in our astrophysical simulation code \genasis~\citep{cardall_etal_2012a,cardall_etal_2012b}, we derive \emph{conservative}, multidimensional, monochromatic moment equations for neutrino radiation transport from first principles, employing the general relativistic framework of \citetalias{cardallMezzacappa_2003}.  
We derive a two-moment model, which evolves angular moments of the neutrino distribution function; i.e., the neutrino energy and momentum (as opposed to only the energy in the flux-limited diffusion approach).  
The radiation moments are functions of spacetime position components associated with a global coordinate basis $x^{\mu}$ and the radiation energy measured by a comoving observer $\epsilon$.  
First we specialize the moment equations to the case of a conformally flat spacetime.  
Then we specialize the equations further to the pseudo-Newtonian \citep[cf.][]{kim_etal_2012} and the Newtonian gravity, $\mathcal{O}(v)$ limits.  
Special relativistic equations are presented in Appendix \ref{app:srMomentEquations}, while we present general relativistic moment equations for spherically symmetric spacetimes in Appendix \ref{app:momentEquationsSphericalSymmetryCFC}.  
(General relativistic moment equations employing the 3+1 formulation have been presented separately in \citealt{cardall_etal_2012}; see also \citealt{shibata_etal_2011}.)  
We elucidate the relationship between the moment equations for neutrino four-momentum and neutrino number in general \citep[see also][]{cardall_etal_2012}, and we pay special attention to consistency between the conservative moment equations for neutrino four-momentum and the conservative equation for the neutrino number density in the non-relativistic limits.  
Detailed knowledge of this relationship may be useful when constructing consistent numerical methods for neutrino radiation transport, where lepton number conservation (in addition to total energy conservation) must also be considered.  
Our conservative moment equations correspond to similar \emph{non-conservative} moment equations presented by other authors \citep[e.g.,][]{buchler_1979,buchler_1983,buchler_1986,kaneko_etal_1984,munierWeaver_1986b,muller_etal_2010}.  

The $\mathcal{O}(v)$ limit of the radiation moment equations is not uniquely defined, and care should be taken to ensure consistency between the conservative equations for the radiation four-momentum and the neutrino number equation \citep[see also][for a detailed discussion]{cardall_etal_2005}.  
In the non-relativistic limit, we find that exact consistency between the conservative moment equations for neutrino four-momentum and the conservative neutrino number equation can be obtained by adopting different orders of $v$ for the radiation energy and momentum equations, respectively.  
As an illustrative example, in the laboratory frame the neutrino number density $\mathcalmcd{E}{N}$ and the neutrino energy density $\mathcal{E}$ and momentum density $\fourvectoru{F}{i}$ are to $\mathcal{O}(v)$ related by
\begin{equation}
  \epsilon\,\mathcalmcd{E}{N}
  =\mathcal{E}-\threevelocityEd{i}\,\fourvectoru{F}{i},
\end{equation}
where $\threevelocityEd{i}$ is the fluid three-velocity.  
The evolution equations for the respective quantities are similarly related.  
By approximating the conservative (lab-frame) radiation energy and momentum equations to $\mathcal{O}(v)$ and $\mathcal{O}(1)$, respectively, we obtain exact consistency with the conservative $\mathcal{O}(v)$ neutrino number equation, and we avoid the introduction of `extra' $\mathcal{O}(v^{2})$ terms.  
However, because we use an $\mathcal{O}(1)$ momentum equation, this description is formally only accurate to $\mathcal{O}(1)$ (we call this the $\mathcal{O}(v)$-minus approximation to indicate that velocity-dependent terms are included, and to distinguish it from `traditional' $\mathcal{O}(1)$ descriptions, where no moving fluid effects are included).  
We obtain a description that is formally accurate to $\mathcal{O}(v)$ by adopting $\mathcal{O}(v^{2})$ and $\mathcal{O}(v)$ radiation energy and momentum equations, respectively.  
This description is exactly consistent with the conservative $\mathcal{O}(v^{2})$ neuterino number equation.  
We refer to this system of moment equations as the $\mathcal{O}(v)$-plus radiation moment equations.  

We have organized the paper as follows: 
we briefly summarize general relativistic kinetic equations in Section \ref{sec:kineticEquations}.  
Angular moments of the distribution are defined in Section \ref{sec:angularMoments}, and we derive conservative evolution equations for the so-called zeroth- and first-order angular moments in Section \ref{sec:momentEquationsDerivation}.  
(We derive conservative monochromatic equations for the neutrino number density \emph{and} four-momentum.)  
We also elucidate the relationship between the stress-energy equation and the number equation in Section \ref{sec:momentEquationsDerivation}.  
In Section \ref{sec:momentEquationsCFC} we specialize the moment equations for a sufficiently general (conformally flat) spacetime metric to accommodate the pseudo-Newtonian radiation moment equations; the Newtonian gravity, $\mathcal{O}(v)$ equations; the special relativistic moment equations; and the general relativistic moment equations for spherically symmetric spacetimes.  
The pseudo-Newtonian moment equations are presented in Section \ref{sec:pseudoNewtonian}.  
We introduce the Newtonian gravity, $\mathcal{O}(v)$-plus and $\mathcal{O}(v)$-minus approximations in Section \ref{sec:newtonianOrderV}.  
A summary and discussion is given in Section \ref{sec:summaryAndDiscussion}.  
Special relativistic moment equations are presented in Appendix \ref{app:srMomentEquations}.  
General relativistic moment equations for spherically symmetric spacetimes are presented in Appendix \ref{app:momentEquationsSphericalSymmetryCFC}, where we also compare the equations with those solved by \citet[][]{muller_etal_2010}.  

\section{GENERAL RELATIVISTIC KINETIC EQUATIONS}
\label{sec:kineticEquations}

In this section we briefly summarize general relativistic kinetic equations, and define the variables associated with the equations \citep[see for example][for detailed treatments of relativistic kinetic theory]{lindquist_1966,ehlers_1971,israel_1972}.  
The kinetic equations form the basis for our derivation of angular moment equations for neutrino radiation transport.  
To this end, we consider a single species of massless, electrically neutral particles (i.e., classical neutrinos).  
We adopt a `geometrized' unit system in which the vacuum speed of light, the gravitational constant, and the Planck constant are unity, and we adopt the usual Einstein summation convention and let repeated Greek indices run from $0$ to $3$, and repeated Latin indices run from $1$ to $3$.  

The classical neutrino distribution function $f(x^{\mu},\puh{\mu})$---the phase space particle density---is governed by the Boltzmann equation.
In its geometric form, it states that the change in $f$ along a phase space trajectory, parametrized by $\lambda$, equals the density $\collision{f}$ of point-like collisions that add or remove particles from the trajectory; i.e.,
\begin{equation}
  \f{df}{d\lambda}=\collision{f}.
  \label{eq:BoltzmannGeometric}
\end{equation}
For practical computations it is necessary to introduce phase space coordinates.  
By introducing spacetime coordinates $x^{\mu}$, and momentum space coordinates $\puh{\imath}$, Equation (\ref{eq:BoltzmannGeometric}) becomes
\begin{equation}
  \f{dx^{\mu}}{d\lambda}\pderiv{f}{x^{\mu}}
  +\f{d\puh{\imath}}{d\lambda}\pderiv{f}{\puh{\imath}}
  =\collision{f}.
  \label{eq:BoltzmannCoordinates} 
\end{equation}
(By the mass shell constraint $\pdh{\mu}\,\puh{\mu}=0$, only three of the four-momentum components are independent, and the contraction over momentum coordinates in Equation (\ref{eq:BoltzmannCoordinates}) only runs over three indices.)  
The geodesic equations describing the trajectory can be written as
\begin{eqnarray}
  \f{dx^{\mu}}{d\lambda}
  &=&\ludh{\mu}{\mu}\,\puh{\mu}, \label{eq:geodesicEquation1} \\
  \f{d\puh{\mu}}{d\lambda}
  &=&-\cuhdhdh{\mu}{\nu}{\rho}\,\puh{\nu}\,\puh{\rho}. \label{eq:geodesicEquation2}
\end{eqnarray}
Then, the general relativistic Boltzmann equation reads (e.g., \citet[][]{lindquist_1966,mezzacappaMatzner_1989}; \citetalias{cardallMezzacappa_2003})
\begin{equation}
  \ludh{\mu}{\mu}\,\puh{\mu}\,\pderiv{f}{x^{\mu}}
  -\cuhdhdh{\imath}{\nu}{\rho}\,\puh{\nu}\,\puh{\rho}\,\pderiv{f}{\puh{\imath}}
  =\collision{f}. \label{eq:boltzmannEquation}
\end{equation}
In the above we have used our freedom in choosing distinct spacetime and momentum space coordinates: 
$\{x^{\mu}\}=\{t,\,x^{i}\}$ are spacetime position components in a \emph{global coordinate (holonomic) basis}, while $\{\puh{\mu}\}=\epsilon\,\{1,\,\nuh{\imath}\}$ are four-momentum components in a \emph{comoving orthonormal basis}, where $\nuh{\imath}$ are components of the spatial unit vector parallel to $\puh{\imath}$, with basis tangent to the spatial coordinate lines in a comoving orthonormal basis.  

A `comoving-frame' at a spacetime point is an inertial reference frame whose four-velocity instantaneously coincides with that of a moving fluid element at that point, and is the frame where interactions between the radiation field and the fluid are most easily described \citep{mihalasMihalas_1999}.  
In the comoving frame, the four-velocity of a comoving observer is $\{\fourvelocityLuh{\mu}\}=\{1,0,0,0\}$.  
Quantities with unaccented indices are defined with respect to the global coordinate basis, while quantities whose indices are accented with a hat (comoving-frame quantities) are defined with respect to the orthonormal comoving basis.  
The composite transformation $\ludh{\mu}{\mu}\equiv\tetudb{\mu}{\mu}\,\boostubdh{\mu}{\mu}$ transforms four-vectors defined with respect to the orthonormal comoving basis into four-vectors defined with respect to the global coordinate basis.  
Thus, the four-velocity of a comoving observer with respect to the global coordinate basis is $\fourvelocityLu{\mu}=\ludh{\mu}{\mu}\,\fourvelocityLuh{\mu}$.  
The composite transformation consists of a Lorentz transformation $\boostubdh{\mu}{\mu}$ from the orthonormal comoving basis (comoving-frame) to an in general noncomoving orthonormal tetrad basis (the orthonormal lab-frame; e.g., $\fourvelocityLub{\mu}=\boostubdh{\mu}{\mu}\,\fourvelocityLuh{\mu}$), followed by a transformation $\tetudb{\mu}{\mu}$ from the orthonormal tetrad basis to the global coordinate basis (e.g., $\fourvelocityLu{\mu}=\tetudb{\mu}{\mu}\,\fourvelocityLub{\mu}$).  
Quantities with indices accented with a bar (lab-frame quantities) are defined with respect to the orthonormal tetrad basis.  
The ``tetrad transformation" locally transforms the metric tensor $\gdd{\mu}{\nu}$ into the Minkowski metric tensor; $\tetudb{\mu}{\mu}\,\tetudb{\nu}{\nu}\,\gdd{\mu}{\nu}=\etadbdb{\mu}{\nu}\equiv\mbox{diag}[-1,1,1,1]$.  
(The Minkowski metric is preserved by the Lorentz transformation, and the composite transformation is also a tetrad transformation.)
The inverse composite transformation is denoted $\luhd{\mu}{\mu}\equiv\boostuhdb{\mu}{\mu}\,\tetubd{\mu}{\mu}$; i.e., $\ludh{\mu}{\mu}\,\luhd{\mu}{\nu}=\deltaud{\mu}{\nu}$, where $\deltaud{\mu}{\nu}$ is the Kronecker delta.  

The `connection coefficients' associated with the orthonormal comoving basis $\cuhdhdh{\mu}{\nu}{\rho}$ (cf. Equation (\ref{eq:boltzmannEquation})) are expressed in terms of the connection coefficients associated with the global spacetime coordinate basis as
\begin{equation}
  \cuhdhdh{\mu}{\nu}{\rho}
  =
  \luhd{\mu}{\mu}\,\ludh{\nu}{\nu}\,\ludh{\rho}{\rho}\,\cudd{\mu}{\nu}{\rho}
  +\luhd{\mu}{\mu}\,\ludh{\rho}{\rho}\,\pderiv{\ludh{\mu}{\nu}}{x^{\rho}}
  \label{eq:connectionCoefficientsComovingBasis}
\end{equation}
where the coordinate basis connection coefficients are given in terms of the spacetime metric $\gdd{\mu}{\nu}$ as
\begin{equation}
  \cudd{\mu}{\nu}{\rho}=\f{1}{2}\guu{\mu}{\sigma}
  \Big(
    \pderiv{\gdd{\nu}{\sigma}}{x^{\rho}}
    +\pderiv{\gdd{\sigma}{\rho}}{x^{\nu}}
    -\pderiv{\gdd{\nu}{\rho}}{x^{\sigma}}
  \Big),
  \label{eq:connectionCoefficientsEulerianBasis}
\end{equation}
and $\guu{\mu}{\nu}$ is the contravariant metric tensor; $\guu{\mu}{\nu}\,\gdd{\nu}{\rho}=\deltaud{\mu}{\rho}$.  
Note now that the factors outside the momentum space derivative of $f$ in the second term on the left-hand side of Equation (\ref{eq:boltzmannEquation}) are proportional to space and time derivatives of the fluid three-velocity \emph{and} components of the metric tensor.  
These terms account for Doppler and Einstein shifts.  
See for example \citet[][]{lentz_etal_2012a} for a general discussion of these terms and a demonstration of their importance in simulations of neutrino transport in core-collapse supernovae.  

The collision operator $\collision{f}$ describes the momentum space evolution of the distribution function $f$ due to neutrino-matter interactions.  
Neutrino-matter interactions crucial to realistic simulations of CCSNe include particle creation and destruction via emission and absorption, scattering (both elastic and inelastic), and neutrino pair creation and annihilation.  
We will not discuss neutrino-matter interactions in any detail in this paper.  
See, however, \citet{bruenn_1985,burrowsThompson_2004}, and the references therein for general details.
See also \citet{lentz_etal_2012b} for a recent discussion of neutrino opacities in core-collapse supernova simulations.  

Our choice of phase space coordinates is well suited for numerical solution of the Boltzmann equation (or equations based on angular moments of the Boltzmann equation).  
In Equation (\ref{eq:boltzmannEquation}), the particle distribution function is parametrized in terms of spacetime position components in a global coordinate basis $\{x^{\mu}\}$ and momentum components in an orthonormal basis comoving with the fluid $\{\puh{\imath}\}$.  
This specific choice is motivated by our intent to develop numerical methods for computer simulations of neutrino transport.  
Neutrino-matter interactions are on the one hand most easily handled computationally in the frame comoving with the fluid, where material properties are isotropic \citep[e.g.,][]{mihalasMihalas_1999}.  
On the other hand, when integrated over momentum space, the Boltzmann equation expresses conservation of particle number and energy in the laboratory frame.  
Accurate accounting of total lepton number and energy in numerical simulations of core-collapse supernovae is important and extremely challenging.  
Convenient treatment of neutrino-matter interactions \emph{and} global conservation are naturally expressed in this phase space coordinate basis.  
However, Equation (\ref{eq:boltzmannEquation}) is not in conservative form, and, from a practical standpoint, may not be the best starting point for designing numerical methods to be implemented in simulation codes.  
Numerical methods based on the solution of the Boltzmann equation as it is expressed in Equation (\ref{eq:boltzmannEquation}) will in general not conserve lepton number \emph{or} energy.  
Even for numerical methods based on a number-conservative formulation of the Boltzmann equation, care must be taken in the discretization process in order to ensure total lepton number and energy conservation within acceptable limits, so that firm conclusions can be drawn from the simulations \citep{liebendorfer_etal_2004}.  

To help facilitate the realization of lepton number and energy conservation in multidimensional simulations of neutrino transport, \citetalias{cardallMezzacappa_2003} derived \emph{conservative} general relativistic formulations of kinetic theory.  
In particular, their number-conservative reformulation of Equation (\ref{eq:boltzmannEquation}) \citepalias[cf. Eq. (168) in][]{cardallMezzacappa_2003} reads
\begin{equation}
  \mathbb{S}_{\scriptscriptstyle N}[f]+\mathbb{M}_{\scriptscriptstyle N}[f]=\collision{f},
  \label{eq:boltzmannEquationNumber}
\end{equation}
where the spacetime divergence is
\begin{equation}
  \mathbb{S}_{\scriptscriptstyle N}[f]
  =\f{1}{\mdet}\f{\partial}{\partial x^{\mu}}\Big(\mdet\,\ludh{\mu}{\mu}\,\puh{\mu}\,f\Big), 
\end{equation}
and the momentum space divergence is
\begin{eqnarray}
  & &
  \mathbb{M}_{\scriptscriptstyle N}[f]
  =-\pAbs\Big|\det{\Big[\pderiv{\vect{p}}{\vect{u}}\Big]}\Big|^{-1}\times \nonumber \\
  & & \hspace{0.0in}
  \times
  \pderiv{}{\uuh{\imath}}
  \Big(
    \f{1}{\pAbs}\Big|\det{\Big[\pderiv{\vect{p}}{\vect{u}}\Big]}\Big|
    \cuhdhdh{\jmath}{\mu}{\nu}\,\pderiv{\uuh{\imath}}{\puh{\jmath}}\,\puh{\mu}\,\puh{\nu}\,f
  \Big).
\end{eqnarray}
The determinant of the metric tensor $\gdd{\mu}{\nu}$ is denoted $g$.  
Similarly, the four-momentum-conservative reformulation of Equation (\ref{eq:boltzmannEquation}) \citepalias[cf. Eq. (169) in][]{cardallMezzacappa_2003} reads
\begin{equation}
  \mathbb{S}_{\scriptscriptstyle T}^{\mu}[f]+\mathbb{M}_{\scriptscriptstyle T}^{\mu}[f]=\ludh{\mu}{\mu}\,\puh{\mu}\,\collision{f},
  \label{eq:boltzmannEquationStressEnergy}
\end{equation}
where the spacetime divergence is
\begin{eqnarray}
  \mathbb{S}_{\scriptscriptstyle T}^{\mu}[f]
  &=&\f{1}{\mdet}\pderiv{}{x^{\nu}}
  \Big(
    \mdet\,\ludh{\mu}{\mu}\,\ludh{\nu}{\nu}\,\puh{\mu}\,\puh{\nu}\,f
  \Big) \nonumber \\
  & & \hspace{0.15in}
  +\cudd{\mu}{\rho}{\nu}\,\ludh{\rho}{\rho}\,\ludh{\nu}{\nu}\,\puh{\rho}\,\puh{\nu}\,f, 
\end{eqnarray}
and the momentum space divergence is
\begin{eqnarray}
  & &
  \mathbb{M}_{\scriptscriptstyle T}^{\mu}[f]
  =-\pAbs\Big|\det{\Big[\pderiv{\vect{p}}{\vect{u}}\Big]}\Big|^{-1}\times \nonumber \\
  & & \hspace{0.0in}
  \times\pderiv{}{\uuh{\imath}}
  \Big(
    \f{1}{\pAbs}\Big|\det{\Big[\pderiv{\vect{p}}{\vect{u}}\Big]}\Big|
    \cuhdhdh{\jmath}{\nu}{\rho}\,\pderiv{\uuh{\imath}}{\puh{\jmath}}\,\ludh{\mu}{\mu}\puh{\mu}\,\puh{\nu}\,\puh{\rho}\,f
  \Big).
\end{eqnarray}
Equations (\ref{eq:boltzmannEquationNumber}) and (\ref{eq:boltzmannEquationStressEnergy}) are conservative in the sense that, when integrated over momentum space, the momentum space derivatives vanish, and the resulting equations---expressing number and four-momentum balance, respectively---are familiar from position space conservation law theory \citepalias[see][for a detailed discussion]{cardallMezzacappa_2003}.  
Note that Equations (\ref{eq:boltzmannEquationNumber}) and (\ref{eq:boltzmannEquationStressEnergy}) are merely conservative (nontrivial) reformulations of Equation (\ref{eq:boltzmannEquation}).  
Thus, Equations (\ref{eq:boltzmannEquationNumber}) and (\ref{eq:boltzmannEquationStressEnergy}) are, of course, not independent.  
While a numerical method is based on solving one of the equations, the numerical solution should ideally remain consistent with both.  
A formidable challenge for future efforts will be to construct a discrete representation of the multidimensional Boltzmann equation (Equation (\ref{eq:boltzmannEquationNumber}) or Equation (\ref{eq:boltzmannEquationStressEnergy})) that is both number \emph{and} energy conservative (i.e., simultaneously consistent with Eqs. (\ref{eq:boltzmannEquationNumber}) and (\ref{eq:boltzmannEquationStressEnergy}))---as was done by \citet{liebendorfer_etal_2004} in the spherically symmetric case.  
The conservative formulations provided by Equations (\ref{eq:boltzmannEquationNumber}) and (\ref{eq:boltzmannEquationStressEnergy}) may be helpful for this task.  
Numerical methods based on solving for angular moments of the distribution function should also retain this ``dual consistency."  
In this paper we will discuss requirements and prospects for constructing numerical methods for neutrino transport in CCSNe based on moment models that simultaneously conserve total lepton number and total energy.  

In Equations (\ref{eq:boltzmannEquationNumber}) and (\ref{eq:boltzmannEquationStressEnergy}), a change to spherical momentum space coordinates $\{\uuh{\imath}\}=\{\epsilon,\,\vartheta,\,\varphi\}$ has been facilitated.  
The Cartesian momentum components are expressed in terms of the spherical momentum space coordinates by $\{\puh{\imath}\}=\epsilon\,\{n^{\hat{1}},\,n^{\hat{2}},\,n^{\hat{3}}\}=\epsilon\,\{\cos\vartheta,\,\sin\vartheta\cos\varphi,\,\sin\vartheta\sin\varphi\}$.  The Jacobian matrix associated with the transformation is 
\begin{eqnarray}
  \pderiv{\puh{\imath}}{\uuh{\jmath}}
  =
  \left(\,
    \begin{array}{ccccc}
    \cos\vartheta\,&&\,-\epsilon\sin\vartheta\,&&\,0 \\
    \sin\vartheta\cos\varphi\,&&\,\epsilon\cos\vartheta\cos\varphi\,&&\,-\epsilon\sin\vartheta\sin\varphi \\
    \sin\vartheta\sin\varphi\,&&\,\epsilon\cos\vartheta\sin\varphi\,&&\,\epsilon\sin\vartheta\cos\varphi
    \end{array}
  \,\right)
  \label{eq:jacobianDpDu}
\end{eqnarray}
whose inverse is
\begin{eqnarray}
  \pderiv{\uuh{\imath}}{\puh{\jmath}}
  =
  \f{1}{\epsilon}
  \left(\,
    \begin{array}{ccccc}
    \epsilon\cos\vartheta\,&&\,\epsilon\sin\vartheta\cos\varphi\,&&\,\epsilon\sin\vartheta\sin\varphi \\
    -\sin\vartheta\,&&\,\cos\vartheta\cos\varphi\,&&\,\cos\vartheta\sin\varphi \\
    0\,&&\,-\sin\varphi/\sin\vartheta\,&&\,\cos\varphi/\sin\vartheta
    \end{array}
  \,\right).  
  \label{eq:jacobianDuDp}
\end{eqnarray}
The particle energy measured by a comoving observer (assuming massless neutrinos) is $\pAbs=\epsilon$, and the determinant of the Jacobian matrix $(\partial\puh{\imath}/\partial \uuh{\jmath})$ is
\begin{equation}
  \Big|\det\Big[\pderiv{\vect{p}}{\vect{u}}\Big]\Big|=\epsilon^{2}\sin\vartheta.  
\end{equation}

This concludes our summary of general relativistic kinetic equations.  

\section{ANGULAR MOMENTS OF THE DISTRIBUTION FUNCTION}
\label{sec:angularMoments}

The distribution function $f(x^{\mu},\puh{\imath})$ provides a detailed statistical description of the particle momentum at a given spacetime location.  
However, as a seven-dimensional object it is in general computationally prohibitive to compute \emph{with sufficient phase space resolution} in multidimensional supernova simulations using currently available computer hardware.  
The solution space must be reduced.  
Instead of solving the Boltzmann equation directly for the particle distribution function, solving equations for moments of the distribution function has become a popular method of reducing the dimensionality of the transport problem to a computationally tractable one \citep[e.g.,][]{lindquist_1966,andersonSpiegel_1972,castor_1972,thorne_1981,munierWeaver_1986b,mihalasMihalas_1999}.  
Variants of so-called moment models (with various degrees of sophistication) have been used extensively for simulation of neutrino transport in core-collapse supernovae \citep[e.g.,][]{bruenn_1985,herant_etal_1994,ramppJanka_2002,bruenn_etal_2006,buras_etal_2006a,burrows_etal_2006,burrows_etal_2007,bruenn_etal_2009,marekJanka_2009,swestyMyra_2009,muller_etal_2010,obergaulingerJanka_2011,bruenn_etal_2012}.  
Energy-dependent, general relativistic angular moment equations are presented in this paper \citep[see also][]{muller_etal_2010,shibata_etal_2011,cardall_etal_2012}.  

The neutrino heating rate in the gain region (i.e., the region behind the shock where the neutrino heating rate exceeds the neutrino cooling rate) depends quadratically on the neutrino energy spectrum \cite[e.g.,][]{mezzacappa_2005}, and mainly for this reason do we retain the energy dependence, while we integrate over the momentum directions.  
The dimensionality of the transport problem is then reduced to five (four spacetime dimensions, and one momentum space dimension).  
In particular, we wish to derive equations for the lab-frame neutrino number density and four-momentum density expressed in terms of angular moments of the distribution function defined in the comoving-frame.  
The invariant momentum space volume element is
\begin{equation}
  \f{d^{3}\vect{p}}{\pAbs}
  =\Big|\det\Big(\pderiv{\vect{p}}{\vect{u}}\Big)\Big|\f{d^{3}\vect{u}}{\pAbs}
  =\epsilon\sin\vartheta\,d\epsilon\,d\vartheta\,d\varphi.  
  \label{eq:invariantMomentumSpaceVolumeElement}
\end{equation}
It is expressed in terms of spherical comoving-frame momentum space coordinates on the far right side.  

The number-flux four-vector (integrated over the entire momentum space $V_{\vect{p}}$) is defined in terms of moments of the distribution function as \citep[e.g.,][]{lindquist_1966}
\begin{equation}
  N^{\mu}
  =\int_{V_{\vect{p}}}\pu{\mu}\,f\,\f{d^{3}\vect{p}}{\pAbs}
  =\ludh{\mu}{\mu}\int_{0}^{\infty}\mathcaluh{N}{\mu}\,\epsilon^{2}\,d\epsilon, 
  \label{eq:numberFluxFourVector}
\end{equation}
where the \emph{monochromatic} number-flux four-vector in the comoving-frame is defined in terms of angular moments of the distribution function as
\begin{equation}
  \mathcaluh{N}{\mu}
  =\f{1}{\epsilon}\int_{\Omega}\puh{\mu}\,f\,d\Omega.  
  \label{eq:numberFluxFourVectorComovingFrame}
\end{equation}
The solid angle element is $d\Omega=\sin\vartheta\,d\vartheta\,d\varphi$, and the angular integration is carried out over the unit sphere; i.e., $\int_{\Omega}\ldots\,d\Omega=\int_{0}^{2\pi}\int_{0}^{\pi}\ldots\sin\vartheta\,d\vartheta\,d\varphi$.  
The monochromatic lab-frame number-flux four-vector is related to the corresponding comoving-frame four-vector by
\begin{equation}
  \fourvectoru{N}{\mu}
  =\ludh{\mu}{\mu}\,\fourvectoruh{N}{\mu}
  =\ludh{\mu}{0}\,\mathcalmcd{J}{N}+\ludh{\mu}{\imath}\,\fourvectoruha{H}{\imath}{N}, 
  \label{eq:numberFluxFourVectorLabFrame}
\end{equation}
where the comoving-frame neutrino number density and number flux density are denoted $\mathcalmcd{J}{N}$ and $\fourvectoruha{H}{\imath}{N}$, respectively.  
We use the calligraphic font to distinguish energy-dependent radiation quantities from energy-integrated (grey) radiation quantities (e.g., $\fourvectoru{N}{\mu}$ vs. $\Fourvectoru{N}{\mu}$).  
Note that what we here call the lab-frame number-flux four-vector is a function of the coordinate basis spacetime position components $x^{\mu}$ and the neutrino energy measured by a comoving observer $\epsilon$; i.e., $\mathcalu{N}{\mu}=\mathcalu{N}{\mu}(x^{\mu},\epsilon)$.  
This parametrization of radiation transport variables was suggested by \citetalias{cardallMezzacappa_2003} \citep[see also][]{riffert_1986,mezzacappaMatzner_1989}.  
This is also the parametrization used by \citet{shibata_etal_2011} and \citet{cardall_etal_2012}.  

In a similar manner, the momentum space integrated symmetric stress-energy tensor is defined in terms of moments of the distribution function as \citep[e.g.,][]{lindquist_1966}
\begin{equation}
  \Tensoruu{T}{\mu}{\nu}
  =\int_{V_{\vect{p}}}\pu{\mu}\,\pu{\nu}\,f\,\f{d^{3}\vect{p}}{\pAbs}
  =\ludh{\mu}{\mu}\,\ludh{\nu}{\nu}
  \int_{0}^{\infty}\tensoruhuh{T}{\mu}{\nu}\,\epsilon^{2}\,d\epsilon,
  \label{eq:stressEnergyTensor}
\end{equation}
where the \emph{monochromatic} stress-energy tensor in the comoving-frame is defined in terms of angular moments of the distribution function as
\begin{equation}
  \tensoruhuh{T}{\mu}{\nu}
  =\f{1}{\epsilon}\int_{\Omega}\puh{\mu}\,\puh{\nu}\,f\,d\Omega.  
  \label{eq:stressEnergyTensorComovingFrame}
\end{equation}
The monochromatic lab-frame stress-energy tensor $\tensoruu{T}{\mu}{\nu}=\ludh{\mu}{\mu}\,\ludh{\nu}{\nu}\,\mathcaluhuh{T}{\mu}{\nu}$ is also a function of the neutrino energy measured by a comoving observer; i.e., $\tensoruu{T}{\mu}{\nu}=\tensoruu{T}{\mu}{\nu}(x^{\mu},\epsilon)$.  
Furthermore, we define the components of the monochromatic comoving-frame stress-energy tensor as
\begin{equation}
  \left(\,
    \begin{array}{ccc}
    \tensoruhuh{T}{0}{0} &\,\,& \tensoruhuh{T}{0}{\jmath} \\
    \tensoruhuh{T}{\imath}{0} &\,\,& \tensoruhuh{T}{\imath}{\jmath}
    \end{array}
  \,\right)
  =
  \left(\,
    \begin{array}{ccc}
    \mathcal{J} &\,\,& \fourvectoruh{H}{\jmath} \\
    \fourvectoruh{H}{\imath} &\,\,& \tensoruhuh{K}{\imath}{\jmath}
    \end{array}
  \,\right) \label{eq:energyFluxPressureComovingFrame}
\end{equation}
(i.e., in terms of the comoving-frame radiation energy density $\mathcal{J}$, momentum density $\fourvectoruh{H}{\imath}$, and stress $\tensoruhuh{K}{\imath}{\jmath}$, respectively).  
In terms of the monochromatic energy density, momentum density, and stress defined in Equation (\ref{eq:energyFluxPressureComovingFrame}), the monochromatic lab-frame stress-energy tensor can now be written as
\begin{equation}
  \tensoruu{T}{\mu}{\nu}
  =\ludh{\mu}{0}\,\ludh{\nu}{0}\,\mathcal{J}
  +\Big(\ludh{\mu}{\imath}\,\ludh{\nu}{0}+\ludh{\nu}{\imath}\,\ludh{\mu}{0}\Big)\fourvectoruh{H}{\imath}
  +\ludh{\mu}{\imath}\,\ludh{\nu}{\jmath}\,\tensoruhuh{K}{\imath}{\jmath}.
  \label{eq:labFrameStressEnergyInTermsOfComovingFrameMoments}
\end{equation}
We can also write $\tensoruhuh{K}{\imath}{\jmath}=\eddingtontensoruhuh{k}{\imath}{\jmath}\,\mathcal{J}$, and define the monochromatic symmetric rank-two ``variable Eddington tensor" as
\begin{equation}
  \eddingtontensoruhuh{k}{\imath}{\jmath}
  =\f{\int_{\Omega} \nuh{\imath}\,\nuh{\jmath}\,f\,d\Omega}{\int_{\Omega} f\,d\Omega}.  
  \label{eq:variableEddingtonTensorRankTwo}
\end{equation}

For the moment equations we derive in the next section it is useful to define the completely symmetric third-order moment
\begin{eqnarray}
  \Tensoruuu{U}{\mu}{\nu}{\rho}
  &=&\int_{V_{\vect{p}}}\pu{\mu}\,\pu{\nu}\,\pu{\rho}\,f\,\f{d^{3}\vect{p}}{\pAbs} \\
  &=&\ludh{\mu}{\mu}\,\ludh{\nu}{\nu}\,\ludh{\rho}{\rho}\int_{0}^{\infty}\tensoruhuhuh{U}{\mu}{\nu}{\rho}\,\epsilon^{2}\,d\epsilon, 
\end{eqnarray}
where the \emph{monochromatic} third-order comoving-frame moments are given by
\begin{equation}
  \tensoruhuhuh{U}{\mu}{\nu}{\rho}
  =\f{1}{\epsilon}\int_{\Omega}\puh{\mu}\,\puh{\nu}\,\puh{\rho}\,f\,d\Omega, 
  \label{eq:thirdOrderMomentComovingFrame}
\end{equation}
with $\tensoruuu{U}{\mu}{\nu}{\rho}=\ludh{\mu}{\mu}\,\ludh{\nu}{\nu}\,\ludh{\rho}{\rho}\,\tensoruhuhuh{U}{\mu}{\nu}{\rho}$.  
Note in particular that
\begin{equation}
  \tensoruhuhuh{U}{0}{\mu}{\nu}
  =\tensoruhuhuh{U}{\mu}{0}{\nu}
  =\tensoruhuhuh{U}{\mu}{\nu}{0}
  =\epsilon\,\tensoruhuh{T}{\mu}{\nu},
\end{equation}
and that we can expand the monochromatic third-order lab-frame moments in terms of the comoving-frame moments as
\begin{eqnarray}
  &&
  \f{1}{\epsilon}\,\tensoruuu{U}{\mu}{\nu}{\rho}
  =
  \mathcal{J}\,\ludh{\mu}{0}\,\ludh{\nu}{0}\,\ludh{\rho}{0}
  +\ludh{\mu}{\imath}\,\fourvectoruh{H}{\imath}\,\ludh{\nu}{0}\,\ludh{\rho}{0} \nonumber \\
  &&
  +\ludh{\nu}{\imath}\,\fourvectoruh{H}{\imath}\,\ludh{\mu}{0}\,\ludh{\rho}{0}
  +\ludh{\rho}{\imath}\,\fourvectoruh{H}{\imath}\,\ludh{\mu}{0}\,\ludh{\nu}{0}
  +\ludh{\mu}{\imath}\,\ludh{\nu}{\jmath}\,\tensoruhuh{K}{\imath}{\jmath}\,\ludh{\rho}{0} \nonumber \\
  &&
  +\ludh{\mu}{\imath}\,\ludh{\rho}{\jmath}\,\tensoruhuh{K}{\imath}{\jmath}\,\ludh{\nu}{0}
  +\ludh{\nu}{\imath}\,\ludh{\rho}{\jmath}\,\tensoruhuh{K}{\imath}{\jmath}\,\ludh{\mu}{0}
  +\ludh{\mu}{\imath}\,\ludh{\nu}{\jmath}\,\ludh{\rho}{k}\,\tensoruhuhuh{L}{\imath}{\jmath}{k},\,\,\,
  \label{eq:labFrameThirdOrderTensorInTermsOfComovingFrameMoments}
\end{eqnarray}
where $\tensoruhuhuh{L}{\imath}{\jmath}{k}=\eddingtontensoruhuhuh{l}{\imath}{\jmath}{k}\,\mathcal{J}$, and we have introduced the completely symmetric rank-three variable Eddington tensor
\begin{equation}
  \eddingtontensoruhuhuh{l}{\imath}{\jmath}{k}
  =\f{\int_{\Omega} \nuh{\imath}\,\nuh{\jmath}\,\nuh{k}\,f\,d\Omega}{\int_{\Omega} f\,d\Omega}.  
  \label{eq:variableEddingtonTensorRankThree}
\end{equation}

We have defined the lab-frame number-flux four-vector, the lab-frame stress-energy tensor, and the third-order lab-frame moments (a rank-three tensor).  
They are expressed in terms of angular moments of the distribution function in the comoving-frame (cf. Equations (\ref{eq:numberFluxFourVectorLabFrame}), (\ref{eq:labFrameStressEnergyInTermsOfComovingFrameMoments}), and (\ref{eq:labFrameThirdOrderTensorInTermsOfComovingFrameMoments})).  
From the definitions in Equations (\ref{eq:numberFluxFourVectorComovingFrame}) and (\ref{eq:stressEnergyTensorComovingFrame}), we clearly have $\epsilon\,\mathcaluh{N}{\mu}=\mathcaluhuh{T}{\mu}{0}$.  
This expression can be generalized to a useful covariant expression relating the number-flux four-vector and the stress-energy tensor by noting that $\tensoruhuh{T}{\mu}{0}=-\fourvelocityLdh{0}\,\tensoruhuh{T}{\mu}{0}=-\fourvelocityLdh{\nu}\,\tensoruhuh{T}{\mu}{\nu}$ \citep{cardall_etal_2012}.  
Thus, we have
\begin{equation}
  \fourvectoru{N}{\mu}
  =-\f{1}{\epsilon}\,\fourvelocityLd{\nu}\,\tensoruu{T}{\mu}{\nu}, 
  \label{eq:numberStressEnergyRelation}
\end{equation}
which is a covariant expression (valid in any frame; although $\epsilon$ is always the neutrino energy measured by a comoving observer).  
Similarly, we have
\begin{equation}
  \tensoruu{T}{\mu}{\nu}
  =-\f{1}{\epsilon}\,\fourvelocityLd{\rho}\,\tensoruuu{U}{\mu}{\nu}{\rho}, 
  \label{eq:stressEnergyThirdOrderMomentRelation}
\end{equation}
which is also a covariant expression relating the stress-energy tensor and the third-order tensor.  

\section{GENERAL RELATIVISTIC ANGULAR MOMENT EQUATIONS}
\label{sec:momentEquationsDerivation}

We derive general relativistic angular moment equations in this section.  
The main results are the number-conservative monochromatic lab-frame number equation (Equation (\ref{eq:numberMoment})), which is based on Equation (\ref{eq:boltzmannEquationNumber}), and the four-momentum-conservative monochromatic lab-frame stress-energy equation (Equation (\ref{eq:stressEnergyMoment})), which is based on Equation (\ref{eq:boltzmannEquationStressEnergy}).  
The steps are straightforward.  

\subsection{Number Equation}

From the definition of the monochromatic number-flux four-vector in Equations (\ref{eq:numberFluxFourVectorComovingFrame}) and (\ref{eq:numberFluxFourVectorLabFrame}), and the expression inside the spacetime derivative in Equation (\ref{eq:boltzmannEquationNumber}), it is clear that we obtain the evolution equation for the lab-frame neutrino number density by integrating Equation (\ref{eq:boltzmannEquationNumber}) over the unit sphere $\Omega$ and dividing by $\epsilon$.  
For the spacetime divergence we find
\begin{equation}
  \f{1}{\mdet}\pderiv{}{x^{\mu}}\Big(\mdet\,\fourvectoru{N}{\mu}\Big).  
\end{equation}

With spherical momentum space coordinates, the momentum space divergence in Equation (\ref{eq:boltzmannEquationNumber}) becomes
\begin{eqnarray}
  &&-\f{1}{\epsilon\sin\vartheta}\pderiv{}{\uuh{\imath}}
  \Big(
    \epsilon\,\sin\vartheta\,\cuhdhdh{\jmath}{\mu}{\nu}\,\pderiv{\uuh{\imath}}{\puh{\jmath}}\,\puh{\mu}\,\puh{\nu}\,f
  \Big) \nonumber \\
  &&
  =-\f{1}{\epsilon}
  \pderiv{}{\epsilon}
  \Big(
    \epsilon\,\cuhdhdh{\jmath}{\mu}{\nu}\,\pderiv{\uuh{1}}{\puh{\jmath}}\,\puh{\mu}\,\puh{\nu}\,f
  \Big) \nonumber \\
  &&\hspace{0.15in}
  -\f{1}{\sin\vartheta}
  \pderiv{}{\vartheta}
  \Big(
    \sin\vartheta\,\cuhdhdh{\jmath}{\mu}{\nu}\,\pderiv{\uuh{2}}{\puh{\jmath}}\,\puh{\mu}\,\puh{\nu}\,f
  \Big) \nonumber \\
  & &\hspace{0.15in}
  -\pderiv{}{\varphi}
  \Big(
    \cuhdhdh{\jmath}{\mu}{\nu}\,\pderiv{\uuh{3}}{\puh{\jmath}}\,\puh{\mu}\,\puh{\nu}\,f
  \Big).
  \label{eq:specificNumberMomSpcDiv}
\end{eqnarray}
We only focus on the first term (containing the derivative with respect to energy $\epsilon$) on the right-hand side of Equation (\ref{eq:specificNumberMomSpcDiv}).  
The second and third terms on the right-hand side of Equation (\ref{eq:specificNumberMomSpcDiv}) vanish upon integration over momentum space angles ($\vartheta$,$\varphi$).  
In particular, integrating over angles and dividing by $\epsilon$, the momentum space divergence becomes
\begin{equation}
  -\f{1}{\epsilon^{2}}\pderiv{}{\epsilon}
  \Big(
    \epsilon\,\cuhdhdh{0}{\mu}{\nu}\int_{\Omega}\puh{\mu}\,\puh{\nu}f\,d\Omega
  \Big), 
  \label{eq:energyDerivativeNumberEquation}
\end{equation}
where we have used the fact that
\begin{equation}
  \epsilon\,\cuhdhdh{\imath}{\mu}{\nu}\,\pderiv{\uuh{1}}{\puh{\imath}}\,\puh{\mu}\,\puh{\nu}
  =\cuhdhdh{\imath}{\mu}{\nu}\,\pdh{\imath}\,\puh{\mu}\,\puh{\nu}
  =\epsilon\,\cuhdhdh{0}{\mu}{\nu}\,\puh{\mu}\,\puh{\nu}
  \label{eq:relationWithDuDp}
\end{equation}
(cf. Equation (\ref{eq:jacobianDuDp})).  
The rightmost expression in Equation (\ref{eq:relationWithDuDp}) is derived from the mass-shell constraint ($\pdh{\mu}\,\puh{\mu}=0$; i.e., $\puh{0}$ is considered a function of the independent momentum space coordinates $\{\puh{\imath}\}$) and Equation (\ref{eq:geodesicEquation2}); cf. Equation (136) in \citetalias[][]{cardallMezzacappa_2003}.  
Then, writing the expression inside the energy derivative in Equation (\ref{eq:energyDerivativeNumberEquation}) in terms of comoving-frame angular moments of the distribution function, we have
\begin{eqnarray}
  &&
  \epsilon\,\cuhdhdh{0}{\mu}{\nu}\int_{\Omega}\puh{\mu}\,\puh{\nu}\,f\,d\Omega
  =\epsilon^{2}\,\cuhdhdh{0}{\mu}{\nu}\,\tensoruhuh{T}{\mu}{\nu} \nonumber \\
  && \hspace{0.15in}
  =-\epsilon^{2}\,\tensoruu{T}{\mu}{\nu}\,\covderiv{\mu}\luhd{0}{\nu}
  =\epsilon^{2}\,\tensoruu{T}{\mu}{\nu}\,\covderiv{\mu}u_{\nu},
  \label{eq:energySpaceNumberFlux}
\end{eqnarray}
where we have eliminated the connection coefficients associated with the orthonormal comoving basis with Equation (\ref{eq:connectionCoefficientsComovingBasis}), and introduced the (coordinate basis) four-velocity of a comoming observer $\fourvelocityLd{\nu}=\luhd{\nu}{\nu}\fourvelocityLdh{\nu}=-\luhd{0}{\nu}$ (cf. Section \ref{sec:kineticEquations}).  
The covariant derivative of a vector $A_{\nu}$ is \citep[cf.][]{landauLifshitz_1975}
\begin{equation}
  \covderiv{\mu}A_{\nu}
  =\pderiv{A_{\nu}}{x^{\mu}}-\cudd{\rho}{\nu}{\mu}A_{\rho}.  
\end{equation}

Thus, the number-conservative monochromatic lab-frame number equation, expressed explicitly in terms of angular moments of the particle distribution function in the comoving-frame (via Equations (\ref{eq:numberFluxFourVectorLabFrame}) and (\ref{eq:labFrameStressEnergyInTermsOfComovingFrameMoments})), becomes
\begin{eqnarray}
  &&
  \f{1}{\mdet}
  \pderiv{}{x^{\mu}}
  \Big(\mdet\,\fourvectoru{N}{\mu}\Big) \nonumber \\
  &&\hspace{0.15in}
  -\f{1}{\epsilon^{2}}\pderiv{}{\epsilon}
  \Big(\epsilon^{2}\,\tensoruu{T}{\mu}{\nu}\,\covderiv{\mu}\fourvelocityLd{\nu}\Big)
  =\f{1}{\epsilon}\int_{\Omega}\collision{f}\,d\Omega.  
  \label{eq:numberMoment}
\end{eqnarray}
Indeed, an integration over energy shells results in
\begin{equation}
  \f{1}{\mdet}\pderiv{}{x^{\mu}}
  \Big(\mdet\,N^{\mu}\Big)
  =\int_{V_{\vect{p}}}\collision{f}\f{d^{3}\vect{p}}{\pAbs}
  \label{eq:number}
\end{equation}
\citep[cf.][]{lindquist_1966}.  

\subsection{Stress-Energy Equation}

Similarly, from the definition of the stress-energy tensor in Equation (\ref{eq:stressEnergyTensorComovingFrame}) and the expression inside the spacetime derivative in Equation (\ref{eq:boltzmannEquationStressEnergy}), it is clear that we obtain the evolution equation for the monochromatic lab-frame neutrino four-momentum by integrating Equation (\ref{eq:boltzmannEquationStressEnergy}) over the full solid angle $\Omega$ and dividing by $\epsilon$.  
The spacetime divergence becomes
\begin{equation}
  \f{1}{\mdet}\pderiv{}{x^{\nu}}
  \Big(\mdet\,\tensoruu{T}{\mu}{\nu}\Big)
  +\cudd{\mu}{\rho}{\nu}\,\tensoruu{T}{\rho}{\nu}.  
\end{equation}

The momentum space divergence in Equation (\ref{eq:boltzmannEquationStressEnergy}) is
\begin{eqnarray}
  &&-\f{1}{\epsilon\sin\vartheta}\pderiv{}{\uuh{\imath}}
  \Big(
    \epsilon\,\sin\vartheta\,\cuhdhdh{\jmath}{\nu}{\rho}\,\pderiv{\uuh{\imath}}{\puh{\jmath}}\,\ludh{\mu}{\mu}\,\puh{\mu}\,\puh{\nu}\,\puh{\rho}\,f
  \Big) \nonumber \\
  &&=
  -\f{1}{\epsilon}\pderiv{}{\epsilon}
  \Big(
    \epsilon\,\cuhdhdh{\jmath}{\nu}{\rho}\,\pderiv{\uuh{1}}{\puh{\jmath}}\,\ludh{\mu}{\mu}\,\puh{\mu}\,\puh{\nu}\,\puh{\rho}\,f
  \Big) \nonumber \\
  &&\hspace{0.15in}
  -\f{1}{\sin\vartheta}\pderiv{}{\vartheta}
  \Big(
    \sin\vartheta\,\cuhdhdh{\jmath}{\nu}{\rho}\,\pderiv{\uuh{2}}{\puh{\jmath}}\,\ludh{\mu}{\mu}\,\puh{\mu}\,\puh{\nu}\,\puh{\rho}\,f
  \Big) \nonumber \\
  &&\hspace{0.15in}
  -\pderiv{}{\varphi}
  \Big(
    \cuhdhdh{\jmath}{\nu}{\rho}\,\pderiv{\uuh{3}}{\puh{\jmath}}\,\ludh{\mu}{\mu}\,\puh{\mu}\,\puh{\nu}\,\puh{\rho}\,f
  \Big).  
  \label{eq:specificStressEnergyMomSpcDiv}
\end{eqnarray}
Again, we focus only on the term with the energy derivative on the right-hand side of Equation (\ref{eq:specificStressEnergyMomSpcDiv}).  
After integrating over the unit sphere and dividing by $\epsilon$, the momentum space divergence reduces to
\begin{equation}
  -\f{1}{\epsilon^{2}}\pderiv{}{\epsilon}
  \Big(
    \epsilon\,\ludh{\mu}{\mu}\,\cuhdhdh{0}{\nu}{\rho}\int_{\Omega}\puh{\mu}\,\puh{\nu}\,\puh{\rho}\,f\,d\Omega
  \Big), 
  \label{eq:energyDerivativeStressEnergyEquation}
\end{equation}
where the relations in Equation (\ref{eq:relationWithDuDp}) have been used.  
We seek to express the momentum space divergence in terms of comoving-frame angular moments of the distribution function, and write
\begin{eqnarray}
  &&
  \epsilon\,\ludh{\mu}{\mu}\,\cuhdhdh{0}{\nu}{\rho}\int_{\Omega}\puh{\mu}\,\puh{\nu}\,\puh{\rho}\,f\,d\Omega
  =\epsilon^{2}\,\ludh{\mu}{\mu}\,\cuhdhdh{0}{\nu}{\rho}\,\tensoruhuhuh{U}{\mu}{\nu}{\rho} \nonumber \\
  && \hspace{0.15in}
  =-\epsilon^{2}\,\tensoruuu{U}{\mu}{\nu}{\rho}\,\covderiv{\nu}\luhd{0}{\rho}
  =\epsilon^{2}\,\tensoruuu{U}{\mu}{\nu}{\rho}\,\covderiv{\nu}\fourvelocityLd{\rho}, 
  \label{eq:energySpaceStressEnergyFlux}
\end{eqnarray}
where we have taken steps similar to those used to arrive at Equation (\ref{eq:energySpaceNumberFlux}).  

Then, combining all the terms, the monochromatic lab-frame stress-energy equation, expressed in terms of comoving-frame angular moments of the distribution function (via Equations (\ref{eq:labFrameStressEnergyInTermsOfComovingFrameMoments}) and (\ref{eq:labFrameThirdOrderTensorInTermsOfComovingFrameMoments})), becomes
\begin{eqnarray}
  &&\f{1}{\mdet}\pderiv{}{x^{\nu}}\Big(\mdet\,\tensoruu{T}{\mu}{\nu}\Big)
  +\cudd{\mu}{\rho}{\nu}\,\tensoruu{T}{\rho}{\nu} \nonumber \\
  &&\hspace{0.15in}
  -\f{1}{\epsilon^{2}}\pderiv{}{\epsilon}\Big(\epsilon^{2}\,\tensoruuu{U}{\mu}{\nu}{\rho}\,\covderiv{\nu}\fourvelocityLd{\rho}\Big)
  =\f{1}{\epsilon}\int_{\Omega}\pu{\mu}\,\collision{f}\,d\Omega.  
  \label{eq:stressEnergyMoment}
\end{eqnarray}
Equation (\ref{eq:stressEnergyMoment}) corresponds to Equation (3.18) in \citet{shibata_etal_2011}, which was derived from the moment formalism of \citet{thorne_1981}.  
Integrating Equation (\ref{eq:stressEnergyMoment}) over comoving-frame energy shells gives the familiar result
\begin{equation}
  \f{1}{\mdet}\pderiv{}{x^{\nu}}\Big(\mdet\,\Tensoruu{T}{\mu}{\nu}\Big)
  +\cudd{\mu}{\rho}{\nu}\Tensoruu{T}{\rho}{\nu}
  =\int_{V_{\vect{p}}}\pu{\mu}\collision{f}\f{d^{3}\vect{p}}{\pAbs}
  \label{eq:stressEnergy}
\end{equation}
\citep[cf.][]{lindquist_1966}.  

By invoking the transformation from the orthonormal tetrad basis to the coordinate basis $\tetudb{\mu}{\mu}$ we can rewrite Equation (\ref{eq:stressEnergyMoment}) in an equivalent form in terms of the stress-energy tensor associated with the orthonormal tetrad basis
\begin{eqnarray}
  & &
  \f{1}{\mdet}\pderiv{}{x^{\nu}}
  \Big(
    \mdet\,\tetudb{\nu}{\nu}\,\tensorubub{T}{\mu}{\nu}
  \Big)
  +\cubdbdb{\mu}{\rho}{\nu}\,\tensorubub{T}{\rho}{\nu} \nonumber \\
  & & \hspace{0.15in}
  -\f{1}{\epsilon^{2}}\pderiv{}{\epsilon}
  \Big(
    \epsilon^{2}\,\tensorububub{U}{\mu}{\nu}{\rho}\,\covderiv{\bar{\nu}}\fourvelocityLdb{\rho}
  \Big)
  =\f{1}{\epsilon}\int_{\Omega}\pub{\mu}\,\collision{f}\,d\Omega, 
  \label{eq:stressEnergyMomentTetrad}
\end{eqnarray}
where the stress-energy tensor $\tensorubub{T}{\mu}{\nu}$ is related to the comoving-frame angular moments through the Lorentz transformation (Appendix \ref{app:srMomentEquations}), the covariant derivative of a four-vector $A_{\bar{\rho}}$ is
\begin{equation}
  \covderiv{\bar{\nu}}A_{\bar{\rho}}
  =\tetudb{\nu}{\nu}\pderiv{A_{\bar{\rho}}}{x^{\nu}}
  -\cubdbdb{\mu}{\rho}{\nu}A_{\bar{\mu}},
\end{equation}
and the connection coefficients associated with the orthonormal tetrad basis are
\begin{equation}
  \cubdbdb{\mu}{\rho}{\nu}
  =
  \tetubd{\mu}{\mu}\,\tetudb{\rho}{\rho}\,\tetudb{\nu}{\nu}\,\cudd{\mu}{\rho}{\nu}
  +\tetubd{\mu}{\mu}\,\tetudb{\nu}{\nu}\,\pderiv{\tetudb{\mu}{\rho}}{x^{\nu}}.  
  \label{eq:connectionCoefficientsTetradBasis}
\end{equation}

Going even one step further, by invoking the Lorentz transformation, we can rewrite Equation (\ref{eq:stressEnergyMomentTetrad}) in an equivalent form in terms of the stress-energy tensor associated with the orthonormal comoving basis
\begin{eqnarray}
  & &
  \f{1}{\mdet}\pderiv{}{x^{\nu}}
  \Big(
    \mdet\,\ludh{\nu}{\nu}\,\tensoruhuh{T}{\mu}{\nu}
  \Big)
  +\cuhdhdh{\mu}{\rho}{\nu}\,\tensoruhuh{T}{\rho}{\nu} \nonumber \\
  & & \hspace{0.15in}
  -\f{1}{\epsilon^{2}}\pderiv{}{\epsilon}
  \Big(
    \epsilon^{2}\,\tensoruhuhuh{U}{\mu}{\nu}{\rho}\,\covderiv{\hat{\nu}}\fourvelocityLdh{\rho}
  \Big)
  =\f{1}{\epsilon}\int_{\Omega}\puh{\mu}\,\collision{f}\,d\Omega, 
  \label{eq:stressEnergyMomentComoving}
\end{eqnarray}
where the covariant derivative of a four-vector $A_{\hat{\rho}}$ is
\begin{equation}
  \covderiv{\hat{\nu}}A_{\hat{\rho}}
  =\ludh{\nu}{\nu}\pderiv{A_{\hat{\rho}}}{x^{\nu}}
  -\cuhdhdh{\mu}{\rho}{\nu}A_{\hat{\mu}},
\end{equation}
and the connection coefficients associated with the orthonormal comoving basis are given in terms of connection coefficients associated with the orthonormal tetrad basis by
\begin{equation}
  \cuhdhdh{\mu}{\rho}{\nu}
  =
  \boostuhdb{\mu}{\mu}\,\boostubdh{\rho}{\rho}\,\boostubdh{\nu}{\nu}\,\cubdbdb{\mu}{\rho}{\nu}
  +\boostuhdb{\mu}{\mu}\,\boostubdh{\nu}{\nu}\,\tetudb{\nu}{\nu}\,\pderiv{\boostubdh{\mu}{\rho}}{x^{\nu}}, 
  \label{eq:connectionCoefficientsComovingBasisFromTetradBasis}
\end{equation}
or in terms of the connection coefficients associated with the coordinate basis in Equation (\ref{eq:connectionCoefficientsComovingBasis}).  

The covariant nature of the general relativistic moment equations is illustrated by Equations (\ref{eq:stressEnergyMoment}), (\ref{eq:stressEnergyMomentTetrad}), and (\ref{eq:stressEnergyMomentComoving}).  
Although they are analytically equivalent, Equation (\ref{eq:stressEnergyMoment}) may be preferred over Equations (\ref{eq:stressEnergyMomentTetrad}) and (\ref{eq:stressEnergyMomentComoving}) as the basis for developing a numerical method to evolve the neutrino radiation field in full general relativity employing the so-called 3+1 decomposition \citep{shibata_etal_2011,cardall_etal_2012}.  
In Section \ref{sec:momentEquationsCFC} we use Equation (\ref{eq:stressEnergyMoment}) to obtain moment equations valid for conformally flat spacetimes.  
Equations (\ref{eq:stressEnergyMoment}) and (\ref{eq:stressEnergyMomentTetrad}) are conservative equations for the lab-frame four-momentum.  
In the absence of gravity and neutrino-matter interactions, Equations (\ref{eq:stressEnergyMoment}) and (\ref{eq:stressEnergyMomentTetrad}) express exact conservation of neutrino energy \emph{and}---if Cartesian coordinates are used---momentum (Appendix \ref{app:srMomentEquations}).  
Equation (\ref{eq:stressEnergyMomentComoving}) is expressed in terms of the comoving-frame stress-energy tensor.  
Recently, \citet{muller_etal_2010} presented numerical methods for neutrino radiation transport based on equations that are closely related to Equation (\ref{eq:stressEnergyMomentComoving}) (see Appendix {\ref{app:momentEquationsSphericalSymmetryCFC}).  
Equation (\ref{eq:stressEnergyMomentComoving}) is nonconservative---even in the absence of gravity and neutrino-matter interactions---since the connection coefficients $\cuhdhdh{\mu}{\nu}{\rho}$ depend on time and space derivatives of the fluid three-velocity (cf. Equation (\ref{eq:connectionCoefficientsComovingBasisFromTetradBasis})).  
In Appendix \ref{app:momentEquationsSphericalSymmetryCFC} we use Equation (\ref{eq:stressEnergyMoment}) to obtain conservative general relativistic moment equations valid for spherically symmetric spacetimes \citep[i.e., similar to][but in conservative form]{muller_etal_2010}.  

\subsection{The Relationship Between the Number Equation and the Stress-Energy Equation}
\label{sec:numberStressEnergyEquationRelation}

In this subsection we illustrate the relationship between the moment equations for the neutrino four-momentum density and the moment equation for the neutrino number density; Equations (\ref{eq:stressEnergyMoment}) and (\ref{eq:numberMoment}), respectively.  
From Equation (\ref{eq:numberStressEnergyRelation}), it is evident how the neutrino number equation is related to the stress-energy equation.  
Thus, contracting $-\epsilon^{-1}\,\fourvelocityLd{\mu}$ with Equation (\ref{eq:stressEnergyMoment}) results in
\begin{eqnarray}
  & &
  \f{1}{\mdet}\pderiv{}{x^{\mu}}
  \Big(
    \mdet\,\fourvectoru{N}{\mu}
  \Big)
  -\f{1}{\epsilon^{2}}\pderiv{}{\epsilon}
  \Big(\epsilon^{2}\,
    \tensoruu{T}{\mu}{\nu}\,\covderiv{\mu}\fourvelocityLd{\nu}
  \Big) \nonumber \\
  & & \hspace{0.0in}
  +\f{1}{\epsilon}
  \Big(
    \tensoruu{T}{\nu}{\rho}
    +\f{1}{\epsilon}\,\fourvelocityLd{\mu}\,\tensoruuu{U}{\mu}{\nu}{\rho}
  \Big)\covderiv{\nu}\fourvelocityLd{\rho}
  =\f{1}{\epsilon}\int_{\Omega}\collision{f}\,d\Omega.
  \label{eq:stressEnergyMomentContracted}
\end{eqnarray}
To arrive at Equation (\ref{eq:stressEnergyMomentContracted}), we have used Equations (\ref{eq:numberStressEnergyRelation}) and (\ref{eq:stressEnergyThirdOrderMomentRelation}) in the first two terms on the left-hand side, and the covariant relation $\fourvelocityLd{\mu}\,\pu{\mu}=\fourvelocityLdh{\mu}\,\puh{\mu}=-\epsilon$ to obtain the collision term on the right-hand side.  
The third term on the left-hand side of Equation (\ref{eq:stressEnergyMomentContracted}) consists of the `leftover' terms after bringing $-\epsilon^{-1}\,\fourvelocityLd{\mu}$ inside the spacetime and energy derivatives, and vanishes exactly by virtue of Equation (\ref{eq:stressEnergyThirdOrderMomentRelation}).  
Thus, as expected, Equation (\ref{eq:stressEnergyMomentContracted})---the contraction of $-\epsilon^{-1}\,\fourvelocityLd{\mu}$ with Equation (\ref{eq:stressEnergyMoment})---reduces to Equation (\ref{eq:numberMoment}); i.e., the number-conservative monochromatic number equation.  
The solution to Equation (\ref{eq:stressEnergyMoment}) can thus be used to construct the solution to Equation (\ref{eq:numberMoment}) via Equation (\ref{eq:numberStressEnergyRelation}).  

When carrying out the steps to obtain the number-conservative monochromatic neutrino number equation from the four-momentum-conservative monochromatic stress-energy equation, we observe that terms emanating from the spacetime divergences and the geometry sources cancel with terms emanating from the momentum space divergences.  
The remaining terms constitute the left-hand side of the number equation.  
Such detailed knowledge about how the conservative number equation is `built into' the four-momentum conservative stress-energy equation may be useful when developing numerical methods.  
In particular, when discretizing the stress-energy equation to develop a two-moment model for neutrino transport, care should be taken to mimic the cancellations that occur in the continuum limit in order to ensure that the resulting numerical solution is also consistent with a discrete version of the conservative number equation, and thereby facilitating lepton number conservation as a `built-in' property of the numerical method.  
We will elaborate further on this issue with specific examples in Sections \ref{sec:momentEquationsCFC} and \ref{sec:newtonianOrderV}, and in Appendix \ref{app:srMomentEquations}.  

We have derived conservative general relativistic moment equations for the number density $\fourvectoru{N}{0}$ (Equation (\ref{eq:numberMoment})) and the four-momentum density $\tensoruu{T}{\mu}{0}$ (e.g., Equation (\ref{eq:stressEnergyMoment})).  
The equations are in conservative form in the sense that (modulo neutrino-matter interactions and the geometry source terms in the stress-energy equation) the time rates of change of $\fourvectoru{N}{0}$ and $\tensoruu{T}{\mu}{0}$ are governed by space and momentum space divergences.  
When integrated over comoving-frame energy bins (with $\epsilon^{2}\,d\epsilon$ as the integration measure), the equations reduce to familiar position space conservation laws (Equations (\ref{eq:number}) and (\ref{eq:stressEnergy}), respectively).  
The equations govern the evolution of quantities defined with respect to the global coordinate basis, which are explicitly expressed in terms of comoving-frame angular moments, $\mathcaluh{N}{\mu}$ and $\mathcaluhuh{T}{\mu}{\nu}$ via Equations (\ref{eq:numberFluxFourVectorLabFrame}) and (\ref{eq:labFrameStressEnergyInTermsOfComovingFrameMoments}).  
Such equations may be suitable for implementation of neutrino transport capabilities in numerical codes intended for simulations of core-collapse supernovae, where acceptable conservation of total energy and lepton number is critical.  
We have also illustrated the relationship between the number equation and the stress-energy equation.  

In terms of a hierarchy of moment equations, we have truncated the hierarchy at the level of the so-called first-order moment (the comoving-frame energy flux density $\fourvectoruh{H}{\imath}$; the comoving-frame energy density $\mathcal{J}$ is proportional to the zeroth-order moment), which results in a two-moment model for neutrino transport.  
When truncating at this particular level, the resulting moment equations involve higher-order moments (or ratios of higher-order moments to the zeroth order moment; i.e., the variable Eddington tensors) $\eddingtontensoruhuh{k}{\imath}{\jmath}$ and $\eddingtontensoruhuhuh{l}{\imath}{\jmath}{k}$, which must be determined in order to obtain a closed system of equations.  
Due to symmetry, six unique components of $\eddingtontensoruhuh{k}{\imath}{\jmath}$ and ten unique components of $\eddingtontensoruhuhuh{l}{\imath}{\jmath}{k}$ must be determined (sixteen in total) per energy group.  
The procedure to determine these higher-order moments in a closure prescription that relates the higher moments to the first two is referred to as the moment closure problem.  
(We do not address the moment closure problem in this paper.)  
Common approaches to the moment closure problem include analytic closure \citep[e.g.,][]{minerbo_1978,levermore_1984}, entropy-based closure \citep[e.g.,][]{cernohorskyBludman_1994,levermore_1996,smit_etal_2000,brunnerHolloway_2001,hauckMacclarren_2010}, and closure based on the solution of an approximate (simplified) Boltzmann equation \citep[e.g.,][]{ramppJanka_2002,muller_etal_2010}.  

\section{MOMENT EQUATIONS FOR CONFORMALLY FLAT SPACETIMES}
\label{sec:momentEquationsCFC}

The moment equations derived in Section \ref{sec:momentEquationsDerivation} are valid for general spacetime coordinate systems.  
In this section we present moment equations suitable for simulations assuming a simplified spacetime metric.  
In particular, we impose the conformal flatness condition on the spatial metric \citep[cf.][]{wilson_etal_1996,isenberg_2008}.  
The equations derived in this section are sufficiently general to accommodate the pseudo-Newtonian moment equations (Section \ref{sec:pseudoNewtonian}), the Newtonian gravity, $\mathcal{O}(v)$ approximation of the radiation moment equations (Section \ref{sec:newtonianOrderV}), the special relativistic limit (Appendix \ref{app:srMomentEquations}), and the fully general relativistic case for spherically symmetric spacetimes (Appendix \ref{app:momentEquationsSphericalSymmetryCFC}).  
\citep[See][for moment equations for radiation transport employing the full 3+1 formulation of general relativity.]{shibata_etal_2011,cardall_etal_2012}
To this end, we adopt the 3+1 form of the spacetime metric
\begin{equation}
  \gdd{\mu}{\nu}=
  \left(
    \begin{array}{ccc}
      -\alpha^{2}+\betad{k}\betau{k} &\,\,& \betad{j} \\
      \betad{i} &\,\,& \gmdd{i}{j}
    \end{array}
  \right),
  \label{eq:metricCFC}
\end{equation}
where $\alpha$, $\betau{i}$, and $\gmdd{i}{j}$ are the lapse function, the shift vector, and the spatial three-metric, respectively.  
The conformal flatness condition is imposed by performing a conformal transformation of the three-metric, $\gmdd{i}{j}=\psi^{4}\,\gmbdd{i}{j}$, where $\psi(x^{\mu})$ is the conformal factor, and insisting that the conformally related metric $\gmbdd{i}{j}$ is diagonal.  
In particular, we set $\gmbdd{i}{j}=\mbox{diag}\left[\,1,\,a^{2}(x^{1}),\,b^{2}(x^{1})\,c^{2}(x^{2})\,\right]$, where the metric functions (or scale factors) $a$, $b$, and $c$ are sufficiently general to accommodate Cartesian, spherical, or cylindrical coordinates \citep[see for example Table I in][]{cardall_etal_2005}.  
The scale factors, and hence the conformally related metric, may depend on the spatial position coordinate, but are taken here to be \emph{independent of time}.  
The lapse function, the shift vector, and the conformal factor completely determine the spacetime metric, and can be obtained by solving a system of nonlinear elliptic equations \citep{wilson_etal_1996}.  
The four-metric and its inverse lower and raise indices on four-vectors and tensors, while the spatial metric $\gmdd{i}{j}$ and its inverse $\gmuu{i}{j}$ can be used to lower and raise spatial indices of purely spatial vectors and tensors.  
The determinant of the spacetime metric is denoted $g=-\alpha^{2}\,\gamma$, where $\gamma=\psi^{12}\,(abc)^{2}$ is the determinant of the spatial metric.  

In the 3+1 formulation of general relativity, the four-dimensional spacetime is sliced into a ``stack" of spatial (spacelike) hypersurfaces of constant (global) time coordinate \citep[see for example the recent book by][]{baumgarteShapiro_2010}.  
In particular, the unit normal to the spacelike hypersurfaces is denoted $\fourvelocityEu{\mu}=\left(\,\alpha^{-1},\,-\alpha^{-1}\,\betau{i}\,\right)$.  
With the metric specified in Equation (\ref{eq:metricCFC}) we have $\fourvelocityEd{\mu}=\gdd{\mu}{\nu}\,\fourvelocityEu{\nu}=\left(\,-\alpha,\,0\,\right)$.  
Observers whose four-velocity coincides with $\fourvelocityEu{\mu}$ are at rest with respect to the spacelike slice, and are referred to as Eulerian observers \citep[e.g.,][]{banyuls_etal_1997}.  
For the transformation from the orthonormal tetrad basis to the global coordinate basis we can use
\begin{equation}
  \tetudb{\mu}{\mu}=
  \left(
  \begin{array}{cc}
    \alpha^{-1} & 0 \\
    -\alpha^{-1}\,\betau{i} & \tetudb{i}{\imath}
  \end{array}
  \right), 
  \label{eq:tetradNewtonianGravity}
\end{equation}
where $\tetudb{i}{\imath}=\psi^{-2}\,\mbox{diag}\left[\,1,\,a^{-1},\,b^{-1}c^{-1}\,\right]$ (i.e., $\tetudb{\mu}{\mu}\,\tetudb{\nu}{\nu}\,\gdd{\mu}{\nu}=\etadbdb{\mu}{\nu}$).  
With $\tetudb{\mu}{\mu}$ given by Equation (\ref{eq:tetradNewtonianGravity}), we have $\tetudb{\mu}{0}\,n_{\mu}=-1$.  
That is, the spacetime orientation of the orthonormal tetrad basis has been chosen so that $\tetudb{\mu}{0}$ is aligned with the unit normal of the spacelike hypersurfaces (i.e., $\tetudb{\mu}{0}=\fourvelocityEu{\mu}$).  
The four-velocity of a comoving observer with respect to the lab-frame coodinate basis can now be computed.  
The result is
\begin{equation}
  \fourvelocityLu{\mu}
  =\ludh{\mu}{\mu}\,\fourvelocityLuh{\mu}
  =W\left(\,\fourvelocityEu{\mu}+\threevelocityEu{\mu}\,\right), 
  \label{eq:fourvelocityComovingObserver}
\end{equation}
where $\threevelocityEu{\mu}=\left(\,0,\,\threevelocityEu{i}\,\right)$ and $\threevelocityEu{i}=\tetudb{i}{\imath}\,\vbub{\imath}$ is the coordinate basis three-velocity of the observer comoving with the fluid (cf. Appendix \ref{app:srMomentEquations}).  
Note that $\fourvelocityLd{\mu}\,\fourvelocityLu{\mu}=-W^{2}\left(\,1-\threevelocityEd{i}\,\threevelocityEu{i}\,\right)\to W^{2}=\left(\,1-\threevelocityEd{i}\,\threevelocityEu{i}\,\right)^{-1}$.  

Following \cite{cardall_etal_2012}, we form `Eulerian' and `Lagrangian' decompositions of four-vectors and tensors.  
In particular, the Eulerian and Lagrangian decompositions of the coordinate basis number-flux four-vector are
\begin{eqnarray}
  \fourvectoru{N}{\mu}
  &=&\mathcalmcd{E}{N}\,\fourvelocityEu{\mu}+\fourvectorua{F}{\mu}{N}, \label{eq:numberFluxEulerian} \\
  &=&\mathcalmcd{J}{N}\,\fourvelocityLu{\mu}+\fourvectorua{H}{\mu}{N}, \label{eq:numberFluxLagrangian}
\end{eqnarray}
where the lab-frame (or Eulerian-frame) number density and flux (the `Eulerian projections' of the number-flux four-vector) are $\mathcalmcd{E}{N}=\fourvectorub{N}{0}$ and $\fourvectorua{F}{\mu}{N}=\tetudb{\mu}{\imath}\,\fourvectorub{N}{\imath}$, respectively.  
The comoving-frame number density and flux (the `Lagrangian projections' of the number-flux four-vector) are $\mathcalmcd{J}{N}=\fourvectoruh{N}{0}$ and $\fourvectorua{H}{\mu}{N}=\ludh{\mu}{\imath}\,\fourvectoruh{N}{\imath}$ (cf. Equation (\ref{eq:numberFluxFourVectorLabFrame})).  
Note that $\fourvectorua{F}{\mu}{N}$ is orthogonal to the Eulerian observer's four-velocity ($\fourvelocityEd{\mu}\,\fourvectorua{F}{\mu}{N}=0$), while $\fourvectorua{H}{\mu}{N}$ is orthogonal to the four-velocity of the comoving observer ($\fourvelocityLd{\mu}\,\fourvectorua{H}{\mu}{N}=0$).  

Similarly, the Eulerian and Lagrangian decompositions of the coordinate basis stress-energy tensor are
\begin{eqnarray}
  \tensoruu{T}{\mu}{\nu}
  &=&\mathcal{E}\,\fourvelocityEu{\mu}\,\fourvelocityEu{\nu}
  +\fourvectoru{F}{\mu}\,\fourvelocityEu{\nu}+\fourvectoru{F}{\nu}\,\fourvelocityEu{\mu}
  +\tensoruu{S}{\mu}{\nu}, \label{eq:stressEnergyTensorEulerian} \\
  &=&\mathcal{J}\,\fourvelocityLu{\mu}\,\fourvelocityLu{\nu}
  +\fourvectoru{H}{\mu}\,\fourvelocityLu{\nu}+\fourvectoru{H}{\nu}\,\fourvelocityLu{\mu}
  +\tensoruu{K}{\mu}{\nu}, \label{eq:stressEnergyTensorLagrangian}
\end{eqnarray}
where the monochromatic lab-frame radiation energy density, flux, and stress (the Eulerian projections of the stress-energy tensor) are related to the components of the stress-energy tensor in the orthonormal tetrad basis by
\begin{equation}
  \mathcal{E}
  =\tensorubub{T}{0}{0}
  \,\mbox{,}\,
  \fourvectoru{F}{\mu}
  =\tetudb{\mu}{\imath}\,\tensorubub{T}{\imath}{0},
  \,\mbox{and }\,
  \tensoruu{S}{\mu}{\nu}
  =\tetudb{\mu}{\imath}\,\tetudb{\nu}{\jmath}\,\tensorubub{T}{\imath}{\jmath}, 
\end{equation}
respectively.  
Note that $\fourvelocityEd{\mu}\,\fourvectoru{F}{\mu}=\fourvelocityEd{\mu}\,\tensoruu{S}{\mu}{\nu}=\fourvelocityEd{\nu}\,\tensoruu{S}{\mu}{\nu}=0$.  
The monochromatic comoving-frame radiation energy density, flux, and stress (the Lagrangian projections of the stress-energy tensor) are related to the components of the stress-energy tensor in the orthonormal comoving basis by (cf. Equation (\ref{eq:labFrameStressEnergyInTermsOfComovingFrameMoments}))
\begin{equation}
  \mathcal{J}
  =\tensoruhuh{T}{0}{0}
  \,\mbox{,}\,
  \fourvectoru{H}{\mu}
  =\ludh{\mu}{\imath}\,\tensoruhuh{T}{\imath}{0},
  \,\mbox{and }\,
  \tensoruu{K}{\mu}{\nu}
  =\ludh{\mu}{\imath}\,\ludh{\nu}{\jmath}\,\tensoruhuh{T}{\imath}{\jmath}, 
\end{equation}
where $\fourvelocityLd{\mu}\,\fourvectoru{H}{\mu}=\fourvelocityLd{\mu}\,\tensoruu{K}{\mu}{\nu}=\fourvelocityLd{\nu}\,\tensoruu{K}{\mu}{\nu}=0$.  

Finally, the Eulerian and Lagrangian decompositions of the third-order moment (rank-three tensor) are
\begin{eqnarray}
  \tensoruuu{U}{\mu}{\nu}{\rho}
  &=&
  \mathcal{G}\,\fourvelocityEu{\mu}\,\fourvelocityEu{\nu}\,\fourvelocityEu{\rho}
  +\fourvectoru{I}{\mu}\,\fourvelocityEu{\nu}\,\fourvelocityEu{\rho}
  +\fourvectoru{I}{\nu}\,\fourvelocityEu{\mu}\,\fourvelocityEu{\rho}
  +\fourvectoru{I}{\rho}\,\fourvelocityEu{\mu}\,\fourvelocityEu{\nu} \nonumber \\
  && \hspace{0.15in}
  +\tensoruu{P}{\mu}{\nu}\,\fourvelocityEu{\rho}
  +\tensoruu{P}{\mu}{\rho}\,\fourvelocityEu{\nu}
  +\tensoruu{P}{\nu}{\rho}\,\fourvelocityEu{\mu}
  +\tensoruuu{Q}{\mu}{\nu}{\rho} \label{eq:thirdOrderTensorEulerian} \\
  &=&
  \epsilon\,
  \left(\,
    \mathcal{J}\,\fourvelocityLu{\mu}\,\fourvelocityLu{\nu}\,\fourvelocityLu{\rho}
    +\fourvectoru{H}{\mu}\,\fourvelocityLu{\nu}\,\fourvelocityLu{\rho}
    +\fourvectoru{H}{\nu}\,\fourvelocityLu{\mu}\,\fourvelocityLu{\rho}
    +\fourvectoru{H}{\rho}\,\fourvelocityLu{\mu}\,\fourvelocityLu{\nu}
  \,\right. \nonumber \\
  && \hspace{0.15in}
  \left.\,
    +\tensoruu{K}{\mu}{\nu}\,\fourvelocityLu{\rho}
    +\tensoruu{K}{\mu}{\rho}\,\fourvelocityLu{\nu}
    +\tensoruu{K}{\nu}{\rho}\,\fourvelocityLu{\mu}
    +\tensoruuu{L}{\mu}{\nu}{\rho}
  \,\right), \label{eq:thirdOrderTensorLagrangian}
\end{eqnarray}
where $\mathcal{G}=\tensorububub{U}{0}{0}{0}$, $\fourvectoru{I}{\mu}=\tetudb{\mu}{\imath}\,\tensorububub{U}{\imath}{0}{0}$, $\tensoruu{P}{\mu}{\nu}=\tetudb{\mu}{\imath}\tetudb{\nu}{\jmath}\,\tensorububub{U}{\imath}{\jmath}{0}$, $\tensoruuu{Q}{\mu}{\nu}{\rho}=\tetudb{\mu}{\imath}\tetudb{\nu}{\jmath}\tetudb{\rho}{k}\,\tensorububub{U}{\imath}{\jmath}{k}$, and $\tensoruuu{L}{\mu}{\nu}{\rho}=\ludh{\mu}{\imath}\ludh{\nu}{\jmath}\ludh{\rho}{k}\,\tensoruhuhuh{L}{\imath}{\jmath}{k}$.  
Again, $\fourvectoru{I}{\mu}$, $\tensoruu{P}{\mu}{\nu}$, and $\tensoruuu{Q}{\mu}{\nu}{\rho}$ are purely spatial and orthogonal to the four-velocity of the Eulerian observer, while $\tensoruuu{L}{\mu}{\nu}{\rho}$ is orthogonal to the four-velocity of the comoving observer.  
The Eulerian and Lagrangian decompositions of the number-flux four-vector, the radiation stress-energy tensor, and the third-order moments are useful when deriving conservative moment equations for the radiation field---in particular, when seeking conservative evolution equations for lab-frame radiation quantities, expressed in terms of comoving-frame moments.  

The radiation moment equations for conformally flat spacetimes can be obtained by specializing the 3+1 moment equations in \citet{cardall_etal_2012}, or by adopting the spacetime metric in Equation (\ref{eq:metricCFC}), with $\gmdd{i}{j}=\psi^{4}\,\gmbdd{i}{j}$, and computing connection coefficients directly.  
Here we have adopted the latter approach, but see also \citet{cardall_etal_2012}.  
We obtain the evolution equation for the monochromatic lab-frame radiation energy density by contracting Equation (\ref{eq:stressEnergyMoment}) with $-n_{\mu}$.  
The result is
\begin{eqnarray}
  & &
  \f{1}{\mdet}\pderiv{}{t}\Big(\smdet\,\mathcal{E}\Big)
  +\f{1}{\mdet}\pderiv{}{x^{i}}\Big(\smdet\Big[\,\alpha\,\fourvectoru{F}{i}-\betau{i}\,\mathcal{E}\Big]\Big) \nonumber \\
  & & \hspace{0.15in}
  +\fourvectoru{F}{i}\pderiv{\ln\alpha}{x^{i}} 
  +\tensorud{S}{i}{i}\pderiv{\ln\psi^{2}}{\tau}
  -\f{\tensoruu{S}{i}{j}}{\alpha}\Big(\covderivb{i}\betad{j}-\pderiv{\ln\psi^{4}}{x^{i}}\betad{j}\Big) \nonumber \\
  & & \hspace{0.15in}
  -\f{1}{\epsilon^{2}}\pderiv{}{\epsilon}
  \Big(\epsilon^{2}\,\mathcal{F}^{\epsilon}\Big)
  =-\fourvelocityEd{\mu}\f{1}{\epsilon}\int_{\Omega}\pu{\mu}\,\collision{f}\,d\Omega, 
  \label{eq:energyEquationCFC}
\end{eqnarray}
where we have defined the energy space energy flux
\begin{eqnarray}
  \mathcal{F}^{\epsilon}
  &=&
  -\fourvelocityEd{\mu}\,\tensoruuu{U}{\mu}{\nu}{\rho}\,\covderiv{\nu}\fourvelocityLd{\rho} \nonumber \\
  &=&W
  \Big\{
    \fourvectord{I}{i}\pderiv{\threevelocityEu{i}}{\tau}
    +\tensordu{P}{i}{j}\pderiv{\threevelocityEu{i}}{x^{j}}
    +\f{1}{2}\psi^{4}\tensoruu{P}{i}{j}\threevelocityEu{k}\pderiv{\gmbdd{i}{j}}{x^{k}} \nonumber \\
  && \hspace{0.0in}
    +\Big(\fourvectoru{I}{i}-\mathcal{G}\threevelocityEu{i}\Big)\pderiv{\ln\alpha}{x^{i}}
    +\tensorud{P}{i}{i}
    \Big(\pderiv{\ln\psi^{2}}{\tau}+\threevelocityEu{j}\pderiv{\ln\psi^{2}}{x^{j}}\Big) \nonumber \\
  && \hspace{0.0in}
    -\f{\tensoruu{P}{i}{j}}{\alpha}\Big(\covderivb{i}\betad{j}-\pderiv{\ln\psi^{4}}{x^{i}}\betad{j}\Big)
    +\threevelocityEu{i}\fourvectord{I}{j}\f{1}{\alpha}\pderiv{\betau{j}}{x^{i}}
  \Big\} \nonumber \\
  && \hspace{0.0in}
  +\Big(\fourvectord{I}{i}\threevelocityEu{i}-\mathcal{G}\Big)\pderiv{W}{\tau}
  +\Big(\tensordu{P}{i}{j}\threevelocityEu{i}-\fourvectoru{I}{j}\Big)\pderiv{W}{x^{j}}.  
  \label{eq:energySpaceEnergyFlux}
\end{eqnarray}
We have also defined the covariant derivative with respect to the conformally related three-metric
\begin{equation}
  \covderivb{i}\betad{j}
  =\pderiv{\betad{j}}{x^{i}}-\cbudd{k}{j}{i}\,\betad{k}, 
\end{equation}
where
\begin{equation}
  \cbudd{k}{i}{j}
  =
  \f{1}{2}\gmbuu{k}{l}
  \left(\,
    \pderiv{\gmbdd{l}{i}}{x^{j}}+\pderiv{\gmbdd{l}{j}}{x^{i}}-\pderiv{\gmbdd{i}{j}}{x^{l}}
  \,\right), 
\end{equation}
and the proper time derivative along constant coordinate lines
\begin{equation}
  \pderiv{}{\tau}
  =\f{1}{\alpha}\pderiv{}{t}-\f{\betau{i}}{\alpha}\pderiv{}{x^{i}}.  
\end{equation}

We obtain the monochromatic radiation momentum equation by contracting Equation (\ref{eq:stressEnergyMoment}) with the projection operator $\gmdd{i}{\mu}=\gdd{i}{\mu}+\fourvelocityEd{i}\,\fourvelocityEd{\mu}$.  
The result is
\begin{eqnarray}
  & &
  \f{1}{\mdet}\pderiv{}{t}\Big(\smdet\,\fourvectord{F}{i}\Big)
  +\f{1}{\mdet}\pderiv{}{x^{j}}\Big(\smdet\Big[\alpha\,\tensorud{S}{j}{i}-\betau{j}\,\fourvectord{F}{i}\Big]\Big) \nonumber \\
  & & \hspace{0.15in}
  -\psi^{4}\f{\tensoruu{S}{j}{k}}{2}\pderiv{\gmbdd{j}{k}}{x^{i}} 
  +\mathcal{E}\pderiv{\ln\alpha}{x^{i}}
  -\tensorud{S}{j}{j}\pderiv{\ln\psi^{2}}{x^{i}}
  -\fourvectord{F}{j}\f{1}{\alpha}\pderiv{\betau{j}}{x^{i}} \nonumber \\
  & & \hspace{0.15in}
  -\f{1}{\epsilon^{2}}\pderiv{}{\epsilon}
  \Big(\epsilon^{2}\,\mathcal{S}_{~i}^{\epsilon}\Big)
  =\gmdd{i}{\mu}\f{1}{\epsilon}\int_{\Omega}\pu{\mu}\,\collision{f}\,d\Omega, 
  \label{eq:momentumEquationCFC}
\end{eqnarray}
where we have defined the energy space momentum flux
\begin{eqnarray}
  \mathcal{S}_{~i}^{\epsilon}
  &=&
  \gmdd{i}{\mu}\,\tensoruuu{U}{\mu}{\nu}{\rho}\,\covderiv{\nu}\fourvelocityLd{\rho} \nonumber \\
  &=&W
  \Big\{
    \tensordd{P}{i}{j}\pderiv{\threevelocityEu{j}}{\tau}
    +\tensorddu{Q}{i}{j}{k}\pderiv{\threevelocityEu{j}}{x^{k}}
    +\f{1}{2}\psi^{4}\tensorduu{Q}{i}{j}{k}\threevelocityEu{l}\pderiv{\gmbdd{j}{k}}{x^{l}} \nonumber \\
  && \hspace{0.0in}
    +\Big(\tensordu{P}{i}{j}-\fourvectord{I}{i}\threevelocityEu{j}\Big)\pderiv{\ln\alpha}{x^{j}} 
    +\tensordud{Q}{i}{j}{j}\Big(\pderiv{\ln\psi^{2}}{\tau}+\threevelocityEu{k}\pderiv{\ln\psi^{2}}{x^{k}}\Big)
     \nonumber \\
  && \hspace{0.0in}
    -\f{\tensorduu{Q}{i}{j}{k}}{\alpha}\Big(\covderivb{j}\betad{k}-\pderiv{\ln\psi^{4}}{x^{j}}\betad{k}\Big)
    +\threevelocityEu{j}\tensordd{P}{i}{k}\f{1}{\alpha}\pderiv{\betau{k}}{x^{j}}
  \Big\} \nonumber \\
  && \hspace{0.0in}
  +\Big(\tensordd{P}{i}{j}\threevelocityEu{j}-\fourvectord{I}{i}\Big)\pderiv{W}{\tau}
  +\Big(\tensorddu{Q}{i}{j}{k}\threevelocityEu{j}-\tensordu{P}{i}{k}\Big)\pderiv{W}{x^{k}}.  
  \label{eq:energySpaceMomentumFlux}
\end{eqnarray}
Equations (\ref{eq:energyEquationCFC}) and (\ref{eq:momentumEquationCFC}) are conservative equations for the lab-frame radiation energy density and momentum density.  
The independent variables are the coordinate basis spacetime position components $x^{\mu}$ and the neutrino energy measured by a comoving observer $\epsilon$.  
When integrated over the comoving-frame energy, the equations reduce to familiar position space conservation laws.  
They express exact energy and momentum conservation in the absence of neutrino-matter interactions and geometry sources (due to gravity and curvilinear coordinates).  
The terms in the energy derivatives (cf. Equations (\ref{eq:energySpaceEnergyFlux}) and (\ref{eq:energySpaceMomentumFlux})) result in changes in the radiation energy spectrum measured in the comoving-frame from gravitational redshifts and acceleration of the comoving observer.  
Note that Equations (\ref{eq:energyEquationCFC}) and (\ref{eq:momentumEquationCFC}) differ from the 3+1 general relativistic moment equations given by \citet{shibata_etal_2011,cardall_etal_2012}, in that they are not expressed in terms of the extrinsic curvature $\Tensordd{K}{i}{j}$ \citep[an evolved quantity in 3+1 general relativity; e.g.,][]{baumgarteShapiro_2010}.  

For the terms inside the energy derivatives in Equations (\ref{eq:energyEquationCFC}) and (\ref{eq:momentumEquationCFC}), we can use Equation (\ref{eq:stressEnergyThirdOrderMomentRelation}), and write $\fourvelocityLd{\mu}=W\left(\,\fourvelocityEd{\mu}+\threevelocityEd{\mu}\,\right)$, to relate Eulerian projections of the third-order moments (Equation (\ref{eq:thirdOrderTensorEulerian})) in terms of Eulerian projections of the stress-energy tensor \citep[Equation (\ref{eq:stressEnergyTensorEulerian});][]{cardall_etal_2012}.  
In particular, we have
\begin{eqnarray}
  &&
  W\fourvectoru{I}{i}
  =\epsilon
  \Big\{
    \fourvectoru{F}{i}+\tensoruu{S}{i}{j}\threevelocityEd{j}+\f{W}{\epsilon}\tensoruuu{Q}{i}{j}{k}\threevelocityEd{j}\threevelocityEd{k}
  \Big\}, \\
  &&
  W\tensoruu{P}{i}{j}
  =\epsilon
  \Big\{
    \tensoruu{S}{i}{j}+\f{W}{\epsilon}\tensoruuu{Q}{i}{j}{k}\threevelocityEd{k}
  \Big\}, \\
  &&
  W\Big(\fourvectoru{I}{i}-\mathcal{G}\threevelocityEu{i}\Big)
  =\epsilon
  \Big\{
    \Big(\fourvectoru{F}{i}-\mathcal{E}\threevelocityEu{i}\Big)
    +\Big(\tensoruu{S}{i}{j}-\threevelocityEu{i}\fourvectoru{F}{j}\Big)\threevelocityEd{j} \nonumber \\
  & & \hspace{0.15in}
    +\Big(\f{W}{\epsilon}\tensoruuu{Q}{i}{j}{k}-\threevelocityEu{i}\tensoruu{S}{j}{k}\Big)\threevelocityEd{j}\threevelocityEd{k}
    -\threevelocityEu{i}\f{W}{\epsilon}\tensoruuu{Q}{j}{k}{l}\threevelocityEd{j}\threevelocityEd{k}\threevelocityEd{l}
  \Big\}, \\
  &&
  W\Big(\tensoruu{P}{i}{j}-\fourvectoru{I}{i}\threevelocityEu{j}\Big)
  =\epsilon
  \Big\{
    \Big(\tensoruu{S}{i}{j}-\fourvectoru{F}{i}\threevelocityEu{j}\Big) \nonumber \\
  && \hspace{0.15in}
    +\Big(\f{W}{\epsilon}\tensoruuu{Q}{i}{j}{k}-\tensoruu{S}{i}{k}\threevelocityEu{j}\Big)\threevelocityEd{k}
    -\threevelocityEu{j}\f{W}{\epsilon}\tensoruuu{Q}{i}{k}{l}\threevelocityEd{k}\threevelocityEd{l}
  \Big\}, \\
  &&
  W\Big(\threevelocityEd{i}\fourvectoru{I}{i}-\mathcal{G}\Big)
  =-\epsilon\,\mathcal{E}, \\
  &&
  W\Big(\threevelocityEd{j}\tensoruu{P}{i}{j}-\fourvectoru{I}{i}\Big)
  =-\epsilon\,\fourvectoru{F}{i}, \\
  &&
  W\Big(\threevelocityEd{k}\tensoruuu{Q}{i}{j}{k}-\tensoruu{P}{i}{j}\Big)
  =-\epsilon\,\tensoruu{S}{i}{j}.  
\end{eqnarray}

The neutrino number equation can be obtained from Equations (\ref{eq:energyEquationCFC}) and (\ref{eq:momentumEquationCFC}).  From Equation (\ref{eq:numberStressEnergyRelation}) we have
\begin{eqnarray}
  \mathcalmcd{E}{N}
  &=&\f{W}{\epsilon}\Big(\,\mathcal{E}-\threevelocityEu{i}\,\fourvectord{F}{i}\,\Big), 
  \label{numberDensityFromEnergyAndFluxNewtonian} \\
  \fourvectorua{F}{i}{N}
  &=&\f{W}{\epsilon}\Big(\,\fourvectoru{F}{i}-\threevelocityEu{j}\,\tensorud{S}{i}{j}\,\Big). 
  \label{numberFluxFromFluxAndStressNewtonian}
\end{eqnarray}
As was done in Section \ref{sec:numberStressEnergyEquationRelation}, we obtain a conservative equation for the monochromatic lab-frame neutrino number density by adding $\epsilon^{-1}W$ times Equation (\ref{eq:energyEquationCFC}) and $-\epsilon\,W\,v^{i}$ contracted with Equation (\ref{eq:momentumEquationCFC})---and subsequently bringing the necessary terms inside the time, space, and energy derivatives, respectively.  
The result is
\begin{eqnarray}
  & &
  \f{1}{\mdet}\pderiv{}{t}\Big(\smdet\,\mathcalmcd{E}{N}\Big)
  +\f{1}{\mdet}\pderiv{}{x^{i}}\Big(\smdet\Big[\alpha\,\fourvectorua{F}{i}{N}-\betau{i}\,\mathcalmcd{E}{N}\Big]\Big) \nonumber \\
  & & \hspace{0.15in}
  -\f{1}{\epsilon^{2}}\pderiv{}{\epsilon}
  \Big(\epsilon^{2}\,\mathcal{F}_{{\scriptscriptstyle\mathcal{N}}}^{\epsilon}\Big)
  =\f{1}{\epsilon}\int_{\Omega}\collision{f}\,d\Omega, 
  \label{eq:numberEquationCFC}
\end{eqnarray}
where the energy space number flux is
\begin{eqnarray}
  \mathcal{F}_{{\scriptscriptstyle\mathcal{N}}}^{\epsilon}
  &=&
  \tensoruu{T}{\mu}{\nu}\,\covderiv{\mu}\fourvelocityLd{\nu}
  =
  \f{W}{\epsilon}\Big(\,\mathcal{F}^{\epsilon}-\threevelocityEu{i}\,\mathcal{S}_{~i}^{\epsilon}\,\Big) \nonumber \\
  &=&W
  \Big\{
      \fourvectord{F}{i}\pderiv{\threevelocityEu{i}}{\tau}
      +\tensordu{S}{i}{j}\pderiv{\threevelocityEu{i}}{x^{j}}
      +\f{1}{2}\psi^{4}\tensoruu{S}{i}{j}\threevelocityEu{k}\pderiv{\gmbdd{i}{j}}{x^{k}} \nonumber \\
    & & \hspace{0.0in}
      +\Big(\fourvectoru{F}{i}-\mathcal{E}\threevelocityEu{i}\Big)\pderiv{\ln\alpha}{x^{i}}
      +\tensorud{S}{i}{i}
      \Big(\pderiv{\ln\psi^{2}}{\tau}+\threevelocityEu{j}\pderiv{\ln\psi^{2}}{x^{j}}\Big) \nonumber \\
    && \hspace{0.0in}
      -\f{\tensoruu{S}{i}{j}}{\alpha}\Big(\covderivb{i}\betad{j}-\pderiv{\ln\psi^{4}}{x^{i}}\betad{j}\Big)
      +\threevelocityEu{i}\fourvectord{F}{j}\f{1}{\alpha}\pderiv{\betau{j}}{x^{i}}
    \Big\} \nonumber \\
    && \hspace{0.0in}
    -\Big(\mathcal{E}-\threevelocityEu{i}\fourvectord{F}{i}\Big)\pderiv{W}{\tau}
    -\Big(\fourvectoru{F}{j}-\threevelocityEu{i}\tensordu{S}{i}{j}\Big)\pderiv{W}{x^{j}}.  
    \label{eq:energySpaceNumberFluxCFC}
\end{eqnarray}
When deriving Equation (\ref{eq:numberEquationCFC}) from Equations (\ref{eq:energyEquationCFC}) and (\ref{eq:momentumEquationCFC}), terms emanating from the time derivatives, space derivatives, and the geometry sources cancel exactly with terms emanating from the energy derivatives \citep[see also][]{cardall_etal_2012}.  
Ideally, a discrete representation of the neutrino energy and momentum equations can be constructed so that a conservative neutrino number equation can be analogously obtained in the discrete limit.  
The discretization of the left-hand sides of the energy and momentum equations is then consistent with neutrino number conservation.  
(A similar consistency must also be considered for the right-hand side ot the equations.)  
Clearly, in order to construct such a discretization, \emph{the individual terms in the neutrino energy and momentum equations cannot be discretized independently}.  

\section{PSEUDO-NEWTONIAN MOMENT EQUATIONS FOR NEUTRINO RADIATION TRANSPORT}
\label{sec:pseudoNewtonian}

In this section, we specialize the moment equations presented in Section \ref{sec:momentEquationsCFC} to the `pseudo-Newtonian' limit by adopting a spacetime metric consistent with the line element \citep[cf.][]{kim_etal_2009,kim_etal_2012}
\begin{equation}
  ds^{2}
  =-\left(\,1+2\,\Phi\,\right)dt^{2}
  +\left(\,1+2\,\Phi\,\right)^{-1}\gmbdd{i}{j}\,dx^{i}\,dx^{j}, 
  \label{eq:metricNewtonian}
\end{equation}
where $\Phi\ll1$ is a pseudo-Newtonian gravitational potential.  
In this approximation, all orders in the fluid velocity $v$ are retained, and the pseudo-Newtonian gravitational potential is obtained by solving a Poisson equation in which the trace of the stress-energy tensor---not just the rest-mass density---contributes to the source on the right-hand side.  
We allow for common curvilinear spatial coordinates (i.e., Cartesian, spherical, and cylidrical) by setting $\gmbdd{i}{j}=\mbox{diag}[\,1,\,a^{2}(x^{1}),\,b^{2}(x^{1})\,c^{2}(x^{2})\,]$ (Section \ref{sec:momentEquationsCFC}).  
The spatial metric $\gmdd{i}{j}=(1+2\,\Phi)^{-1}\,\gmbdd{i}{j}$ and its inverse are used to lower and raise indices on (spatial) vectors and tensors.  
The determinant of the spacetime metric becomes
\begin{equation}
  \mdet=(\,1+2\,\Phi\,)^{-1}\smbdet.  
  \label{eq:metricDeterminantNewtonian}
\end{equation}
Similarly, we have $\smdet=(1+2\,\Phi)^{-3/2}\smbdet$, where $\smbdet=abc$.  
Note that the spacetime metric in Equation (\ref{eq:metricNewtonian}) differs slightly from the Newtonian one given by \citet{misner_etal_1973,schutz_1985}, but agrees to $\mathcal{O}(\Phi)$; i.e., $(1+2\Phi)^{-1}=(1-2\Phi)+\mathcal{O}(\Phi^{2})$.  
Using the spacetime metric in Equation (\ref{eq:metricNewtonian}) turns out to be algebraically advantageous.  
Note in particular that $\mdet=\alphat\,\smdet$, where $\alphat=(1+2\,\Phi)^{1/2}$.  
We note in passing that \citet{kim_etal_2009,kim_etal_2012} recently combined relativistic hydrodynamics with the weak-field (Newtonian) limit of general relativity (cf. Equation (\ref{eq:metricNewtonian})) to study equilibrium solutions of rotating relativistic stars and found good agreement with general relativistic computations in the reported cases.  
\citep[See also][who adopted a similar approach in magneto-rotational core-collapse simulations.]{takiwaki_etal_2009}

We obtain the pseudo-Newtonian radiation moment equations from Equations (\ref{eq:energyEquationCFC}) and (\ref{eq:momentumEquationCFC}) by setting $\alpha=\alphat\equiv(1+2\,\Phi)^{1/2}$, $\beta^{i}=0$, and $\psi^{2}=\psit^{2}\equiv(1+2\,\Phi)^{-1/2}$, and retaining all terms of $\mathcal{O}(\Phi)$.  
We state the results for easy reference.  
The radiation energy equation becomes
\begin{eqnarray}
  &&
  \f{1}{\mdet}\pderiv{}{t}\Big(\smdet\,\mathcal{E}\Big)
  +\f{1}{\mdet}\pderiv{}{x^{i}}\Big(\mdet\,\fourvectoru{F}{i}\Big) 
  +\fourvectoru{F}{i}\pderiv{\Phi}{x^{i}} \nonumber \\
  && \hspace{0.15in}
  -\tensorud{S}{i}{i}\pderiv{\Phi}{t}
  -\f{1}{\epsilon^{2}}\pderiv{}{\epsilon}
  \Big(\epsilon^{2}\,\mathcal{F}^{\epsilon}\Big)
  =\f{\alphat}{\epsilon}\int_{\Omega}\pu{0}\,\collision{f}\,d\Omega, 
  \label{eq:energyEquationPseudoNewtonian}
\end{eqnarray}
where the lab-frame radiation energy density, momentum density, and stress ($\mathcal{E}$, $\fourvectoru{F}{i}$, and $\tensoruu{S}{i}{j}$, respectively) are related to the comoving-frame moments in Equations (\ref{eq:eulerianEnergyLagrangianProjections})-(\ref{eq:eulerianStressLagrangianProjections}).  
The energy space energy flux (cf. Equation (\ref{eq:energySpaceEnergyFlux})) reduces to
\begin{eqnarray}
  \mathcal{F}^{\epsilon}
  &=&W
  \Big\{
    \fourvectord{I}{i}\f{1}{\alphat}\pderiv{\threevelocityEu{i}}{t}
    +\tensordu{P}{i}{j}\pderiv{\threevelocityEu{i}}{x^{j}}
    +\f{1}{2}\psit^{4}\tensoruu{P}{i}{j}\threevelocityEu{k}\pderiv{\gmbdd{i}{j}}{x^{k}} \nonumber \\
  && \hspace{0.25in}
    +\Big(\fourvectoru{I}{i}-\mathcal{G}\threevelocityEu{i}\Big)\pderiv{\Phi}{x^{i}}
    -\tensorud{P}{i}{i}
    \Big(\pderiv{\Phi}{t}+\threevelocityEu{j}\pderiv{\Phi}{x^{j}}\Big)
  \Big\} \nonumber \\
  && \hspace{0.0in}
  +\Big(\fourvectord{I}{i}\threevelocityEu{i}-\mathcal{G}\Big)\f{1}{\alphat}\pderiv{W}{t}
  +\Big(\tensordu{P}{i}{j}\threevelocityEu{i}-\fourvectoru{I}{j}\Big)\pderiv{W}{x^{j}}.
  \label{eq:energySpaceEnergyFluxPseudoNewtonian}
\end{eqnarray}
Similarly, the radiation momentum equation becomes
\begin{eqnarray}
  & &
  \f{1}{\mdet}\pderiv{}{t}\Big(\smdet\,\fourvectord{F}{i}\Big)
  +\f{1}{\mdet}\pderiv{}{x^{j}}\Big(\mdet\,\tensorud{S}{j}{i}\Big) \nonumber \\
  & & \hspace{0.15in}
  -\psit^{4}\f{\tensoruu{S}{j}{k}}{2}\pderiv{\gmbdd{j}{k}}{x^{i}}
  +\mathcal{E}\pderiv{\Phi}{x^{i}}
  +\tensorud{S}{j}{j}\pderiv{\Phi}{x^{i}}
  -\f{1}{\epsilon^{2}}\pderiv{}{\epsilon}
  \Big(\epsilon^{2}\,\mathcal{S}_{~i}^{\epsilon}\Big) \nonumber \\
  && \hspace{0.0in}
  =\gmdd{i}{j}\f{1}{\epsilon}\int_{\Omega}\pu{j}\,\collision{f}\,d\Omega, 
  \label{eq:momentumEquationPseudoNewtonian}
\end{eqnarray}
where the energy space momentum flux is given by (cf. Equation (\ref{eq:energySpaceMomentumFlux}))
\begin{eqnarray}
  \mathcal{S}_{~i}^{\epsilon}
  &=&W
  \Big\{
    \tensordd{P}{i}{j}\f{1}{\alphat}\pderiv{\threevelocityEu{j}}{t}
    +\tensorddu{Q}{i}{j}{k}\pderiv{\threevelocityEu{j}}{x^{k}}
    +\f{1}{2}\psit^{4}\tensorduu{Q}{i}{j}{k}\threevelocityEu{l}\pderiv{\gmbdd{j}{k}}{x^{l}} \nonumber \\
  && \hspace{0.25in}
    +\Big(\tensordu{P}{i}{j}-\fourvectord{I}{i}\threevelocityEu{j}\Big)\pderiv{\Phi}{x^{j}} 
    -\tensordud{Q}{i}{j}{j}\Big(\pderiv{\Phi}{t}+\threevelocityEu{k}\pderiv{\Phi}{x^{k}}\Big)
  \Big\} \nonumber \\
  && \hspace{0.0in}
  +\Big(\tensordd{P}{i}{j}\threevelocityEu{j}-\fourvectord{I}{i}\Big)\f{1}{\alphat}\pderiv{W}{t}
  +\Big(\tensorddu{Q}{i}{j}{k}\threevelocityEu{j}-\tensordu{P}{i}{k}\Big)\pderiv{W}{x^{k}}.  
\end{eqnarray}

Equations (\ref{eq:energyEquationPseudoNewtonian}) and (\ref{eq:momentumEquationPseudoNewtonian}) are valid to all orders in $v$, but limited to weak gravitational fields ($\Phi\ll1$).  
In the case of no gravitational fields ($\Phi\equiv0$) they reduce to the special relativistic moment equations (cf. Equations (\ref{eq:energyMomentSR}) and (\ref{eq:momentumMomentSR}) in Appendix \ref{app:srMomentEquations}).  
Moreover, Equations (\ref{eq:energyEquationPseudoNewtonian}) and (\ref{eq:momentumEquationPseudoNewtonian}) are consistent with the conservative neutrino number equation given by
\begin{eqnarray}
  & &
  \f{1}{\mdet}\pderiv{}{t}\Big(\smdet\,\mathcalmcd{E}{N}\Big)
  +\f{1}{\mdet}\pderiv{}{x^{i}}\Big(\mdet\,\fourvectorua{F}{i}{N}\Big) \nonumber \\
  & & \hspace{0.15in}
  -\f{1}{\epsilon^{2}}\pderiv{}{\epsilon}
  \Big(\epsilon^{2}\,\mathcal{F}_{{\scriptscriptstyle\mathcal{N}}}^{\epsilon}\Big)
  =\f{1}{\epsilon}\int_{\Omega}\collision{f}\,d\Omega, 
  \label{eq:numberEquationPseudoNewtonian}
\end{eqnarray}
where the energy space number flux is (cf. Equation (\ref{eq:energySpaceNumberFluxCFC}))
\begin{eqnarray}
  \mathcal{F}_{{\scriptscriptstyle\mathcal{N}}}^{\epsilon}
  &=&W
  \Big\{
      \fourvectord{F}{i}\f{1}{\alphat}\pderiv{\threevelocityEu{i}}{t}
      +\tensordu{S}{i}{j}\pderiv{\threevelocityEu{i}}{x^{j}}
      +\f{1}{2}\psit^{4}\tensoruu{S}{i}{j}\threevelocityEu{k}\pderiv{\gmbdd{i}{j}}{x^{k}} \nonumber \\
    & & \hspace{0.25in}
      +\Big(\fourvectoru{F}{i}-\mathcal{E}\threevelocityEu{i}\Big)\pderiv{\Phi}{x^{i}}
      -\tensorud{S}{i}{i}
      \Big(\pderiv{\Phi}{t}+\threevelocityEu{j}\pderiv{\Phi}{x^{j}}\Big)
    \Big\} \nonumber \\
    && \hspace{0.0in}
    -\Big(\mathcal{E}-\threevelocityEu{i}\fourvectord{F}{i}\Big)\f{1}{\alphat}\pderiv{W}{t}
    -\Big(\fourvectoru{F}{j}-\threevelocityEu{i}\tensordu{S}{i}{j}\Big)\pderiv{W}{x^{j}}.  
\end{eqnarray}
The lab-frame number density and number flux are related to the corresponding comoving-frame moments in Equations (\ref{eq:eulerianNumberDensityLagrangianProjections}) and (\ref{eq:eulerianNumberFluxLagrangianProjections}), respectively.  

Equations (\ref{eq:energyEquationPseudoNewtonian}), (\ref{eq:momentumEquationPseudoNewtonian}), and (\ref{eq:numberEquationPseudoNewtonian}) simplify even further when slowly varying gravitational fields are considered \citep[$\partial\Phi/\partial t=0$; cf.][]{kim_etal_2012}.  

\section{NON-RELATIVISTIC SELF-GRAVITATING NEUTRINO RADIATION HYDRODYNAMICS}
\label{sec:newtonianOrderV}

In this section we detail the full set of neutrino radiation hydrodynamics equations intended for our planned non-relativistic simulations of core-collapse supernovae and related systems.  
The equations are deduced from the relativistic equations derived in previous sections.  
Per our initial discussion in Section \ref{sec:introduction}, in order to formulate a system of moment equations that are consistent with a conservative equation for the neutrino number density, we adopt different orders of $v$ for the radiation energy and momentum equations.  
We consider two possible cases: the `$\mathcal{O}(v)$-plus' and the `$\mathcal{O}(v)$-minus' moment equations.  
The $\mathcal{O}(v)$-plus moment equations consist of an $\mathcal{O}(v^{2})$ energy equation and an $\mathcal{O}(v)$ momentum equation, and are consistent with the conservative $\mathcal{O}(v^{2})$ neutrino number equation.  
Similarly, the $\mathcal{O}(v)$-minus moment equations consist of an $\mathcal{O}(v)$ energy equation and an $\mathcal{O}(1)$ momentum equation, and are consistent with the conservative $\mathcal{O}(v)$ neutrino number equation.  
Dimensional analysis suggests that gravitational effects on the radiation field are formally $\mathcal{O}(v^{2})$.  
Therefore, in the $\mathcal{O}(v)$-plus moment equations for neutrino radiation transport, we also retain terms due to a Newtonian gravitational potential.  
These `gravitational redshift' terms may play a non-negligible role in the post-bounce supernova environment \citep[e.g.,][]{bruenn_etal_2001}.  
Moreover, by including gravitational redshift effects, the $\mathcal{O}(v)$-plus moment equations are conceptually more similar to the fully relativistic moment equations \citep[e.g.,][]{shibata_etal_2011,cardall_etal_2012}.  

\subsection{$\mathcal{O}(v)$-Plus Moment Equations for Neutrino Transport}
\label{sec:momentsOrderVPlus}

In this subsection we detail radiation moment equations formally valid to $\mathcal{O}(v)$, including effects due to a Newtonian gravitational field.  
We obtain these Newtonian, $\mathcal{O}(v)$-plus radiation moment equations from the pseudo-Newtonian equations in Section \ref{sec:pseudoNewtonian}---Equations (\ref{eq:energyEquationPseudoNewtonian}) and (\ref{eq:momentumEquationPseudoNewtonian})---by considering non-relativistic fluid velocities.  
In the radiation moment equations presented in this section, we also omit terms containing the time derivative of the gravitational potential.  
Furthermore, we retain at least all terms that are linear in the fluid velocity.  
We retain $\mathcal{O}(v^{2})$ terms in the radiation energy equation, while we retain $\mathcal{O}(v)$ terms in the radiation momentum equation.  
As discussed in the introduction, and in detail below, we adopt this `mixed' order of the radiation energy and momentum equations in order to achieve satisfactory consistency with the conservative neutrino number equation.  
Since we assume $v^{2}\ll1$, we set $W=\lorentzApp\equiv1+\f{1}{2}v^{2}$, where $v^{2}=\threevelocityEu{i}\threevelocityEd{i}$.  
Then, the monochromatic $\mathcal{O}(v^{2})$ radiation energy equation becomes (cf. Equation (\ref{eq:energyEquationPseudoNewtonian}))
\begin{eqnarray}
  &&
  \f{1}{\mdet}\pderiv{}{t}\Big(\smdet\,\mathcalt{E}\Big)
  +\f{1}{\mdet}\pderiv{}{x^{i}}\Big(\mdet\,\fourvectortu{F}{i}\Big) \nonumber \\
  && \hspace{0.15in}
  +\fourvectoru{F}{i}\,\pderiv{\Phi}{x^{i}}
  -\f{1}{\epsilon^{2}}\pderiv{}{\epsilon}\Big(\epsilon^{2}\,\tilde{\mathcal{F}}^{\epsilon}\Big)
  =\collisionEnergyEquationt, 
  \label{eq:radiationEnergyEquationNewtonianOrderVV}
\end{eqnarray}
where the $\mathcal{O}(v^{2})$ lab-frame radiation energy density and momentum density are related to the comoving-frame moments ($\mathcal{J}$, $\fourvectoru{H}{i}$, and $\tensoruu{K}{i}{j}$) by
\begin{eqnarray}
  \mathcalt{E}
  &=&\mathcal{E}+\threevelocityEd{i}\Big(\,\threevelocityEu{i}\,\mathcal{J}+\threevelocityEd{j}\,\tensoruu{K}{i}{j}\,\Big),
  \label{eq:eulerianEnergyLagrangianProjectionsOrderVV} \\
  \fourvectortu{F}{i}
  &=&\fourvectoru{F}{i}+\threevelocityEd{j}\Big(\,\threevelocityEu{i}\,\fourvectoru{H}{j}+\f{1}{2}\,\fourvectoru{H}{i}\,\threevelocityEu{j}\,\Big),
  \label{eq:eulerianMomentumLagrangianProjectionsOrderVV}
\end{eqnarray}
and the $\mathcal{O}(v)$ radiation energy density, momentum density, and stress are related to the comoving-frame moments by
\begin{eqnarray}
  \mathcal{E}
  &=&\mathcal{J}+2\,\threevelocityEd{i}\,\fourvectoru{H}{i}, 
  \label{eq:eulerianEnergyLagrangianProjectionsOrderV} \\
  \fourvectoru{F}{i}
  &=&\fourvectoru{H}{i}+\threevelocityEu{i}\,\mathcal{J}+\threevelocityEd{j}\,\tensoruu{K}{i}{j}, 
  \label{eq:eulerianMomentumLagrangianProjectionsOrderV} \\
  \tensoruu{S}{i}{j}
  &=&\tensoruu{K}{i}{j}+\threevelocityEu{i}\,\fourvectoru{H}{j}+\fourvectoru{H}{i}\,\threevelocityEu{j}
  \label{eq:eulerianStressLagrangianProjectionsOrderV}
\end{eqnarray}
(cf. Equations (\ref{eq:eulerianEnergyLagrangianProjections})-(\ref{eq:eulerianStressLagrangianProjections})).  
Radiation quantities adorned with a tilde (e.g., $\mathcalt{E}$) are accurate to $\mathcal{O}(v^{2})$.  
To $\mathcal{O}(v)$ we also have
\begin{eqnarray}
  \fourvectoru{H}{i}
  &=&\tetudb{i}{\imath}\,\deltaubdh{\imath}{\imath}\,\fourvectoruh{H}{\imath}, 
  \label{eq:lagrangianMomentumComovingFrameMomentsOrderV} \\
  \tensoruu{K}{i}{j}
  &=&\tetudb{i}{\imath}\,\tetudb{j}{\jmath}\,\deltaubdh{\imath}{\imath}\,\deltaubdh{\jmath}{\jmath}\,\tensoruhuh{K}{\imath}{\jmath}, 
  \label{eq:lagrangianStressComovingFrameMomentsOrderV} \\
  \tensoruuu{L}{i}{j}{k}
  &=&\tetudb{i}{\imath}\,\tetudb{j}{\jmath}\,\tetudb{k}{k}\,\deltaubdh{\imath}{\imath}\,\deltaubdh{\jmath}{\jmath}\,\deltaubdh{k}{k}\,\tensoruhuhuh{L}{\imath}{\jmath}{k}, 
  \label{eq:lagrangianThirdOrderMomentsComovingFrameMomentsOrderV}
\end{eqnarray}
(cf. Equations (\ref{eq:lagrangianMomentumComovingFrameMoments}), (\ref{eq:lagrangianStressComovingFrameMoments}), and (\ref{eq:lagrangianThirdOrderMomentsComovingFrameMoments})).  
The comoving-frame moments $\mathcal{J}$, $\fourvectoruh{H}{\imath}$, $\tensoruhuh{K}{\imath}{\jmath}$, and $\tensoruhuhuh{L}{\imath}{\jmath}{k}$ are defined in Section \ref{sec:angularMoments}.  
The energy space energy flux, written in terms of comoving-frame moments, is (cf. Equation (\ref{eq:energySpaceEnergyFluxPseudoNewtonian}))
\begin{eqnarray}
  \tilde{\mathcal{F}}^{\epsilon}
  &=&
  \mathcal{F}^{\epsilon}
  +\epsilon\,
  \Big\{
      \Big(
        \mathcal{J}\,\threevelocityEd{j}+2\,\threevelocityEu{i}\,\tensordd{K}{i}{j}
      \Big)\f{1}{\alphat}\pderiv{\threevelocityEu{j}}{t} \nonumber \\
      && \hspace{0.5in}
      +\Big(
        \threevelocityEd{j}\,\fourvectoru{H}{k}+\fourvectord{H}{j}\,\threevelocityEu{k}+\threevelocityEu{i}\,\tensorddu{L}{i}{j}{k}
      \Big)\pderiv{\threevelocityEu{j}}{x^{k}} \nonumber \\
    && \hspace{0.5in}
      +\f{1}{2}\psit^{4}
      \Big(
        \threevelocityEu{j}\,\fourvectoru{H}{k}+\fourvectoru{H}{j}\,\threevelocityEu{k}+\threevelocityEu{i}\,\tensorduu{L}{i}{j}{k}
      \Big)\threevelocityEu{l}\pderiv{\gmbdd{j}{k}}{x^{l}} \nonumber \\
    && \hspace{0.5in}
      +\Big(
        2\,\threevelocityEu{i}\,\tensordu{K}{i}{j}-\tensorud{K}{i}{i}\,\threevelocityEu{j}
      \Big)\pderiv{\Phi}{x^{j}} \nonumber \\
    && \hspace{0.5in}
      -\mathcalt{E}\,\f{1}{\alphat}\pderiv{\ln\lorentzApp}{t}
      -\fourvectortu{F}{i}\,\pderiv{\ln\lorentzApp}{x^{i}}
    \Big\},
    \label{eq:energySpaceEnergyFluxNewtonianOrderVV}
\end{eqnarray}
where the $\mathcal{O}(v)$ and $\mathcal{O}(\Phi)$ terms are defined separately in
\begin{equation}
  \mathcal{F}^{\epsilon}
  =\epsilon\,
  \Big\{\,
    \fourvectord{H}{j}\f{1}{\alphat}\pderiv{\threevelocityEu{j}}{t}
    +\tensordu{K}{j}{k}\pderiv{\threevelocityEu{j}}{x^{k}}
    +\f{1}{2}\psit^{4}\tensoruu{K}{j}{k}\threevelocityEu{l}\pderiv{\gmbdd{j}{k}}{x^{l}}
    +\fourvectoru{H}{j}\pderiv{\Phi}{x^{j}}
  \,\Big\}.
  \label{eq:energySpaceEnergyFluxNewtonianOrderV}
\end{equation}
In Equation (\ref{eq:energySpaceEnergyFluxNewtonianOrderVV}), except for the last two terms, we have dropped terms of order equal to or higher than $\mathcal{O}(v^{3})$ and $\mathcal{O}(v^{2}\,\Phi)$.  
Similarly, the monochromatic $\mathcal{O}(v)$ radiation momentum equation becomes
\begin{eqnarray}
  &&
  \f{1}{\mdet}\pderiv{}{t}\Big(\smdet\,\fourvectord{F}{i}\Big)
  +\f{1}{\mdet}\pderiv{}{x^{j}}\Big(\mdet\,\tensorud{S}{j}{i}\Big)
  -\psit^{4}\f{\tensoruu{S}{j}{k}}{2}\pderiv{\gmbdd{j}{k}}{x^{i}} \nonumber \\
  && \hspace{0.15in}
  +\Big(\mathcal{J}+\tensorud{K}{j}{j}\Big)\pderiv{\Phi}{x^{i}}
  -\f{1}{\epsilon^{2}}\pderiv{}{\epsilon}\Big(\epsilon^{2}\,\tilde{\mathcal{S}}_{~i}^{\epsilon}\Big)
  =\collisionMomentumEquationd{i},
  \label{eq:radiationMomentumEquationNewtonianOrderV}
\end{eqnarray}
where energy space momentum flux is
\begin{equation}
  \tilde{\mathcal{S}}_{~i}^{\epsilon}
  =\mathcal{S}_{~i}^{\epsilon}
  -\epsilon\,
  \Big\{
    \fourvectord{F}{i}\,\f{1}{\alphat}\pderiv{\ln\lorentzApp}{t}
    +\tensordu{S}{i}{j}\,\pderiv{\ln\lorentzApp}{x^{j}}
  \Big\}, 
  \label{eq:energySpaceMomentumFluxNewtonian}
\end{equation}
and the $\mathcal{O}(v)$ and $\mathcal{O}(\Phi)$ terms are contained in
\begin{equation}
  \mathcal{S}_{~i}^{\epsilon}
  =\epsilon\,
  \Big\{
    \tensordd{K}{i}{j}\f{1}{\alphat}\pderiv{\threevelocityEu{j}}{t}
    +\tensorddu{L}{i}{j}{k}\pderiv{\threevelocityEu{j}}{x^{k}}
    +\f{1}{2}\psit^{4}\tensorduu{L}{i}{j}{k}\threevelocityEu{l}\pderiv{\gmbdd{j}{k}}{x^{l}}
    +\tensordu{K}{i}{j}\pderiv{\Phi}{x^{j}}
  \Big\}.
  \label{eq:energySpaceMomentumFluxNewtonianOrderV}
\end{equation}
In Equation (\ref{eq:energySpaceMomentumFluxNewtonian}), except for the last two terms, we have dropped terms of order equal to or higher than $\mathcal{O}(v^{2})$ and $\mathcal{O}(v\,\Phi)$.  
Note that $\tilde{\mathcal{S}}_{~i}^{\epsilon}$ is not accurate to $\mathcal{O}(v^{2})$, but is adorned with a tilde to signify that some higher-order terms have been retained.  
Also note that we have retained the $\Phi$-dependence in the determinant of the spacetime metric (cf. Equation (\ref{eq:metricDeterminantNewtonian})) appearing in the moment equations, and retained factors $\alphat$ and $\psit$, which give rise to higher-order terms in some cases (e.g., Equation (\ref{eq:energySpaceMomentumFluxNewtonianOrderV})).  
These should all be kept for consistency with the conservative $\mathcal{O}(v^{2})$ number equation (see below).  

In Equations (\ref{eq:radiationEnergyEquationNewtonianOrderVV}) and (\ref{eq:radiationMomentumEquationNewtonianOrderV}), the collision terms on the right-hand side are, to $\mathcal{O}(v^{2})$ and $\mathcal{O}(v)$, respectively, 
\begin{eqnarray}
  \collisionEnergyEquationt{}
  &=&\lorentzApp\int_{\Omega}\collision{f}d\Omega
  +\vbdh{\imath}\int_{\Omega}\nuh{\imath}\,\collision{f}d\Omega, 
  \label{eq:collisionTermEnergyEquationNewtonianOrderVV} \\
  \collisionMomentumEquationd{i}
  &=&\threevelocityEd{i}\int_{\Omega}\collision{f}d\Omega
  +\gmdd{i}{j}\,\tetudb{j}{\jmath}\,\deltaubdh{\jmath}{\jmath}\int_{\Omega}\nuh{\jmath}\,\collision{f}d\Omega. 
  \label{eq:collisionTermMomentumEquationNewtonianOrderV}
\end{eqnarray}
($\vbdh{\imath}$ is the fluid three-velocity with respect to the orthonormal tetrad basis; Appendix \ref{app:srMomentEquations}.)  

Equations (\ref{eq:radiationEnergyEquationNewtonianOrderVV}) and (\ref{eq:radiationMomentumEquationNewtonianOrderV}) are \emph{conservative} equations for the monochromatic lab-frame radiation energy density and momentum density, respectively.  
Equation (\ref{eq:radiationEnergyEquationNewtonianOrderVV}) expresses exact conservation of radiation energy in the absence of gravity and neutrino-matter interactions, and---if in addition, Cartesian coordinates are used---Equation (\ref{eq:radiationMomentumEquationNewtonianOrderV}) expresses exact conservation of radiation momentum.  
In the strict $\mathcal{O}(v)$ limit\footnote{By the ``strict" $\mathcal{O}(v)$ limit we mean that all (or most) terms that are at most linear in the fluid velocity have been retained in the radiation energy \emph{and} momentum equations.  As we emphasize in this section, in this strict $\mathcal{O}(v)$ limit, the radiation moment equations are not fully consistent with a \emph{conservative} equation for the neutrino number density.}, Equations (\ref{eq:radiationEnergyEquationNewtonianOrderVV}) and (\ref{eq:radiationMomentumEquationNewtonianOrderV}) can be compared with corresponding equations derived by \citet[][his Equations (9) and (10), respectively]{buchler_1979}.  
See also \citet[][]{kaneko_etal_1984,munierWeaver_1986b,buras_etal_2006a}.  
We obtain the strict $\mathcal{O}(v)$ limit of Equations (\ref{eq:radiationEnergyEquationNewtonianOrderVV}) and (\ref{eq:radiationMomentumEquationNewtonianOrderV}) by setting $\Phi=0$, letting $\mathcalt{E},\fourvectortu{F}{i},\tilde{\mathcal{F}}^{\epsilon},\tilde{\mathcal{S}}_{~i}^{\epsilon}\to\mathcal{E},\fourvectoru{F}{i},\mathcal{F}^{\epsilon},\mathcal{S}_{~i}^{\epsilon}$, and $\collisionEnergyEquationt{}\to\collisionEnergyEquation{}$ by setting $\lorentzApp=1$ in Equation (\ref{eq:collisionTermEnergyEquationNewtonianOrderVV}).  
Buchler's equations evolve the comoving-frame moments and are \emph{non-conservative}.  
However, to $\mathcal{O}(v)$, the terms inside the energy derivatives in Equations (\ref{eq:radiationEnergyEquationNewtonianOrderVV}) and (\ref{eq:radiationMomentumEquationNewtonianOrderV}) agree with the energy derivative terms in Buchler's energy and momentum equations.  
\citet{shibata_etal_2011} also list \emph{conservative} moment equations for neutrino radiation transport in the `slow-motion' limit (cf. their Equations (8.5) and (8.6)).  
They do not include gravitational effects, or any higher-order terms in the fluid velocity.  
They also do not include the terms involving the derivative of the fluid velocity with respect to time in the energy derivative terms, which can be important when $\lambda/\tau\lesssim1$ \citep[e.g., in optically thin regions;][]{buchler_1979}, where $\lambda$ and $\tau$ are typical length and time scales, respectively.  

By integrating Equations (\ref{eq:radiationEnergyEquationNewtonianOrderVV}) and (\ref{eq:radiationMomentumEquationNewtonianOrderV}) over the comoving-frame energy (with $\epsilon^{2}\,d\epsilon$ as the measure of integration) we obtain the energy-integrated (grey) Eulerian-frame radiation energy equation, 
\begin{equation}
  \f{1}{\mdet}\pderiv{\Big(\smdet\,\tilde{E}\Big)}{t}
  +\f{1}{\mdet}\pderiv{\Big(\mdet\,\Fourvectortu{F}{i}\Big)}{x^{i}}
  =-\Fourvectoru{F}{i}\,\pderiv{\Phi}{x^{i}}
  +\CollisionEnergyEquationt, 
  \label{eq:radiationEnergyEquationGreyNewtonianOrderV}
\end{equation}
and the grey Eulerian-frame radiation momentum equation, 
\begin{eqnarray}
  &&
  \f{1}{\mdet}\pderiv{\Big(\smdet\,\Fourvectord{F}{i}\Big)}{t}
  +\f{1}{\mdet}\pderiv{\left(\,\mdet\,\Tensorud{S}{j}{i}\,\right)}{x^{j}}
  -\psit^{4}\f{\Tensoruu{S}{j}{k}}{2}\pderiv{\gmbdd{j}{k}}{x^{i}} \nonumber \\
  && \hspace{0.0in}
  =-\Big(J+\Tensorud{K}{j}{j}\Big)\pderiv{\Phi}{x^{i}}
  +\CollisionMomentumEquationd{i}, 
  \label{eq:radiationMomentumEquationGreyNewtonianOrderV}
\end{eqnarray}
where the grey radiation variables are
\begin{eqnarray}
  &&
  \big\{
    \tilde{E},\Fourvectortu{F}{i},\Fourvectoru{F}{i},\Tensoruu{S}{i}{j},J,\Tensoruu{K}{i}{j},\CollisionEnergyEquationt,\CollisionMomentumEquation{i}
  \big\} \nonumber \\
  && \hspace{0.0in}
  =\int_{0}^{\infty}
  \big\{
    \mathcalt{E},\fourvectortu{F}{i},\fourvectoru{F}{i},\tensoruu{S}{i}{j},\mathcal{J},\tensoruu{K}{i}{j},\collisionEnergyEquationt,\collisionMomentumEquation{i}
  \big\}\,\epsilon^{2}\,d\epsilon.  
\end{eqnarray}
Note that we use the font type to distinguish the energy-dependent radiation variables (denoted with calligraphic font) from the grey radiation variables.  
The Eulerian-frame grey radiation energy density, momentum density, and stress ($E$, $\Fourvectoru{F}{i}$, and $\Tensoruu{S}{i}{j}$, respectively) are related to the corresponding comoving-frame quantities ($J$, $\Fourvectoru{H}{i}$, and $\Tensoruu{K}{i}{j}$) via relations analogous to those listed in Equations (\ref{eq:eulerianEnergyLagrangianProjectionsOrderV})-(\ref{eq:eulerianStressLagrangianProjectionsOrderV}).  
In the strict $\mathcal{O}(v)$ limit, for Cartesian coordinates, and with $\Phi=0$, Equations (\ref{eq:radiationEnergyEquationGreyNewtonianOrderV}) and (\ref{eq:radiationMomentumEquationGreyNewtonianOrderV}) correspond to Equations (32a) and (32b) given by \citet{lowrie_etal_2001}, which they refer to as the `correct' $\mathcal{O}(v)$ Eulerian-frame radiation energy and momentum equations, respectively.  
One requirement for the correct $\mathcal{O}(v)$ limit is that the hyperbolic wave speeds associated with the system of equations is bounded by the speed of light.  
The presence of geometry sources due to the gravitational field or the use of curvilinear coordinates does not destroy the hyperbolic character of the equations, since they do not contain any differential operators acting on the radiation variables \citep{banyuls_etal_1997}.  

Equations (\ref{eq:radiationEnergyEquationNewtonianOrderVV}) and (\ref{eq:radiationMomentumEquationNewtonianOrderV}) are correct to $\mathcal{O}(v^{2})$ and $\mathcal{O}(v)$, respectively.  
We have explicitly retained higher-order (nonlinear) terms in the fluid velocity inside the energy derivative of both equations.  
In the energy equation, we have also retained terms containing the dot product of the fluid three-velocity with the gradient of the gravitational potential.  
However, when combined, the system is only accurate to $\mathcal{O}(v)$.  
Hence, we refer to these equations as the `$\mathcal{O}(v)$-plus' approximation of the radiation moment equations.  
These higher-order terms must (by definition) be small in the $\mathcal{O}(v)$ limit we are interested in here.  
However, by retaining these terms, the solution to Equations (\ref{eq:radiationEnergyEquationNewtonianOrderVV}) and (\ref{eq:radiationMomentumEquationNewtonianOrderV}) becomes consistent with the conservative $\mathcal{O}(v^{2})$ neutrino number equation.  
This consistency may help enable exact lepton number conservation in the two-moment model for neutrino radiation transport in the $\mathcal{O}(v)$ limit.  
We elaborate further on the details here.  

Equation (\ref{numberDensityFromEnergyAndFluxNewtonian}) relates the Eulerian-frame neutrino number density to the corresponding neutrino energy density and momentum density.  
The equations governing their time evolution are similarly related.  
In particular, by adding $\epsilon^{-1}\lorentzApp$ times Equation (\ref{eq:radiationEnergyEquationNewtonianOrderVV}) and $-\epsilon^{-1}\lorentzApp\,\threevelocityEu{i}$ contracted with Equation (\ref{eq:radiationMomentumEquationNewtonianOrderV}) we obtain
\begin{eqnarray}
  &&
  \f{1}{\mdet}\pderiv{}{t}
  \Big(\smdet\,\f{\lorentzApp}{\epsilon}\Big[\mathcalt{E}-\threevelocityEu{i}\,\fourvectord{F}{i}\Big]\Big) \nonumber \\
  && \hspace{0.15in}
  +\f{1}{\epsilon}
  \Big[
    \lorentzApp\fourvectord{F}{i}\f{1}{\alphat}\pderiv{\threevelocityEu{i}}{t}
    -\Big(\mathcalt{E}-\threevelocityEu{i}\,\fourvectord{F}{i}\Big)\f{1}{\alphat}\pderiv{\lorentzApp}{t}
  \Big]
  \label{eq:timeDerivativeTermsNewtonianOrderV}
\end{eqnarray}
from the time derivative terms.  
Remember $\lorentzApp=1+\f{1}{2}v^{2}$.  
Similarly we obtain
\begin{eqnarray}
  &&
  \f{1}{\mdet}\pderiv{}{x^{j}}
  \Big(
    \mdet\,\f{\lorentzApp}{\epsilon}\Big[\fourvectortu{F}{j}-\threevelocityEu{i}\,\tensorud{S}{j}{i}\Big]
  \Big) \nonumber \\
  && \hspace{0.15in}
  +\f{1}{\epsilon}
  \Big[
    \lorentzApp\tensordu{S}{i}{j}\pderiv{\threevelocityEu{i}}{x^{j}}
    -\Big(\fourvectortu{F}{j}-\threevelocityEu{i}\,\tensordu{S}{i}{j}\Big)\pderiv{\lorentzApp}{x^{j}}
  \Big]
  \label{eq:spaceDerivativeTermsNewtonianOrderV}
\end{eqnarray}
from the space derivative terms, while the combination of the geometry sources results in
\begin{equation}
  \f{\lorentzApp}{\epsilon}
  \Big\{
    \f{1}{2}\psit^{4}\tensoruu{S}{i}{j}\threevelocityEu{k}\pderiv{\gmbdd{i}{j}}{x^{k}}
    +\Big(\fourvectoru{H}{j}+\threevelocityEu{i}\tensordu{K}{i}{j}+\tensorud{K}{i}{i}\threevelocityEu{j}\Big)\pderiv{\Phi}{x^{j}}
  \Big\}.
  \label{eq:geometrySourceTermsNewtonianOrderV}
\end{equation}
From the energy derivative terms we obtain
\begin{equation}
  -\f{1}{\epsilon^{2}}\pderiv{}{\epsilon}
  \Big(\epsilon^{2}\,\tilde{\mathcal{F}}_{{\scriptscriptstyle\mathcal{N}}}^{\epsilon}\Big)
  -\f{1}{\epsilon}\,\tilde{\mathcal{F}}_{{\scriptscriptstyle\mathcal{N}}}^{\epsilon}, 
  \label{eq:energyDerivativeTermsNewtonianOrderV}
\end{equation}
where the number-flux in energy space is obtained from
\begin{eqnarray}
  \tilde{\mathcal{F}}_{{\scriptscriptstyle\mathcal{N}}}^{\epsilon}
  &=&
  \f{\lorentzApp}{\epsilon}\Big(\tilde{\mathcal{F}}^{\epsilon}-\threevelocityEu{i}\,\tilde{\mathcal{S}}^{\epsilon}_{~i}\Big) \nonumber \\
  &=&
  \lorentzApp
  \Big\{
    \Big(\fourvectord{H}{i}+\mathcal{J}\threevelocityEd{i}+\threevelocityEu{j}\tensordd{K}{i}{j}\Big)\f{1}{\alphat}\pderiv{\threevelocityEu{i}}{t} \nonumber \\
  && \hspace{0.2in}
  +\Big(\tensordu{K}{i}{j}+\threevelocityEd{i}\fourvectoru{H}{j}+\fourvectord{H}{i}\threevelocityEu{j}\Big)\pderiv{\threevelocityEu{i}}{x^{j}} \nonumber \\
  && \hspace{0.2in}
  +\f{1}{2}\psit^{4}\Big(\tensoruu{K}{i}{j}+\threevelocityEu{i}\fourvectoru{H}{j}+\fourvectoru{H}{i}\threevelocityEu{j}\Big)\threevelocityEu{k}\pderiv{\gmbdd{i}{j}}{x^{k}} \nonumber \\
  && \hspace{0.2in}
  +\Big(\fourvectoru{H}{j}+\threevelocityEu{i}\tensordu{K}{i}{j}+\tensorud{K}{i}{i}\threevelocityEu{j}\Big)\pderiv{\Phi}{x^{j}}
  \Big\} \nonumber \\
  && \hspace{0.0in}
  -\Big(\mathcalt{E}-\threevelocityEu{i}\fourvectord{F}{i}\Big)\f{1}{\alphat}\pderiv{\lorentzApp}{t}
  -\Big(\fourvectortu{F}{j}-\threevelocityEu{i}\tensordu{S}{i}{j}\Big)\pderiv{\lorentzApp}{x^{j}}.
  \label{eq:energySpaceNumberFluxNewtonian}
\end{eqnarray}
Finally, the collision terms combine to give
\begin{eqnarray}
  \collisionNumberEquationt
  &=&\f{\lorentzApp}{\epsilon}\Big(\collisionEnergyEquationt-\threevelocityEu{i}\,\collisionMomentumEquationd{i}\Big) \nonumber \\
  &=&\Big(1-\f{1}{4}v^{4}\Big)\f{1}{\epsilon}\int_{\Omega}\collision{f}d\Omega.
  \label{eq:collisionTermNumberEquationNewtonianOrderV}
\end{eqnarray}
(Note that $\threevelocityEd{i}\,\tetudb{i}{\imath}\,\deltaubdh{\imath}{\imath}=\vbdh{\imath}$; Appendix \ref{app:srMomentEquations}.)  

When adding Equations (\ref{eq:timeDerivativeTermsNewtonianOrderV})-(\ref{eq:energyDerivativeTermsNewtonianOrderV}), which result in the left-hand side of the conservative neutrino number equation (Equation (\ref{eq:numberEquationNewtonianOrderVV}) below), we note that the second term in Equation (\ref{eq:energyDerivativeTermsNewtonianOrderV}) (cf. Equation (\ref{eq:energySpaceNumberFluxNewtonian})) cancels exactly with the `left-over' terms in Equations (\ref{eq:timeDerivativeTermsNewtonianOrderV}) and (\ref{eq:spaceDerivativeTermsNewtonianOrderV}), and Equation (\ref{eq:geometrySourceTermsNewtonianOrderV}).  
In particular, the first term in the second line of Equation (\ref{eq:timeDerivativeTermsNewtonianOrderV}) cancels with terms emanating from the second line on the right-hand side of Equation (\ref{eq:energySpaceNumberFluxNewtonian}).  
Similarly, the first term in the second line of Equation (\ref{eq:spaceDerivativeTermsNewtonianOrderV}) cancels with terms emanating from the third line on the right-hand side of Equation (\ref{eq:energySpaceNumberFluxNewtonian}).  
The terms emanating from the geometry sources (Equation (\ref{eq:geometrySourceTermsNewtonianOrderV})) cancel with terms emanating from the fourth and fifth line on the right-hand side of Equation (\ref{eq:energySpaceNumberFluxNewtonian}).  
Finally, the terms involving time and space derivatives of the approximate Lorentz factor $\lorentzApp$ cancel with the terms emanating from the last line on the right-hand side of Equation (\ref{eq:energySpaceNumberFluxNewtonian}).  
Moreover, when contracted with $-\lorentzApp\epsilon^{-1}\,\threevelocityEu{i}$, several terms in the energy derivative of the radiation momentum equation cancel exactly with some of the $\mathcal{O}(v^{2})$ terms in the energy derivative of the $\lorentzApp\epsilon^{-1}$-multiplied radiation energy equation (these cancellations are only obtained by retaining higher-order terms in Equation (\ref{eq:radiationEnergyEquationNewtonianOrderVV})).  
We also find that the $\mathcal{O}(v^{2})$ terms involving the time and space derivatives of $\lorentzApp$ in Equations (\ref{eq:energySpaceEnergyFluxNewtonianOrderVV}) and (\ref{eq:energySpaceMomentumFluxNewtonian}) are needed to cancel with the corresponding terms in Equations (\ref{eq:timeDerivativeTermsNewtonianOrderV}) and (\ref{eq:spaceDerivativeTermsNewtonianOrderV}), obtained after pulling $\lorentzApp$ inside the time and space derivatives.  

By combining all the terms (Equations (\ref{eq:timeDerivativeTermsNewtonianOrderV})-(\ref{eq:collisionTermNumberEquationNewtonianOrderV})) we obtain
\begin{eqnarray}
  &&
  \f{1}{\mdet}\pderiv{}{t}\Big(\smdet\,\mathcaltmcd{E}{N}\Big)
  +\f{1}{\mdet}\pderiv{}{x^{i}}\Big(\mdet\,\fourvectortua{F}{i}{N}\Big) \nonumber \\
  && \hspace{0.15in}
  -\f{1}{\epsilon^{2}}\pderiv{}{\epsilon}\Big(\epsilon^{2}\,\mathcalt{F}_{{\scriptscriptstyle\mathcal{N}}}^{\epsilon}\Big)
  =\collisionNumberEquationt, 
  \label{eq:numberEquationNewtonianOrderVV}
\end{eqnarray}
where the Eulerian-frame $\mathcal{O}(v^{2})$ number density and number flux density are (cf. Equations (\ref{eq:eulerianNumberDensityLagrangianProjections}) and (\ref{eq:eulerianNumberFluxLagrangianProjections}))
\begin{eqnarray}
  \mathcaltmcd{E}{N}
  &=&\f{\lorentzApp}{\epsilon}\Big(\mathcalt{E}-\threevelocityEu{i}\fourvectord{F}{i}\Big)
  =\lorentzApp\mathcalmcd{J}{N}+\threevelocityEd{i}\fourvectorua{H}{i}{N}+\mathcal{O}(v^{3}), 
  \label{eq:eulerianNumberLagrangianProjectionsOrderVV} \\
  \fourvectortua{F}{i}{N}
  &=&\f{\lorentzApp}{\epsilon}\Big(\fourvectortu{F}{i}-\threevelocityEu{j}\tensorud{S}{i}{j}\Big)
  =\fourvectorua{H}{i}{N}+\threevelocityEu{i}\mathcalmcd{J}{N}+\mathcal{O}(v^{3}), 
  \label{eq:eulerianNumberFluxLagrangianProjectionsOrderVV}
\end{eqnarray}
where we have used Equations (\ref{eq:eulerianEnergyLagrangianProjectionsOrderVV})-(\ref{eq:eulerianMomentumLagrangianProjectionsOrderVV}) and (\ref{eq:eulerianEnergyLagrangianProjectionsOrderV})-(\ref{eq:eulerianStressLagrangianProjectionsOrderV}).  
We have $\fourvectorua{H}{i}{N}=\tetudb{i}{\imath}\,\deltaubdh{\imath}{\imath}\,\fourvectoruha{H}{\imath}{N}$, and the comoving-frame moments $\mathcalmcd{J}{N}$ and $\fourvectoruha{H}{\imath}{N}$ are defined in Section \ref{sec:angularMoments}.  
We have retained the $\mathcal{O}(v^{4})$ term appearing in the collision term on the right-hand side of Equation (\ref{eq:numberEquationNewtonianOrderVV}) (defined in Equation (\ref{eq:collisionTermNumberEquationNewtonianOrderV})), but this term can safely be dropped in practical computations.  

Equation (\ref{eq:numberEquationNewtonianOrderVV}) is a conservative equation for the Eulerian-frame neutrino number density.  
It is valid to $\mathcal{O}(v^{2})$.  
Together with an equation for the electron density, it states that the total lepton number is conserved during lepton number exchange with the fluid (see discussion in Section \ref{sec:conservation}).  
In the absence of neutrino-matter interactions, Equation (\ref{eq:numberEquationNewtonianOrderVV}) states conservation of particle number.  
It is obtained analytically from the monochromatic radiation energy and momentum equations.  
Modulo terms of $\mathcal{O}(v^{3})$ or higher (cf. Equations (\ref{eq:collisionTermNumberEquationNewtonianOrderV}), (\ref{eq:eulerianNumberLagrangianProjectionsOrderVV}), and (\ref{eq:eulerianNumberFluxLagrangianProjectionsOrderVV})), it can also be obtained directly from Equation (\ref{eq:numberEquationCFC}).  

A numerical solution to Equations (\ref{eq:radiationEnergyEquationNewtonianOrderVV}) and (\ref{eq:radiationMomentumEquationNewtonianOrderV}) should also be consistent with Equation (\ref{eq:numberEquationNewtonianOrderVV}) in order to ensure lepton number conservation in simulations of neutrino radiation transport in core-collapse supernovae and related systems (Section \ref{sec:conservation}).  
Ideally, the discretized neutrino energy and momentum equations can be similarly combined to obtain a discrete representation of the conservative neutrino number equation.  
Note that we arrived at Equation (\ref{eq:numberEquationNewtonianOrderVV}) in an analytically exact manner (due to exact cancellations; we did not throw away any $\mathcal{O}(v^{3})$ terms outside the time, space, or energy derivatives).  
By retaining the higher-order terms in Equation (\ref{eq:radiationEnergyEquationNewtonianOrderVV}) and Equation (\ref{eq:radiationMomentumEquationNewtonianOrderV}), the two-moment model of neutrino radiation transport is consistent with the conservative $\mathcal{O}(v^{2})$ neutrino number equation.  
If the higher-order terms in Equations (\ref{eq:radiationEnergyEquationNewtonianOrderVV}) and (\ref{eq:radiationMomentumEquationNewtonianOrderV}) are not retained (i.e., in the strict $\mathcal{O}(v)$ limit), there will be additional $\mathcal{O}(v^{2})$ terms in the number equation derived from the $\mathcal{O}(v)$ energy and momentum equations, and the consistency of the two-moment model with the number equation is only $\mathcal{O}(v)$---which may be acceptable in practical numerical computations.  
However, note that by omitting the $\partial\threevelocityEu{i}/\partial t$-term inside the energy derivative in the energy equation, the consistency with the conservative neutrino number equation can reduce to $\mathcal{O}(1)$.  
Thus, omitting this term in numerical simulations of core-collapse supernovae can potentially result in severe consequences for lepton number conservation.  

By integrating Equation (\ref{eq:numberEquationNewtonianOrderVV}) over the comoving-frame energy we obtain the Eulerian-frame grey neutrino number equation
\begin{equation}
  \f{1}{\mdet}\pderiv{}{t}\Big(\smdet\,\tilde{E}_{\scriptscriptstyle N}\Big)
  +\f{1}{\mdet}\pderiv{}{x^{i}}\Big(\mdet\,\Fourvectortua{F}{i}{N}\Big)
  =\CollisionNumberEquationt{}, 
  \label{eq:numberEquationGreyNewtonianOrderV}
\end{equation}
where
\begin{equation}
  \big\{\,\tilde{E}_{\scriptscriptstyle N},\,\Fourvectortua{F}{i}{N},\,\CollisionNumberEquationt{}\,\big\}
  =\int_{0}^{\infty}
  \big\{\,\mathcaltmcd{E}{N},\,\fourvectortua{F}{i}{N},\,\collisionNumberEquationt{}\,\big\}\,\epsilon^{2}\,d\epsilon.  
\end{equation}

We have presented conservative equations for the monochromatic lab-frame radiation energy density and momentum density---valid to $\mathcal{O}(v^{2})$ and $\mathcal{O}(v)$, respectively---for common curvilinear coordinates (Cartesian, spherical polar, and cylindrical); Equations (\ref{eq:radiationEnergyEquationNewtonianOrderVV}) and (\ref{eq:radiationMomentumEquationNewtonianOrderV}), respectively.  
We have also demonstrated how these equations (and therefore also their solutions) are fully consistent with the conservative equation for the monochromatic $\mathcal{O}(v^{2})$ lab-frame radiation particle density (Equation (\ref{eq:numberEquationNewtonianOrderVV})).  
Note that for a consistent description of neutrino radiation hydrodynamics with the $\mathcal{O}(v)$-plus moment equations, the hydrodynamics equations may also have to be promoted to include higher-order terms in the fluid velocity.  

\subsection{Further Simplifications of the Neutrino Radiation Transport Equations: $\mathcal{O}(v)$-Minus Moment Equations}
\label{sec:momentsOrderVMinus}

In this subsection we further specialize the radiation moment equations presented in Section \ref{sec:momentsOrderVPlus} by introducing the $\mathcal{O}(v)$-minus moment equations.  
Although apparently less complex than the fully relativistic moment equations (cf. \citealt{shibata_etal_2011,cardall_etal_2012}, Section \ref{sec:momentEquationsCFC}), the $\mathcal{O}(v)$-plus moment equations presented in Section \ref{sec:momentsOrderVPlus} are still nontrivial to discretize for numerical computations.  
(Their complexity rivals that of the pseudo-Newtonian moment equations presented in Section \ref{sec:pseudoNewtonian}.)
Therefore, we propose a further simplification, which involves solving the $\mathcal{O}(v)$ energy equation and the $\mathcal{O}(1)$ momentum equation, as a useful intermediate (first) step beyond the $\mathcal{O}(1)$ moment formalism previously used by some to model neutrino radiation transport in core-collapse supernovae \citep[e.g.,][]{burrows_etal_2006}.  
The resulting system is formally valid to $\mathcal{O}(1)$.  
However, the moment equations are consistent with the conservative $\mathcal{O}(v)$ neutrino number equation.  
Moreover, the equations presented here evolve the radiation energy density \emph{and} momentum density, and, with a proper closure prescription, may be an improvement over the equations solved in common $\mathcal{O}(v)$ multigroup flux-limited diffusion approaches \citep[e.g.,][]{swestyMyra_2009,zhang_etal_2012}.  
In the non-relativistic, Newtonian gravity limit, we have $\Phi,c_{s}^{2},v^{2}\ll1$, where $c_{s}$ is the sound speed.  
Thus, we ignore gravitational effects on the radiation field in this subsection (i.e., $\Phi=0$).  

Then, the monochromatic $\mathcal{O}(v)$ lab-frame radiation energy equation becomes (cf. Equation (\ref{eq:radiationEnergyEquationNewtonianOrderVV}))
\begin{eqnarray}
  &&
  \pderiv{\mathcal{E}}{t}
  +\f{1}{\smbdet}\pderiv{}{x^{i}}\Big(\smbdet\,\fourvectoru{F}{i}\Big)
  -\f{1}{\epsilon^{2}}\pderiv{}{\epsilon}\Big(\epsilon^{2}\,\mathcal{F}^{\epsilon}\Big) \nonumber \\
  && \hspace{0.15in}
  =\int_{\Omega}\collision{f}d\Omega+\vbdh{\imath}\int_{\Omega}\nuh{\imath}\,\collision{f}d\Omega
  =\collisionEnergyEquation{},
  \label{eq:radiationEnergyEquationOrderV}
\end{eqnarray}
where the $\mathcal{O}(v)$ lab-frame radiation energy density and momentum density, $\mathcal{E}$ and $\fourvectoru{F}{i}$, are given by Equations (\ref{eq:eulerianEnergyLagrangianProjectionsOrderV}) and (\ref{eq:eulerianMomentumLagrangianProjectionsOrderV}), respectively.  
The energy space energy flux $\mathcal{F}^{\epsilon}$ is given by Equation (\ref{eq:energySpaceEnergyFluxNewtonianOrderV}), with $\Phi=0$.  

Similarly, the monochromatic $\mathcal{O}(1)$ radiation momentum equation becomes (cf. Equation (\ref{eq:radiationMomentumEquationNewtonianOrderV}))
\begin{eqnarray}
  &&
  \pderiv{\fourvectord{H}{i}}{t}
  +\f{1}{\smbdet}\pderiv{}{x^{j}}\Big(\smbdet\,\tensorud{K}{j}{i}\Big)
  -\f{1}{2}\tensoruu{K}{j}{k}\pderiv{\gmbdd{j}{k}}{x^{i}} \nonumber \\
  && \hspace{0.15in}
  =\gmbdd{i}{j}\,\tetudb{j}{\jmath}\,\deltaubdh{\jmath}{\jmath}\int_{\Omega}\nuh{\jmath}\,\collision{f}d\Omega
  =\collisionMomentumEquationOrderOned{i}.  
  \label{eq:radiationMomentumEquationOrderOne}
\end{eqnarray}
In Equation (\ref{eq:radiationMomentumEquationOrderOne}), the energy derivative terms vanish since we have dropped all the velocity-dependent terms in addition to the gravitational terms (cf. Equation (\ref{eq:energySpaceMomentumFluxNewtonian})).  

Equations (\ref{eq:radiationEnergyEquationOrderV}) and (\ref{eq:radiationMomentumEquationOrderOne}) are consistent with the conservative $\mathcal{O}(v)$ neutrino number equation: by adding $\epsilon^{-1}$ times Equation (\ref{eq:radiationEnergyEquationOrderV}) and $-\epsilon^{-1}\threevelocityEu{i}$ contracted with Equation (\ref{eq:radiationMomentumEquationOrderOne}) we obtain
\begin{equation}
  \pderiv{\mathcal{E}_{{\scriptscriptstyle\mathcal{N}}}}{t}
  +\f{1}{\smbdet}\pderiv{}{x^{i}}\Big(\smbdet\,\fourvectorua{F}{i}{N}\Big)
  -\f{1}{\epsilon^{2}}\pderiv{}{\epsilon}\Big(\epsilon^{2}\,\mathcal{F}_{{\scriptscriptstyle\mathcal{N}}}^{\epsilon}\Big)
  =\f{1}{\epsilon}\int_{\Omega}\collision{f}d\Omega,
  \label{eq:numberEquationOrderV}
\end{equation}
where the $\mathcal{O}(v)$ lab-frame number density and number flux density, and the number flux in energy space are
\begin{eqnarray}
  \mathcal{E}_{{\scriptscriptstyle\mathcal{N}}}
  &=&\mathcalmcd{J}{N}+\threevelocityEd{i}\,\fourvectorua{H}{i}{N}, \\
  \fourvectorua{F}{i}{N}
  &=&\fourvectorua{H}{i}{N}+\threevelocityEu{i}\,\mathcalmcd{J}{N}, \\
  \mathcal{F}_{{\scriptscriptstyle\mathcal{N}}}^{\epsilon}
  &=&\fourvectord{H}{i}\,\pderiv{\threevelocityEu{i}}{t}
  +\tensordu{K}{i}{j}\,\pderiv{\threevelocityEu{i}}{x^{j}}
  +\f{1}{2}\,\tensoruu{K}{i}{j}\,\threevelocityEu{k}\,\pderiv{\gmbdd{i}{j}}{x^{k}}.  
\end{eqnarray}

Equation (\ref{eq:numberEquationOrderV}) is the conservative $\mathcal{O}(v)$ number equation, and is obtained in a manner similar to the $\mathcal{O}(v^{2})$ number equation detailed in Section \ref{sec:momentsOrderVPlus}.  
On the left-hand side, the extra terms emanating from pulling the fluid three-velocity inside the time and space derivatives of the momentum equation cancel with the extra terms emanating from pulling $\epsilon^{-1}$ inside the energy derivative of the energy equation.  
On the right-hand side, the velocity-dependent term in the collision term of the energy equation cancels with the contraction of $-\epsilon^{-1}\threevelocityEu{i}$ with the collision term of the momentum equation.  
However, by reducing the order of the energy and momentum equations (to $\mathcal{O}(v)$ and $\mathcal{O}(1)$, respectively) the number of cancellations that occur is dramatically reduced.  
Because of these simplifications, Equations (\ref{eq:radiationEnergyEquationOrderV}) and (\ref{eq:radiationMomentumEquationOrderOne}), which are consistent with the conservative $\mathcal{O}(v)$ neutrino number equation, may be a suitable starting point for developing lepton number conservative numerical methods for neutrino radiation hydrodynamics based on the two-moment model.  

\subsection{Non-Relativistic, Self-Gravitating Hydrodynamics}
\label{sec:hydro}

For self-consistent Newtonian, non-relativistic simulations of self-gravitating neutrino radiation hydrodynamics, the moment equations for the radiation field in Section \ref{sec:momentsOrderVMinus} (one set for each of the neutrino species) must be coupled to the equations of self-gravitating hydrodynamics.  
For completeness we list the non-relativistic hydrodynamics equations including self-gravity in this subsection.  
(For brevity do we not consider nuclear reactions here.  However, see for example \citealt[][]{plewaMuller_1999}.)
We consider a perfect fluid; i.e., we ignore fluid viscosity and thermal conduction.  

The equations describing a self-gravitating perfect fluid \citep[e.g.,][]{landauLifshitz_1959} include the mass conservation equation
\begin{equation}
  \pderiv{\rho}{t}+\f{1}{\smbdet}\pderiv{}{x^{i}}\Big(\smbdet\rho\,\threevelocityEu{i}\Big)=0,
  \label{eq:massConservationEquationNewtonianOrderVV}
\end{equation}
the fluid momentum equation
\begin{equation}
  \pderiv{\Fourvectorda{F}{i}{f}}{t}
  +\f{1}{\smbdet}\pderiv{}{x^{j}}\Big(\smbdet\Tensoruda{S}{j}{i}{f}\Big)
  -\f{\Tensoruua{S}{j}{k}{f}}{2}\pderiv{\gmbdd{j}{k}}{x^{i}}
  =
  -\rho\,\pderiv{\Phi}{x^{i}}
  -\CollisionMomentumEquationOrderOned{i},
  \label{eq:fluidMomentumEquationNewtonianOrderVV}
\end{equation}
and the fluid energy equation
\begin{equation}
  \pderiv{E_{\scriptscriptstyle f}}{t}
  +\f{1}{\smbdet}\pderiv{}{x^{i}}\Big(\smbdet\Big[E_{\scriptscriptstyle f}+p\Big]\threevelocityEu{i}\Big)
  =
  -\rho\threevelocityEu{i}\pderiv{\Phi}{x^{i}}
  -\CollisionEnergyEquation{}.  
  \label{eq:fluidEnergyEquationNewtonianOrderVV}
\end{equation}
For simulations involving a nuclear equation of state, Equations (\ref{eq:massConservationEquationNewtonianOrderVV})-(\ref{eq:fluidEnergyEquationNewtonianOrderVV}) must be supplied with a balance equation for the electron number density
\begin{equation}
  \pderiv{n_{e}}{t}+\f{1}{\smbdet}\pderiv{}{x^{i}}\Big(\smbdet\,n_{e}v^{i}\Big)
  =-\Big(\CollisionNumberEquation{\nue}-\CollisionNumberEquation{\nueb}\Big).  
  \label{eq:electronNumberEquationNewtonianOrderVV}
\end{equation}
In Equations (\ref{eq:massConservationEquationNewtonianOrderVV})-(\ref{eq:electronNumberEquationNewtonianOrderVV}), the fluid energy density, momentum density and stress are $E_{\scriptscriptstyle f}=e+\f{1}{2}\rho\,v_{i}\,v^{i}$, $\Fourvectorda{F}{i}{f}=\rho\,\threevelocityEd{i}$, and $\Tensoruda{S}{i}{j}{f}=\rho\,\threevelocityEu{i}\,\threevelocityEd{j}+\deltaud{i}{j}\,p$, where $\rho=\bar{m}_{b}\,n_{b}$, $\threevelocityEu{i}$, $p$, $e$, and $n_{e}$ denote the mass density, the $i$th component of the fluid three-velocity, the fluid pressure and internal energy density, and the electron density (electrons minus positrons), respectively.  
The average baryon mass and the baryon density are denoted $\bar{m}_{b}$ and $n_{b}$.  
Equations (\ref{eq:massConservationEquationNewtonianOrderVV})-(\ref{eq:electronNumberEquationNewtonianOrderVV}) are closed with the specification of an equation of state $p=p(\rho,T,Y_{e})$, where $T$ and $Y_{e}=n_{e}/n_{b}$ are the fluid temperature and electron fraction, respectively.  
In Equations (\ref{eq:fluidMomentumEquationNewtonianOrderVV}) and (\ref{eq:fluidEnergyEquationNewtonianOrderVV}), the collision terms include energy and momentum exchange with all neutrino species $s$; i.e., 
\begin{equation}
  \big\{\,
    \CollisionEnergyEquation{},\,
    \CollisionMomentumEquationOrderOned{i}{}
  \,\big\}
  =\sum_{s}
  \big\{\,
    \CollisionEnergyEquations,\,
    \CollisionMomentumEquationOrderOneds{i}
  \,\big\}.  
\end{equation}
Here we consider electron lepton number exchange between the fluid and the neutrino radiation field.  
Only interactions involving electron neutrinos ($\nue$) and electron antineutrinos ($\nueb$) contribute to the right-hand side of Equation (\ref{eq:electronNumberEquationNewtonianOrderVV}).  

The Newtonian gravitational potential is obtained by solving Poisson's equation
\begin{equation}
  \f{1}{\smbdet}\pderiv{}{x^{i}}\Big(\smbdet\,\gmbuu{i}{j}\pderiv{\Phi}{x^{j}}\Big)=4\pi\,\rho.  
  \label{eq:poissonEquation}
\end{equation}
Note that in the non-relativistic, Newtonian gravity limit only the mass density contributes to the gravitational field (as opposed to all types of stress-energy in general relativity); i.e., we assume $v^{2}\ll1$ and $e,p,J,\Tensorud{K}{i}{i}\ll\rho$.  

Using Equation (\ref{eq:poissonEquation}), we can rewrite the gravitational force in Equation (\ref{eq:fluidMomentumEquationNewtonianOrderVV}) as
\begin{equation}
  \rho\pderiv{\Phi}{x^{i}}
  =\f{1}{\smbdet}\pderiv{}{x^{j}}\Big(\smbdet\,\Tensoruda{S}{j}{i}{\Phi}\Big)-\f{1}{2}\Tensoruua{S}{j}{k}{\Phi}\pderiv{\gmbdd{j}{k}}{x^{i}}, 
\end{equation}
where the gravitational stress tensor is defined as
\begin{equation}
  \Tensoruda{S}{i}{j}{\Phi}
  =\f{1}{4\pi}\Big(\Phi^{,i}\,\Phi_{,j}-\f{1}{2}\,\Phi_{,k}\,\Phi^{,k}\,\deltaud{i}{j}\Big), 
\end{equation}
with $\Phi_{,i}=\partial\Phi/\partial x^{i}$ and $\Phi^{,i}=\gmbuu{i}{j}\,\Phi_{,j}$.  
We then obtain a conservative fluid momentum equation
\begin{eqnarray}
  &&
  \pderiv{\Fourvectorda{F}{i}{f}}{t}
  +\f{1}{\smbdet}\pderiv{}{x^{j}}\Big(\smbdet\,\Big[\Tensoruda{S}{j}{i}{f}+\Tensoruda{S}{j}{i}{\Phi}\Big]\Big) \nonumber \\
  && \hspace{0.15in}
  -\f{1}{2}\Big(\Tensoruua{S}{j}{k}{f}+\Tensoruua{S}{j}{k}{\Phi}\Big)\pderiv{\gmbdd{j}{k}}{x^{i}}
  =-\CollisionMomentumEquationOrderOned{i}.
  \label{eq:fluidMomentumEquationConservativeNewtonianOrderVV}
\end{eqnarray}
In the absence of neutrino-matter interactions, and if Cartesian coordinates are used, Equation (\ref{eq:fluidMomentumEquationConservativeNewtonianOrderVV}) states that the fluid momentum is conserved.  

Using Equations (\ref{eq:massConservationEquationNewtonianOrderVV}), (\ref{eq:fluidEnergyEquationNewtonianOrderVV}), and (\ref{eq:poissonEquation}), we can write a conservative equation for the energy density $E_{\scriptscriptstyle f}+\f{1}{2}\,\rho\,\Phi$
\begin{equation}
  \pderiv{}{t}\Big(E_{\scriptscriptstyle f}+E_{\scriptscriptstyle \Phi}\Big)
  +\f{1}{\smbdet}\pderiv{}{x^{i}} 
  \Big(\smbdet
    \Big[
      \Big(E_{\scriptscriptstyle f}+p\Big)\threevelocityEu{i}+\Fourvectorua{F}{i}{\Phi}
    \Big]
  \Big)
  =-\CollisionEnergyEquation,
  \label{eq:fluidEnergyEquationConservativeNewtonianOrderVV}
\end{equation}
where $E_{\scriptscriptstyle \Phi}=\f{1}{2}\,\rho\,\Phi$ is the gravitational energy density, 
\begin{equation}
  \Fourvectorua{F}{i}{\Phi}
  =\rho\Phi\threevelocityEu{i}+\f{1}{8\pi}\,\gmbuu{i}{j}\Big(\Phi\,\pderiv{\Phi_{,t}}{x^{j}}-\Phi_{,t}\,\pderiv{\Phi}{x^{j}}\Big)
\end{equation}
is the gravitational energy flux density, and $\Phi_{,t}=\partial \Phi/\partial t$.  
Equation (\ref{eq:fluidEnergyEquationConservativeNewtonianOrderVV}) states that, in the absence of neutrino-matter interactions, the fluid energy (internal plus kinetic) plus the gravitational potential energy is conserved.  

\subsection{Conservation Laws in Non-Relativistic, Self-Gravitating Neutrino Radiation Hydrodynamics}
\label{sec:conservation}

We discuss conservation laws for non-relativistic, self-gravitating neutrino radiation hydrodynamics in this subsection.  
Tracking the evolution of conserved quantities is useful when evaluating the physical reliability of numerical simulations.  
Equations (\ref{eq:radiationEnergyEquationOrderV}) and (\ref{eq:radiationMomentumEquationOrderOne}) (or more precisely, their energy-integrated versions) combined with Equations (\ref{eq:massConservationEquationNewtonianOrderVV})-(\ref{eq:electronNumberEquationNewtonianOrderVV}) state the conservation of several physical quantities in neutrino radiation hydrodynamics.  
Mass conservation is trivially stated by Equation (\ref{eq:massConservationEquationNewtonianOrderVV}).  
Equations (\ref{eq:radiationMomentumEquationOrderOne}) and (\ref{eq:fluidMomentumEquationNewtonianOrderVV}) result in a total (fluid plus radiation) momentum equation
\begin{eqnarray}
  &&
  \pderiv{}{t}\Big(\Fourvectorda{F}{i}{f}+\Fourvectord{H}{i}\Big)
  +\f{1}{\smbdet}\pderiv{}{x^{j}}\Big(\smbdet\Big[\Tensoruda{S}{j}{i}{f}+\Tensorud{K}{j}{i}\Big]\Big) \nonumber \\
  && \hspace{0.15in}
  -\f{1}{2}\Big(\Tensoruua{S}{j}{k}{f}+\Tensoruu{K}{j}{k}\Big)\pderiv{\gmbdd{j}{k}}{x^{i}}
  =-\rho\pderiv{\Phi}{x^{i}}.  
  \label{eq:radHydroMomentumEquationNewtonianOrderV}
\end{eqnarray}
Alternatively, combining Equations (\ref{eq:radiationMomentumEquationOrderOne}) and (\ref{eq:fluidMomentumEquationConservativeNewtonianOrderVV}) results in
\begin{eqnarray}
  &&
  \pderiv{}{t}\Big(\Fourvectorda{F}{i}{f}+\Fourvectord{H}{i}\Big)
  +\f{1}{\smbdet}\pderiv{}{x^{j}}\Big(\smbdet\Big[\Tensoruda{S}{j}{i}{f}+\Tensoruda{S}{j}{i}{\Phi}+\Tensorud{K}{j}{i}\Big]\Big) \nonumber \\
  && \hspace{0.15in}
  -\f{1}{2}\Big(\Tensoruua{S}{j}{k}{f}+\Tensoruua{S}{j}{k}{\Phi}+\Tensoruu{K}{j}{k}\Big)\pderiv{\gmbdd{j}{k}}{x^{i}}
  =0.  
  \label{eq:radHydroMomentumEquationConservativeNewtonianOrderV}
\end{eqnarray}

Equations (\ref{eq:radiationEnergyEquationOrderV}) and (\ref{eq:fluidEnergyEquationNewtonianOrderVV}) result in a conservative equation for the ``total" (internal plus kinetic plus radiation) energy density of a radiating flow
\begin{eqnarray}
  &&
  \pderiv{}{t}\Big(E_{\scriptscriptstyle f}+E\Big)
  +\f{1}{\smbdet}\pderiv{}{x^{i}}\Big(\smbdet\Big[\left(E_{\scriptscriptstyle f}+p\right)\threevelocityEu{i}+\Fourvectoru{F}{i}\Big]\Big) \nonumber \\
  && \hspace{0.15in}
  =-\rho\threevelocityEu{i}\pderiv{\Phi}{x^{i}}.
  \label{eq:radHydroEnergyEquationNewtonianOrderV}
\end{eqnarray}
Combining Equations (\ref{eq:radiationEnergyEquationOrderV}) and (\ref{eq:fluidEnergyEquationConservativeNewtonianOrderVV}) results in a conservative equation for the total (internal plus kinetic plus gravitational plus radiation) energy density of a self-gravitating radiating flow
\begin{eqnarray}
  &&
  \pderiv{}{t}\Big(E_{\scriptscriptstyle f}+E_{\scriptscriptstyle \Phi}+E\Big) \nonumber \\
  && \hspace{0.05in}
  +\f{1}{\smbdet}\pderiv{}{x^{i}}
  \Big(\smbdet
    \Big[
      \left(E_{\scriptscriptstyle f}+p\right)\threevelocityEu{i}
      +\Fourvectorua{F}{i}{\Phi}
      +\Fourvectoru{F}{i}
    \Big]
  \Big)=0.
  \label{eq:radHydroEnergyEquationConservativeNewtonianOrderV}
\end{eqnarray}
In Equations (\ref{eq:radHydroMomentumEquationNewtonianOrderV})-(\ref{eq:radHydroEnergyEquationConservativeNewtonianOrderV}), the neutrino energy density, momentum density, and stress include contributions from all neutrino species; e.g., 
\begin{equation}
  \Big\{\,
    E,\,\Fourvectord{F}{i},\,\Tensorud{S}{j}{i}
  \,\Big\}
  =
  \sum_{s}
  \Big\{\,
    E_{{\scriptscriptstyle s}},\,\Fourvectorda{F}{i}{s},\,\Tensoruda{S}{j}{i}{s}
  \,\Big\}.
\end{equation}

Equations (\ref{eq:radHydroMomentumEquationConservativeNewtonianOrderV}) (assuming Cartesian coordinates are used) and (\ref{eq:radHydroEnergyEquationConservativeNewtonianOrderV}) are exactly (and locally) conservative.  
Therefore, energy and momentum conservation can be useful checks for evaluating the physical reliability of non-relativistic neutrino radiation hydrodynamics simulations based on the $\mathcal{O}(v)$-minus moment equations.  
However, for radiation hydrodynamics based on the $\mathcal{O}(v)$-plus moment equations, or the pseudo-Newtonian moment equations in Section \ref{sec:pseudoNewtonian}, where the gravitational potential is obtained by solving a modified Poisson equation \citep[e.g.,][]{kim_etal_2012} and the hydrodynamics equations are promoted to include more relativistic effects, we are unable to take steps similar to those taken to obtain Equations (\ref{eq:fluidMomentumEquationConservativeNewtonianOrderVV}) and (\ref{eq:fluidEnergyEquationConservativeNewtonianOrderVV}) from Equations (\ref{eq:fluidMomentumEquationNewtonianOrderVV}) and (\ref{eq:fluidEnergyEquationNewtonianOrderVV}), respectively.  
Moreover, there are additional source terms on the right-hand sides of the total momentum and energy equations due to changes in the radiation momentum and energy caused by the gravitational field (e.g., the `bending of light' effect and gravitational redshifts).  
These (locally) non-vanishing gravitational source terms (and the lack of global conservation) limit the usefulness of tracking total energy and momentum as checks on the physical reliability of (self-gravitating) neutrino radiation hydrodynamics simulations.  
\citep[Note however that in 3+1 general relativity, the AMD mass is conserved in asymptotically flat spacetimes; e.g.,][]{baumgarteShapiro_2010}.  

Consistency with the neutrino number equation results in lepton number conservation in (self-gravitating) neutrino radiation hydrodynamics.  
Combining the energy-integrated version of Equation (\ref{eq:numberEquationOrderV}) (for electron neutrinos and electron antineutrinos) with Equation (\ref{eq:electronNumberEquationNewtonianOrderVV}) results in a conservation equation for the total electron lepton number
\begin{eqnarray}
  &&
  \pderiv{}{t}\Big(E_{\scriptscriptstyle N,\nue}-E_{\scriptscriptstyle N,\nueb}+n_{e}\Big) \nonumber \\
  && \hspace{0.05in}
  +\f{1}{\smbdet}\pderiv{}{x^{i}}\Big(\smbdet\Big[\Fourvectoruas{\tilde{F}}{i}{N}{\nue}-\Fourvectoruas{\tilde{F}}{i}{N}{\nueb}+n_{e}\,\threevelocityEu{i}\Big]\Big)
  =0.
\end{eqnarray}
Lepton number conservation in numerical simulations based on solving Equations (\ref{eq:radiationEnergyEquationOrderV}) and (\ref{eq:radiationMomentumEquationOrderOne}), or any of the other radiation moment equations presented in this paper, may depend sensitively on the chosen discretization.  
In particular, the discretized neutrino energy and momentum equations must be consistent (in the sense discussed in Sections \ref{sec:momentsOrderVPlus} and {\ref{sec:momentsOrderVMinus}}) with a discretized version of the conservative neutrino number equation; Equation (\ref{eq:numberEquationOrderV}) in the particular case discussed here.  
When discretizing the monochromatic energy and momentum equations, attention should be paid to the cancellations that occur when deriving the neutrino number equation from the energy and momentum equations, which ideally also occur in the discrete limit.  
To achieve this, \emph{the individual terms in the moment equations for the neutrino four-momentum should be discretized in a coordinated rather than independent fashion}.  
Despite the lack of energy conservation due to gravitational redshifts discussed above, the total lepton number is still conserved.  
Thus, lepton number conservation may serve as an extremely useful gauge on the physical consistency of numerical simulations of neutrino radiation hydrodynamics \citep[e.g.,][]{liebendorfer_etal_2004,lentz_etal_2012a}.  

\section{SUMMARY AND DISCUSSION}
\label{sec:summaryAndDiscussion}

In preparation for development of numerical methods for multidimensional neutrino radiation hydrodynamics, with the eventual goal of simulating the explosion mechanism of core-collapse supernovae, we have derived conservative, monochromatic general relativistic moment equations for the radiation four-momentum (cf. Equation (\ref{eq:stressEnergyMoment})).  
The radiation moment equations are conservative in the sense that (modulo radiation-matter interactions and geometry sources) the time rate of change of the radiation four-momentum is governed by space and momentum space divergences.  
We have used the freedom to choose distinct spacetime and momentum space coordinates \citep[cf.][]{lindquist_1966,riffert_1986,mezzacappaMatzner_1989,cardallMezzacappa_2003}.  
The evolved radiation quantities are functions of the coordinate basis spacetime position components $x^{\mu}$ and the radiation energy $\epsilon$ measured by an observer comoving with the fluid.  
(When integrated over the comoving-frame energy, the equations reduce to familiar position space conservation laws.)  
The specific choice of phase space coordinates is motivated by our intent to develop numerical methods for computer simulations of multidimensional neutrino transport.  
Neutrino-matter interactions are on the one hand most easily handled computationally in the comoving-frame.  
On the other hand, conservation of global quantities (e.g., energy and lepton number) is naturally expressed in the laboratory frame.  
Convenient treatment of neutrino-matter interactions \emph{and} global conservation are naturally handled with the chosen phase space coordinates.  

We have presented radiation moment equations valid for conformally flat spacetimes (Section \ref{sec:momentEquationsCFC}; cf. Equations (\ref{eq:energyEquationCFC}) and (\ref{eq:momentumEquationCFC})).  
We have further specialized the radiation moment equations to the pseudo-Newtonian and the Newtonian gravity, $\mathcal{O}(v)$ limits (Sections \ref{sec:pseudoNewtonian} and \ref{sec:newtonianOrderV}).  
Furthermore, in the $\mathcal{O}(v)$ limit, we have presented the $\mathcal{O}(v)$-plus (Section \ref{sec:momentsOrderVPlus}) and $\mathcal{O}(v)$-minus (Section \ref{sec:momentsOrderVMinus}) moment equations.  
The $\mathcal{O}(v)$-plus radiation energy and momentum equations are given by Equations (\ref{eq:radiationEnergyEquationNewtonianOrderVV}) and (\ref{eq:radiationMomentumEquationNewtonianOrderV}), respectively.  
In the no gravity, strict $\mathcal{O}(v)$ limit, these equations are \emph{conservative} formulations of similar, \emph{non-conservative} equations presented by other authors \citep[e.g.,][]{buchler_1979,kaneko_etal_1984,munierWeaver_1986b}.  
The $\mathcal{O}(v)$-minus radiation energy and momentum equations are given by Equations (\ref{eq:radiationEnergyEquationOrderV}) and (\ref{eq:radiationMomentumEquationOrderOne}), respectively.  
Special relativistic moment equations are given in Appendix \ref{app:srMomentEquations}, while general relativistic moment equations for spherically symmetric spacetimes are given in Appendix \ref{app:momentEquationsSphericalSymmetryCFC}.  
The moment equations given in the appendices are also conservative versions of non-conservative equations presented by other authors \citep[e.g.,][]{castor_1972,mihalas_1980,munierWeaver_1986b,muller_etal_2010}.  
\citep[See][for conservative 3+1 general relativistic moment equations.]{shibata_etal_2011,cardall_etal_2012}  

We have paid special attention to the issue of neutrino number (and lepton number) conservation.  
For numerical methods based on solving moment equations for the neutrino four-momentum, total lepton number conservation will likely serve as a very useful gauge on the physical consistency of simulations of neutrino radiation hydrodynamics.  
To this end, we have exposed the relationship between the equations for the neutrino four-momentum and the neutrino number equation with multiple examples (cf. Sections \ref{sec:numberStressEnergyEquationRelation}, \ref{sec:momentEquationsCFC}, \ref{sec:pseudoNewtonian}, and \ref{sec:newtonianOrderV}, and Appendix \ref{app:srMomentEquations}).  
The lab-frame radiation number density is related to the lab-frame energy and momentum densities by Equation (\ref{numberDensityFromEnergyAndFluxNewtonian}).  
The conservative neutrino number equation and the equations for the neutrino four-momentum are similarly related.  
The non-relativistic limits of the radiation moment equations are not uniquely defined.  
We obtain consistency with the conservative neutrino number equation by adopting different orders of $v$ for the energy and momentum equations (cf. the $\mathcal{O}(v)$-minus and $\mathcal{O}(v)$-plus approximations in Section \ref{sec:newtonianOrderV}).  
In particular, as was detailed in Section \ref{sec:momentsOrderVPlus}, when carrying out the steps to obtain the conservative neutrino number equation from the conservative energy and momentum equations, we observe that terms emanating from the time and space derivatives and the geometry sources cancel with terms emanating from the energy derivatives.  
The remaining terms constitute the left-hand side of the number equation.  
(The right-hand sides of the equations are similarly related.)  
The energy and momentum equations are therefore consistent with the conservative number equation.  
In the $\mathcal{O}(v)$-plus limit, consistency with the conservative $\mathcal{O}(v^{2})$ number equation is obtained by adopting different orders of $v$ \emph{and} retaining some higher-order terms in the energy and momentum equations (accurate to $\mathcal{O}(v^{2})$ and $\mathcal{O}(v)$, respectively).  
In the $\mathcal{O}(v)$-minus limit, we obtain consistency with the conservative $\mathcal{O}(v)$ number equation by adopting energy and momentum equations accurate to $\mathcal{O}(v)$ and $\mathcal{O}(1)$, respectively.  
Ideally, discrete representations of the radiation energy and momentum equations can be constructed so that a conservative neutrino number equation can be analogously obtained in the discrete limit.  
The discretization is then consistent with neutrino number conservation (which is necessary to ensure lepton number conservation; Section \ref{sec:conservation}).  
The realization of this consistency in a numerical method for neutrino radiation transport based on the two-moment model derived here will be the focus of a future study.  

General relativistic effects are important in the core-collapse supernova environment \citep[e.g.,][]{bruenn_etal_2001,muller_etal_2012a}, and definitive simulations elucidating the explosion mechanism of core-collapse supernovae must eventually be performed in full general relativity (possibly employing multi-energy \emph{and} multi-angle neutrino transport).  
To this end, the pseudo-Newtonian and the Newtonian, $\mathcal{O}(v)$-plus equations may represent useful self-consistent approximations beyond the $\mathcal{O}(v)$-minus equations.  
They are closely related to the corresponding equations valid for conformally flat spacetimes, which again are close to the fully general relativistic equations \citep{shibata_etal_2011,cardall_etal_2012}.  
It seems plausible that numerical methods developed for the self-consistent moment equations presented in this paper---which arguably are easier to work with---can be extended in steps of increasing degree of complexity to the fully general relativistic case, as has been done in the case of conservative methods for hydrodynamics \citep[e.g.,][]{font_etal_1994,banyuls_etal_1997}.  

\acknowledgments

This research was supported by the Office of Advanced Scientific Computing Research and the Office of Nuclear Physics, U.S. Department of Energy.  

\appendix

\section{CONSERVATIVE SPECIAL RELATIVISTIC MOMENT EQUATIONS}
\label{app:srMomentEquations}

In this appendix we present conservative, multidimensional, monochromatic moment equations valid in special relativity.  
The equations are obtained from the general relativistic moment equations presented in Sections \ref{sec:momentEquationsDerivation} and \ref{sec:momentEquationsCFC} in the limit of flat spacetime.  
First we list sufficiently general equations to accommodate Cartesian, spherical polar, and cylindrical coordinates.  
Then we specialize the special relativistic moment equations to spherical symmetry and compare with the corresponding equations given by \citet{mihalasMihalas_1999}.  
Monochromatic, special relativistic moment equations have been presented elsewhere \citep[e.g.,][]{castor_1972,mihalas_1980,munierWeaver_1986b,mihalasMihalas_1999}.  
However, the equations presented in this appendix are in fully conservative form.  
We also present the conservative neutrino number equation, and discuss the relationship between the conservative number equation and the conservative radiation energy and momentum equations.  

The invariant line element is
\begin{equation}
  ds^{2}=g_{\mu\nu}\,dx^{\mu}\,dx^{\nu}, 
  \label{eq:srMetric}
\end{equation}
where, for flat spacetime, the covariant metric tensor is diagonal and given by $g_{\mu\nu}=\mbox{diag}[-1,1,a^{2}(x^{1}),b^{2}(x^{1})\,c^{2}(x^{2})]$.  
The contravariant metric tensor is $g^{\mu\nu}=\mbox{diag}[-1,1,a^{-2},b^{-2}\,c^{-2}]$.  
The transformation from the orthonormal tetrad basis to the global coordinate basis may simply be written as $\tetudb{\mu}{\mu}=\mbox{diag}[1,1,a^{-1},b^{-1}\,c^{-1}]$, and the corresponding inverse transformation is $\tetubd{\mu}{\mu}=\mbox{diag}[1,1,a,b\,c]$; i.e., $\tetudb{\alpha}{\mu}\,\tetubd{\mu}{\mu}=\deltaud{\alpha}{\mu}$.  
We also have $\mdet=\smdet=abc$.  
The Lorentz transformation (or boost) from the orthonormal comoving basis to the orthonormal (non-comoving) tetrad basis is given by
\begin{equation}
  \boostubdh{\mu}{\mu}
  =
  \left(
    \begin{array}{cc}
      \boostubdh{0}{0} & \boostubdh{0}{\imath} \\
      \boostubdh{\imath}{0}  & \boostubdh{\imath}{\imath}
    \end{array}
  \right)
  =
  \left(
    \begin{array}{cc}
      W & W\,\vbdh{\imath} \\
      W\,\vbub{\imath} & \deltaubdh{\imath}{\imath}+\f{W^{2}}{W+1}\,\vbub{\imath}\,\vbdh{\imath}
    \end{array}
  \right),  \label{eq:lorentzBoost}
\end{equation}
where $W=(\,1-\vbdb{\imath}\,\vbub{\imath}\,)^{-1/2}$ is the Lorentz factor.  
In Equation (\ref{eq:lorentzBoost}), $\vbub{\imath}$ \emph{and} $\vbdh{\imath}$ are three-velocity components defined with respect to the orthonormal tetrad basis, and are therefore accented with a bar ($\vbdb{\imath}=\vbub{\imath}=\vbuh{\imath}=\vbdh{\imath}$, for $\bar{\imath}=\hat{\imath}$)---they should not be considered spatial components of a four-vector.  
The placement of, and the accent on, the indices of the three-velocity components match those of the Lorentz transformation $\boostubdh{\mu}{\mu}$.  
This results in unambiguous notation when relations between quantities in the different reference frames are made explicit (see for example Equations (\ref{eq:numberDensityTetradBasis})-(\ref{eq:numberFluxTetradBasis}) and (\ref{eq:T00})-(\ref{eq:Tij}) below).  
The Lorentz transformation from the orthonormal tetrad basis to the orthonormal comoving basis, is obtained by reversing the sign on the three-velocity components; i.e., 
\begin{equation}
  \boostuhdb{\mu}{\mu}
  =
  \left(
    \begin{array}{cc}
      \boostuhdb{0}{0} & \boostuhdb{0}{\imath} \\
      \boostuhdb{\imath}{0} & \boostuhdb{\imath}{\imath}
    \end{array}
  \right)
  =
  \left(
    \begin{array}{cc}
      W & -W\,\vbdb{\imath} \\
      -W\,\vbuh{\imath} & \deltauhdb{\imath}{\imath}+\f{W^{2}}{W+1}\,\vbuh{\imath}\,\vbdb{\imath}
    \end{array}
  \right).  \label{eq:inverseLorentzBoost}
\end{equation}

With the specifications of the transformations between the orthonormal tetrad basis and the coordinate basis, and the Lorentz transformations between the orthonormal comoving basis and the orthonormal tetrad basis we can obtain specific expressions for the four-velocity of the comoving observer, Equation (\ref{eq:fourvelocityComovingObserver}).  
We can also relate the Eulerian projections of the number-flux four-vector, the stress-energy tensor, and the third-order moments in terms of the corresponding Lagrangian projections (Section \ref{sec:momentEquationsCFC}).  
In the comoving-frame, the four-velocity of the comoving observer is simply $\fourvelocityLuh{\mu}=(\,1,0\,)$.  
The four-velocity of the comoving observer in the orthonormal tetrad basis is
\begin{equation}
  \fourvelocityLub{\mu}
  =\boostubdh{\mu}{\mu}\,\fourvelocityLuh{\mu}
  =W(\,1,\vbub{\imath}\,), 
\end{equation}
while the coordinate basis four-velocity of the comoving observer is
\begin{equation}
  \fourvelocityLu{\mu}
  =\tetudb{\mu}{\mu}\,\fourvelocityLub{\mu}
  =W(\,1,\threevelocityEu{i}\,), 
  \label{eq:fourvelocityComovingObserverApp}
\end{equation}
where $\threevelocityEu{i}=\tetudb{i}{\imath}\,\vbub{\imath}$.  
Note that $\vbdb{\imath}\,\vbub{\imath}=\threevelocityEd{i}\,\threevelocityEu{i}$.  
Moreover, by contracting Equation (\ref{eq:fourvelocityComovingObserverApp}) with $\fourvelocityLd{\mu}$, we find $W^{2}=(\,1-\threevelocityEd{i}\,\threevelocityEu{i}\,)^{-1}$.  

The Eulerian projections of the number-flux four-vector can now be expressed in terms of the corresponding Lagrangian projections as (cf. Equations (\ref{eq:numberFluxEulerian}) and (\ref{eq:numberFluxLagrangian}))
\begin{eqnarray}
  \mathcalmcd{E}{N}
  &=&W\,\mathcalmcd{J}{N}+\threevelocityEd{i}\,\fourvectorua{H}{i}{N}, \label{eq:eulerianNumberDensityLagrangianProjections} \\
  \fourvectorua{F}{i}{N}
  &=&\fourvectorua{H}{i}{N}+W\,\threevelocityEu{i}\,\mathcalmcd{J}{N}, \label{eq:eulerianNumberFluxLagrangianProjections}
\end{eqnarray}
where $\fourvectorua{H}{i}{N}$ is related to the comoving-frame moments by
\begin{equation}
  \fourvectorua{H}{i}{N}
  =\tetudb{i}{\imath}\,
  \Big(
    \deltaubdh{\imath}{\imath}+\f{W^{2}}{W+1}\,\vbub{\imath}\,\vbdh{\imath}
  \Big)\fourvectoruha{H}{\imath}{N}.  
\end{equation}
The Eulerian projections of the stress-energy tensor are related to the corresponding Lagrangian projections in a similar manner.  
In particular, we have (cf. Equations (\ref{eq:stressEnergyTensorEulerian}) and (\ref{eq:stressEnergyTensorLagrangian}))
\begin{eqnarray}
  \mathcal{E}
  &=&
  W^{2}\,\mathcal{J}
  +2\,W\,\threevelocityEd{i}\,\fourvectoru{H}{i}
  +\threevelocityEd{i}\,\threevelocityEd{j}\,\tensoruu{K}{i}{j}, \label{eq:eulerianEnergyLagrangianProjections} \\
  \fourvectoru{F}{i}
  &=&
  W\,\fourvectoru{H}{i}
  +W^{2}\,\threevelocityEu{i}\,\mathcal{J}
  +\threevelocityEd{j}\,\tensoruu{K}{i}{j}
  +W\,\threevelocityEu{i}\,\threevelocityEd{j}\,\fourvectoru{H}{j}, \label{eq:eulerianMomentumLagrangianProjections} \\
  \tensoruu{S}{i}{j}
  &=&
  \tensoruu{K}{i}{j}
  +W\left(\,\threevelocityEu{i}\,\fourvectoru{H}{j}
  +\fourvectoru{H}{i}\,\threevelocityEu{j}\,\right)
  +W^{2}\,\threevelocityEu{i}\,\threevelocityEu{j}\,\mathcal{J} \label{eq:eulerianStressLagrangianProjections}.
\end{eqnarray}
In Equations (\ref{eq:eulerianEnergyLagrangianProjections})-(\ref{eq:eulerianStressLagrangianProjections}), the Lagrangian projections $\fourvectoru{H}{i}$ and $\tensoruu{K}{i}{j}$ are related to the comoving-frame moments by
\begin{eqnarray}
  \fourvectoru{H}{i}
  &=&
  \tetudb{i}{\imath}\,
  \Big(
    \deltaubdh{\imath}{\imath}+\f{W^{2}}{W+1}\,\vbub{\imath}\,\vbdh{\imath}
  \Big)\fourvectoruh{H}{\imath}, \label{eq:lagrangianMomentumComovingFrameMoments} \\
  \tensoruu{K}{i}{j}
  &=&
  \tetudb{i}{\imath}\,\tetudb{j}{\jmath}\,
  \Big(
    \deltaubdh{\imath}{\imath}\,\deltaubdh{\jmath}{\jmath}
    +\f{W^{2}}{W+1}\Big[\vbub{\imath}\,\vbdh{\imath}\,\deltaubdh{\jmath}{\jmath}+\deltaubdh{\imath}{\imath}\,\vbub{\jmath}\,\vbdh{\jmath}\Big]
    +\f{W^{4}}{(\,W+1\,)^{2}}\,\vbub{\imath}\,\vbub{\jmath}\,\vbdh{\imath}\,\vbdh{\jmath}
  \Big)\tensoruhuh{K}{\imath}{\jmath}. \label{eq:lagrangianStressComovingFrameMoments}
\end{eqnarray}
Note that $\tensoruhuh{K}{\imath}{\jmath}=\eddingtontensoruhuh{k}{\imath}{\jmath}\,\mathcal{J}$, where the rank-two variable Eddington tensor $\eddingtontensoruhuh{k}{\imath}{\jmath}$ is defined in Equation (\ref{eq:variableEddingtonTensorRankTwo}).  
Finally, we express the third-order moments in terms of the Lagrangian projections (cf. Equations (\ref{eq:thirdOrderTensorEulerian}) and (\ref{eq:thirdOrderTensorLagrangian}))
\begin{eqnarray}
  \f{1}{\epsilon}\,\tensoruuu{Q}{i}{j}{k}
  &=&
  \tensoruuu{L}{i}{j}{k}
  +W\,\Big(\threevelocityEu{i}\,\tensoruu{K}{j}{k}+\threevelocityEu{j}\,\tensoruu{K}{i}{k}+\threevelocityEu{k}\,\tensoruu{K}{i}{j}\Big)
  +W^{2}\,
  \Big(
    \threevelocityEu{i}\,\threevelocityEu{j}\,\fourvectoru{H}{k}
    +\threevelocityEu{i}\,\threevelocityEu{k}\,\fourvectoru{H}{j}
    +\threevelocityEu{j}\,\threevelocityEu{k}\,\fourvectoru{H}{i}
  \Big)
  +W^{3}\,\threevelocityEu{i}\,\threevelocityEu{j}\,\threevelocityEu{k}\,\mathcal{J}, \label{eq:eulerianThirdOrderMomentsLagrangianProjections}
\end{eqnarray}
where the Lagrangian projections of the third-order tensor is related to the comoving-frame moments by
\begin{eqnarray}
  \tensoruuu{L}{i}{j}{k}
  &=&
  \tetudb{i}{\imath}\,\tetudb{j}{\jmath}\,\tetudb{k}{k}\,
  \Big(
    \deltaubdh{\imath}{\imath}\,\deltaubdh{\jmath}{\jmath}\,\deltaubdh{k}{k}
    +\f{W^{2}}{W+1}\,
    \Big[
      \vbub{\imath}\,\vbdh{\imath}\,\deltaubdh{\jmath}{\jmath}\,\deltaubdh{k}{k}
      +\vbub{\jmath}\,\vbdh{\jmath}\,\deltaubdh{\imath}{\imath}\,\deltaubdh{k}{k}
      +\vbub{k}\,\vbdh{k}\,\deltaubdh{\imath}{\imath}\,\deltaubdh{\jmath}{\jmath}
    \Big] \nonumber \\
  && \hspace{0.0in}
    +\f{W^{4}}{(\,W+1\,)^{2}}\,
    \Big[
      \vbub{\imath}\,\vbdh{\imath}\,\vbub{\jmath}\,\vbdh{\jmath}\,\deltaubdh{k}{k}
      +\vbub{\imath}\,\vbdh{\imath}\,\vbub{k}\,\vbdh{k}\,\deltaubdh{\jmath}{\jmath}
      +\vbub{\jmath}\,\vbdh{\jmath}\,\vbub{k}\,\vbdh{k}\,\deltaubdh{\imath}{\imath}
    \Big]
    +\f{W^{6}}{(\,W+1\,)^{3}}\,\vbub{\imath}\,\vbub{\jmath}\,\vbub{k}\,\vbdh{\imath}\,\vbdh{\jmath}\,\vbdh{k}
  \Big)\tensoruhuhuh{L}{\imath}{\jmath}{k}, \label{eq:lagrangianThirdOrderMomentsComovingFrameMoments}
\end{eqnarray}
and $\tensoruhuhuh{L}{\imath}{\jmath}{k}=\eddingtontensoruhuhuh{l}{\imath}{\jmath}{k}\,\mathcal{J}$.  
The rank-three variable Eddington tensor $\eddingtontensoruhuhuh{l}{\imath}{\jmath}{k}$ is defined in Equation (\ref{eq:variableEddingtonTensorRankThree}).  

Alternatively, we can write the Eulerian projections of the number-flux four-vector in terms of the number-flux four-vector in the orthonormal tetrad basis, whose components are expressed in terms of comoving-frame angular moments by
\begin{eqnarray}
  \fourvectorub{N}{0}
  &=&W\Big(\mathcalmcd{J}{N}+\vbdh{\imath}\,\fourvectoruha{H}{\imath}{N}\Big), \label{eq:numberDensityTetradBasis} \\
  \fourvectorub{N}{\imath}
  &=&\deltaubdh{\imath}{\imath}\,\fourvectoruha{H}{\imath}{N}
  +W\,\vbub{\imath}\Big(\mathcalmcd{J}{N}+\f{W}{W+1}\,\vbdh{\imath}\,\fourvectoruha{H}{\imath}{N}\Big) \label{eq:numberFluxTetradBasis}.  
\end{eqnarray}
Similarly, The monochromatic radiation energy density, momentum density, and stress in the orthonormal tetrad basis can be expressed in terms of the comoving-frame moments as
\begin{eqnarray}
  \tensorubub{T}{0}{0}
  &=&W^{2}\Big(\mathcal{J}+2\,\vbdh{\imath}\,\fourvectoruh{H}{\imath}+\vbdh{\imath}\,\vbdh{\jmath}\,\mathcaluhuh{K}{\imath}{\jmath}\Big), 
  \label{eq:T00} \\
  \tensorubub{T}{\imath}{0}
  &=&W\,
  \Big[
    \deltaubdh{\imath}{\imath}\,\fourvectoruh{H}{\imath}
    +W\,\vbub{\imath}\,\mathcal{J}
    +\deltaubdh{\imath}{\imath}\,\vbdh{\jmath}\,\tensoruhuh{K}{\imath}{\jmath}
    +\f{W}{W+1}\,\vbub{\imath}\Big(\left[\,2\,W+1\,\right]\,\vbdh{\imath}\,\fourvectoruh{H}{\imath}+W\,\vbdh{\imath}\,\vbdh{\jmath}\,\tensoruhuh{K}{\imath}{\jmath}\Big)
  \Big], 
  \label{eq:Ti0} \\
  \tensorubub{T}{\imath}{\jmath}
  &=&\deltaubdh{\imath}{\imath}\,\deltaubdh{\jmath}{\jmath}\,\tensoruhuh{K}{\imath}{\jmath}
  +W\,\Big(\vbub{\imath}\,\deltaubdh{\jmath}{\jmath}\,\fourvectoruh{H}{\jmath}+\vbub{\jmath}\,\deltaubdh{\imath}{\imath}\,\fourvectoruh{H}{\imath}\Big)
  +W^{2}\,\vbub{\imath}\,\vbub{\jmath}\,\mathcal{J} \nonumber \\
  &&+\f{W^{2}}{W+1}
  \Big(
    \Big[\vbub{\imath}\,\vbdh{\imath}\,\deltaubdh{\jmath}{\jmath}+\deltaubdh{\imath}{\imath}\,\vbub{\jmath}\,\vbdh{\jmath}\Big]\tensoruhuh{K}{\imath}{\jmath}
    +2\,W\,\vbub{\imath}\,\vbub{\jmath}\,\vbdh{\imath}\,\fourvectoruh{H}{\imath}
  \Big)
  +\f{W^{4}}{(\,W+1\,)^{2}}\vbub{\imath}\,\vbub{\jmath}\,\vbdh{\imath}\,\vbdh{\jmath}\,\tensoruhuh{K}{\imath}{\jmath}.  
  \label{eq:Tij}
\end{eqnarray}
Equations (\ref{eq:T00}) and (\ref{eq:Ti0}) correspond to Equations (91.10) and (91.11) in \citet{mihalasMihalas_1999} \citep[see also Equations (182) and (183) in][]{munierWeaver_1986a}.  
Equation (\ref{eq:Tij}) corresponds to Equation (91.12) in \citet{mihalasMihalas_1999} \citep[see also Equation (184) in][]{munierWeaver_1986a}.  
We also have
\begin{eqnarray}
  \f{1}{\epsilon}\,\tensorububub{U}{\imath}{\jmath}{k}
  &=&
  \deltaubdh{\imath}{\imath}\,\deltaubdh{\jmath}{\jmath}\,\deltaubdh{k}{k}\,\tensoruhuhuh{L}{\imath}{\jmath}{k}
  +W
  \Big(
    \vbub{k}\,\deltaubdh{\imath}{\imath}\,\deltaubdh{\jmath}{\jmath}\,\tensoruhuh{K}{\imath}{\jmath}
    +\vbub{\jmath}\,\deltaubdh{\imath}{\imath}\,\deltaubdh{k}{k}\,\tensoruhuh{K}{\imath}{k}
    +\vbub{\imath}\,\deltaubdh{\jmath}{\jmath}\,\deltaubdh{k}{k}\,\tensoruhuh{K}{\jmath}{k}
  \Big) \nonumber \\
  &&
  +W^{2}
  \Big(
    \vbub{\jmath}\,\vbub{k}\,\deltaubdh{\imath}{\imath}\,\fourvectoruh{H}{\imath}
    +\vbub{\imath}\,\vbub{k}\,\deltaubdh{\jmath}{\jmath}\,\fourvectoruh{H}{\jmath}
    +\vbub{\imath}\,\vbub{\jmath}\,\deltaubdh{k}{k}\,\fourvectoruh{H}{k}
  \Big)
  +\f{W^{2}}{W+1}
  \Big(
    \deltaubdh{\imath}{\imath}\,\deltaubdh{\jmath}{\jmath}\,\vbub{k}\vbdh{k}
    +\deltaubdh{\imath}{\imath}\,\deltaubdh{k}{k}\,\vbub{\jmath}\vbdh{\jmath}
    +\deltaubdh{\jmath}{\jmath}\,\deltaubdh{k}{k}\,\vbub{\imath}\vbdh{\imath}
  \Big)\tensoruhuhuh{L}{\imath}{\jmath}{k} \nonumber \\
  &&
  +W^{3}\vbub{\imath}\,\vbub{\jmath}\,\vbub{k}\,\mathcal{J}
  +\f{W^{3}}{W+1}
  \Big[
    \Big(
      \deltaubdh{\imath}{\imath}\,\vbub{\jmath}\vbdh{\jmath}
      +\vbub{\imath}\vbdh{\imath}\,\deltaubdh{\jmath}{\jmath}
    \Big)\vbub{k}\tensoruhuh{K}{\imath}{\jmath}
    +
    \Big(
      \deltaubdh{\imath}{\imath}\,\vbub{k}\vbdh{k}
      +\vbub{\imath}\vbdh{\imath}\,\deltaubdh{k}{k}
    \Big)\vbub{\jmath}\tensoruhuh{K}{\imath}{k}
    +
    \Big(
      \deltaubdh{\jmath}{\jmath}\,\vbub{k}\vbdh{k}
      +\vbub{\jmath}\vbdh{\jmath}\,\deltaubdh{k}{k}
    \Big)\vbub{\imath}\tensoruhuh{K}{\jmath}{k}
  \Big] \nonumber \\
  &&
  +\f{3W^{4}}{W+1}\vbub{\imath}\,\vbub{\jmath}\,\vbub{k}\,\vbdh{\imath}\,\fourvectoruh{H}{\imath}
  +\f{W^{4}}{(W+1)^{2}}
  \Big(
    \deltaubdh{\imath}{\imath}\,\vbub{\jmath}\vbdh{\jmath}\,\vbub{k}\vbdh{k}
    +\deltaubdh{\jmath}{\jmath}\,\vbub{\imath}\vbdh{\imath}\,\vbub{k}\vbdh{k}
    +\deltaubdh{k}{k}\,\vbub{\imath}\vbdh{\imath}\,\vbub{\jmath}\vbdh{\jmath}
  \Big)\tensoruhuhuh{L}{\imath}{\jmath}{k} \nonumber \\
  &&
  +\f{3W^{5}}{(1+W)^{2}}\vbub{\imath}\,\vbub{\jmath}\,\vbub{k}\,\vbdh{\imath}\,\vbdh{\jmath}\,\tensoruhuh{K}{\imath}{\jmath}
  +\f{W^{6}}{(W+1)^{3}}\vbub{\imath}\vbub{\jmath}\vbub{k}\,\vbdh{\imath}\vbdh{\jmath}\vbdh{k}\,\tensoruhuhuh{L}{\imath}{\jmath}{k}.
\end{eqnarray}
We can now list the conservative special relativistic radiation moment equations.  

The conservative, monochromatic special relativistic radiation moment equations are obtained from Equations (\ref{eq:energyEquationCFC}) and (\ref{eq:momentumEquationCFC}) by setting $\alpha=\psi=1$ and $\betau{i}=0$.  
The radiation energy equation becomes
\begin{equation}
  \pderiv{\mathcal{E}}{t}
  +\f{1}{\smdet}\pderiv{}{x^{i}}\Big(\smdet\,\fourvectoru{F}{i}\Big)
  -\f{1}{\epsilon^{2}}\pderiv{}{\epsilon}\Big(\epsilon^{2}\,\mathcal{F}^{\epsilon}\Big)
  =\f{1}{\epsilon}\int_{\Omega}\pu{0}\,\collision{f}\,d\Omega, 
  \label{eq:energyMomentSR}
\end{equation}
where the energy space energy flux has been defined as
\begin{eqnarray}
  \mathcal{F}^{\epsilon}
  &=&
  \epsilon\,
  \Big\{
    \Big(
      \fourvectord{F}{i}
      +\tensordu{S}{i}{j}\,\threevelocityEd{j}
      +W\,\f{1}{\epsilon}\,\tensorduu{Q}{i}{j}{k}\,\threevelocityEd{j}\,\threevelocityEd{k}
      -W^{2}\,\mathcal{E}\,\threevelocityEd{i}
    \Big)\pderiv{\threevelocityEu{i}}{t} \nonumber \\
    && \hspace{0.25in}
    +
    \Big(
      \tensordu{S}{i}{j}
      +W\,\f{1}{\epsilon}\,\tensorduu{Q}{i}{j}{k}\,\threevelocityEd{k}
      -W^{2}\,\threevelocityEd{i}\,\fourvectoru{F}{j}
    \Big)\pderiv{\threevelocityEu{i}}{x^{j}}
    +\f{1}{2}
    \Big(
      \tensoruu{S}{i}{j}
      +W\,\f{1}{\epsilon}\,\tensoruuu{Q}{i}{j}{k}\,\threevelocityEd{k}
    \Big)\threevelocityEu{l}\,\pderiv{\gmdd{i}{j}}{x^{l}}
  \Big\}.  
  \label{eq:energySpaceEnergyFluxSR}
\end{eqnarray}
We have written $\partial\ln W/\partial t=W^{2}\,\threevelocityEd{i}\,\partial\threevelocityEu{i}/\partial t$ and $\partial\ln W/\partial x^{j}=W^{2}\,\threevelocityEd{i}\,\partial\threevelocityEu{i}/\partial x^{j}$.  
The terms proportional to derivatives of three-velocity components account for Doppler shifts of the radiation energy spectrum as measured by the comoving observer.  
The radiation momentum equation becomes
\begin{equation}
  \pderiv{\fourvectord{F}{i}}{t}
  +\f{1}{\smdet}\pderiv{}{x^{j}}\Big(\smdet\,\tensorud{S}{j}{i}\Big)
  -\f{1}{2}\tensoruu{S}{j}{k}\,\pderiv{\gmdd{j}{k}}{x^{i}}
  -\f{1}{\epsilon^{2}}\pderiv{}{\epsilon}\Big(\epsilon^{2}\,\mathcal{S}_{~i}^{\epsilon}\Big)
  =\gmdd{i}{j}\,\f{1}{\epsilon}\int_{\Omega}\pu{j}\,\collision{f}\,d\Omega, 
  \label{eq:momentumMomentSR}
\end{equation}
where we have defined the energy space momentum flux as
\begin{eqnarray}
  \mathcal{S}_{~i}^{\epsilon}
  &=&
  \epsilon\,
  \Big\{
    \Big(
      \tensordd{S}{i}{j}
      +W\,\f{1}{\epsilon}\tensorddu{Q}{i}{j}{k}\,\threevelocityEd{k}
      -W^{2}\,\fourvectord{F}{i}\,\threevelocityEd{j}
    \Big)\pderiv{\threevelocityEu{j}}{t} \nonumber \\
    && \hspace{0.25in}
    +
    \Big(
      W\,\f{1}{\epsilon}\tensorddu{Q}{i}{j}{k}
      -W^{2}\,\tensordu{S}{i}{k}\,\threevelocityEd{j}
    \Big)\pderiv{\threevelocityEu{j}}{x^{k}}
    +\f{1}{2}W\,\f{1}{\epsilon}\tensorduu{Q}{i}{j}{k}\,\threevelocityEu{l}\,\pderiv{\gmdd{j}{k}}{x^{l}}
  \Big\}.  
  \label{eq:energySpaceMomentumFluxSR}
\end{eqnarray}

We specialize the monochromatic radiation energy and momentum equations to spherical polar coordinates; i.e., $(x^{1},x^{2},x^{3})=(r,\theta,\phi)$, $a=b=r$, and $c=\sin\theta$, and impose spherical symmetry ($\partial/\partial\theta=\partial/\partial\phi=0$).  
The only nonzero component of the fluid three-velocity is $\threevelocityEu{1}=\threevelocityEd{1}\equiv v_{r}$.  
In spherical symmetry, the comoving-frame angular moments are
\begin{equation}
  \{\,\mathcal{J},\,\mathcal{H},\,\mathcal{K},\,\mathcal{L}\,\}
  =\{\,\mathcal{J},\,\fourvectoruh{H}{1},\,\tensoruhuh{K}{1}{1},\,\tensoruhuhuh{L}{1}{1}{1}\,\}
  =2\pi\,\epsilon\int_{-1}^{1}f\,\mu^{\{0,1,2,3\}}\,d\mu,
  \label{eq:energyMomentsSphericalSymmetry}
\end{equation}
with $\mu=\cos\vartheta$.  
Then, the monochromatic radiation energy and momentum equations become
\begin{eqnarray}
  \pderiv{\mathcal{E}}{t}
  +\f{1}{r^{2}}\pderiv{}{r}\Big(r^{2}\,\mathcal{F}\Big)
  -\f{1}{\epsilon^{2}}\pderiv{}{\epsilon}\Big(\epsilon^{2}\,\mathcal{F}^{\epsilon}\Big)
  =2\pi\,W\,\Big(\int_{-1}^{1}\collision{f}\,d\mu+v_{r}\,\int_{-1}^{1}\collision{f}\mu\,d\mu\Big), 
  \label{eq:energyMomentSRSphericalSymmetry}
\end{eqnarray}
and
\begin{eqnarray}
  \pderiv{\mathcal{F}}{t}
  +\f{1}{r^{2}}\pderiv{}{r}\Big(r^{2}\,\mathcal{S}\Big)
  -\f{1}{r}\,\Big(\mathcal{J}-\mathcal{K}\Big)
  -\f{1}{\epsilon^{2}}\pderiv{}{\epsilon}\Big(\epsilon^{2}\,\mathcal{S}^{\epsilon}\Big)
  =2\pi\,W\,\Big(v_{r}\,\int_{-1}^{1}\collision{f}\,d\mu+\int_{-1}^{1}\collision{f}\mu\,d\mu\Big), 
  \label{eq:momentumMomentSRSphericalSymmetry}
\end{eqnarray}
respectively.  
With spherical symmetry imposed, we have
\begin{eqnarray}
  \mathcal{E}
  &=&
  W^{2}\left(\,\mathcal{J}+2\,v_{r}\,\mathcal{H}+v_{r}^{2}\,\mathcal{K}\,\right), \\
  \mathcal{F}
  &=&
  W^{2}\left(\,[\,1+v_{r}^{2}\,]\,\mathcal{H}+v_{r}\,[\,\mathcal{J}+\mathcal{K}\,]\,\right), \\
  \mathcal{S}
  &=&W^{2}\left(\,\mathcal{K}+2\,v_{r}\,\mathcal{H}+v_{r}^{2}\,\mathcal{J}\,\right).  
\end{eqnarray}
Moreover, we have $\tensordu{S}{2}{2}=\tensordu{K}{2}{2}=\tensoruhuh{K}{2}{2}=\f{1}{2}\,(\,\mathcal{J}-\mathcal{K}\,)$ and $\tensordu{S}{3}{3}=\tensordu{K}{3}{3}=\tensoruhuh{K}{3}{3}=\f{1}{2}\,(\,\mathcal{J}-\mathcal{K}\,)$.  
The fluxes inside the energy derivatives can be written explicitly in terms of the comoving-frame moments as
\begin{eqnarray}
  \mathcal{F}^{\epsilon}
  &=&
  \epsilon\,W^{2}
  \Big\{
    W^{2}\,
    \Big[
      \left(\,\mathcal{H}+v_{r}\,\mathcal{K}\,\right)
      +v_{r}\,\left(\,\mathcal{K}+v_{r}\,\mathcal{L}\,\right)
    \Big]\pderiv{v_{r}}{t}
    +W^{2}\,
    \Big[
      v_{r}\,\left(\,\mathcal{H}+v_{r}\,\mathcal{K}\,\right)
      +\left(\,\mathcal{K}+v_{r}\,\mathcal{L}\,\right)
    \Big]\pderiv{v_{r}}{r} \nonumber \\
    && \hspace{0.45in}
    +
    \Big[
      \left(\,\mathcal{J}-\mathcal{K}\,\right)
      +v_{r}\,\left(\,\mathcal{H}-\mathcal{L}\,\right)
    \Big]\,\f{v_{r}}{r}
  \Big\}, \\
  \mathcal{S}^{\epsilon}
  &=&
  \epsilon\,W^{2}
  \Big\{
    W^{2}\,
    \Big[
      \left(\,v_{r}\,\mathcal{H}+\mathcal{K}\,\right)
      +v_{r}\,\left(\,v_{r}\,\mathcal{K}+\mathcal{L}\,\right)
    \Big]\pderiv{v_{r}}{t}
    +W^{2}\,
    \Big[
      v_{r}\,\left(\,v_{r}\,\mathcal{H}+\mathcal{K}\,\right)
      +\left(\,v_{r}\,\mathcal{K}+\mathcal{L}\,\right)
    \Big]\pderiv{v_{r}}{r} \nonumber \\
    && \hspace{0.45in}
    +
    \Big[
      v_{r}\,\left(\,\mathcal{J}-\mathcal{K}\,\right)
      +\left(\,\mathcal{H}-\mathcal{L}\,\right)
    \Big]\,\f{v_{r}}{r}
  \Big\}, 
\end{eqnarray}
where we have used
\begin{eqnarray}
  \f{1}{\epsilon}\tensoruuu{Q}{1}{1}{1}
  &=&W^{3}\left(\,\mathcal{L}+3\,v_{r}\,\mathcal{K}+3\,v_{r}^{2}\,\mathcal{H}+v_{r}^{3}\,\mathcal{J}\,\right), \\
  \f{1}{\epsilon}\tensorddu{Q}{1}{2}{2}
  =
  \f{1}{\epsilon}\tensorddu{Q}{1}{3}{3}
  &=&
  \f{1}{2}\,W
  \left(\,
    \left[\,\mathcal{H}-\mathcal{L}\,\right]
    +v_{r}\,\left[\,\mathcal{J}-\mathcal{K}\,\right]
  \,\right).  
\end{eqnarray}
Equations (\ref{eq:energyMomentSRSphericalSymmetry}) and (\ref{eq:momentumMomentSRSphericalSymmetry}) are monochromatic lab-frame radiation energy and momentum equations in conservative form---valid to all orders of $v_{r}$.  
They can be expressed explicitly in terms of the comoving-frame moments, and compared with corresponding \emph{non-conservative} equations in \citet{mihalasMihalas_1999}; their Equations (95.11) and (95.12).  
Equation (\ref{eq:energyMomentSRSphericalSymmetry}) is equivalent to Equation (95.11) plus $v_{r}$ times Equation (95.12) in \citet{mihalasMihalas_1999}.  
Similarly, Equation (\ref{eq:momentumMomentSRSphericalSymmetry}) is equivalent to Equation (95.12) plus $v_{r}$ times Equation (95.11) in \citet{mihalasMihalas_1999}.  

The special relativistic moment equations are consistent with the conservative neutrino number equation, which we obtain by adding $W\epsilon^{-1}$ times Equation (\ref{eq:energyMomentSR}) and $-W\epsilon^{-1}\,\threevelocityEu{i}$ contracted with Equation (\ref{eq:momentumMomentSR}).  
An intermediate result is
\begin{eqnarray}
  &&
  \pderiv{}{t}\Big(\f{1}{\epsilon}\,W\Big[\mathcal{E}-\threevelocityEu{i}\,\fourvectord{F}{i}\Big]\Big)
  +\f{1}{\smdet}\pderiv{}{x^{i}}\Big(\smdet\,\f{1}{\epsilon}\,W\Big[\fourvectoru{F}{i}-\threevelocityEu{j}\,\tensorud{S}{i}{j}\Big]\Big)
  -\f{1}{\epsilon^{2}}\pderiv{}{\epsilon}\Big(\epsilon^{2}\,\f{1}{\epsilon}\,W\Big[\mathcal{F}^{\epsilon}-\threevelocityEu{i}\,\mathcal{S}_{~i}^{\epsilon}\Big]\Big)
  \nonumber \\
  && \hspace{0.15in}
  +\f{1}{\epsilon}W
  \Big\{
    \Big[
      \fourvectord{F}{i}
      -W^{2}\Big(\mathcal{E}\,\threevelocityEd{i}-\threevelocityEu{j}\,\fourvectord{F}{j}\,\threevelocityEd{i}\Big)
    \Big]\pderiv{\threevelocityEu{i}}{t}
    +
    \Big[
      \tensordu{S}{i}{j}
      -W^{2}\,\threevelocityEd{i}\Big(\fourvectoru{F}{j}-\tensorud{S}{j}{k}\,\threevelocityEu{k}\Big)
    \Big]\pderiv{\threevelocityEu{i}}{x^{j}}
    +\tensoruu{S}{i}{j}\,\threevelocityEu{k}\,\pderiv{\gmdd{i}{j}}{x^{k}}
  \Big\}
  \nonumber \\
  && \hspace{0.15in}
  -\f{1}{\epsilon^{2}}W\Big(\mathcal{F}^{\epsilon}-\threevelocityEu{i}\,\mathcal{S}_{~i}^{\epsilon}\Big)
  =\f{1}{\epsilon}\int_{\Omega}\collision{f}\,d\Omega.
  \label{eq:numberMomentIntermediateSR}
\end{eqnarray}
(Note that $\fourvelocityLd{\mu}\,\pu{\mu}=-\epsilon$.)  
From Equations (\ref{eq:energySpaceEnergyFluxSR}) and (\ref{eq:energySpaceMomentumFluxSR}) we have
\begin{eqnarray}
  \f{1}{\epsilon}\Big(\mathcal{F}^{\epsilon}-\threevelocityEu{i}\,\mathcal{S}_{~i}^{\epsilon}\Big)
  &=&
  \Big[
    \fourvectord{F}{i}
    -W^{2}\Big(\mathcal{E}\,\threevelocityEd{i}-\threevelocityEu{j}\,\fourvectord{F}{j}\,\threevelocityEd{i}\Big)
  \Big]\pderiv{\threevelocityEu{i}}{t} \nonumber \\
  &&
  +
  \Big[
    \tensordu{S}{i}{j}
    -W^{2}\threevelocityEd{i}\Big(\fourvectoru{F}{j}-\tensorud{S}{j}{k}\,\threevelocityEu{k}\Big)
  \Big]\pderiv{\threevelocityEu{i}}{x^{j}}
  +\tensoruu{S}{i}{j}\,\threevelocityEu{k}\,\pderiv{\gmdd{i}{j}}{x^{k}}.  
\end{eqnarray}
Thus, by virtue of the cancellation of the terms on the second line with the last term on the left-hand side, Equation (\ref{eq:numberMomentIntermediateSR}) reduces to the conservative monochromatic number equation
\begin{equation}
  \pderiv{\mathcalmcd{E}{N}}{t}
  +\f{1}{\smdet}\pderiv{}{x^{i}}\Big(\smdet\,\fourvectorua{F}{i}{N}\Big)
  -\f{1}{\epsilon^{2}}\pderiv{}{\epsilon}\Big(\epsilon^{2}\,\mathcal{F}_{{\scriptscriptstyle\mathcal{N}}}^{\epsilon}\Big)
  =\f{1}{\epsilon}\int_{\Omega}\collision{f}\,d\Omega, 
  \label{eq:numberMomentSR}
\end{equation}
where the energy space number flux is
\begin{equation}
  \mathcal{F}_{{\scriptscriptstyle\mathcal{N}}}^{\epsilon}
  =W
  \Big\{
    \Big(
      \fourvectord{F}{i}-W^{2}\Big[\mathcal{E}-\fourvectord{F}{j}\,\threevelocityEu{j}\Big]\threevelocityEd{i}
    \Big)\pderiv{\threevelocityEu{i}}{t}
    +
    \Big(
      \tensordu{S}{i}{j}
      -W^{2}\Big[\fourvectoru{F}{j}-\tensorud{S}{j}{k}\,\threevelocityEu{k}\Big]\threevelocityEd{i}
    \Big)\pderiv{\threevelocityEu{i}}{x^{j}}
    +\tensoruu{S}{i}{j}\,\threevelocityEu{k}\,\pderiv{\gmdd{i}{j}}{x^{k}}
  \Big\}.  
  \label{eq:energySpaceNumberFluxSR}
\end{equation}

Equation (\ref{eq:numberMomentSR}) is a conservative equation for the neutrino number density.  
In the absence of neutrino-matter interactions, it states that the neutrino number is conserved.  
In carrying out the derivation of the number equation from the radiation energy and momentum equations, we observe that the leftover terms emanating from bringing $W$ and $W\threevelocityEu{i}$ inside the space and time derivatives in the energy and momentum equations cancel with the leftover terms from bringing $\epsilon^{-1}$ inside the energy derivatives.  
The remaining terms constitute the left-hand side of the conservative number equation (Equation (\ref{eq:numberMomentSR})).  
These cancellations may be key to constructing a numerical scheme for neutrino transport based on the solution of the conservative energy and momentum equations that is also consistent with the conservative number equation.  
Cancellations similar to those occurring in the continuum derivation of the number equation form the energy and momentum equations should also occur in the discrete limit in order to ensure consistency with neutrino number conservation.  
Such consistency may help ensure lepton conservation in simulations of neutrino radiation hydrodynamics.  

\section{GENERAL RELATIVISTIC MOMENT EQUATIONS FOR SPHERICALLY SYMMETRIC SPACETIMES}
\label{app:momentEquationsSphericalSymmetryCFC}

\citet[][hereafter referred to as \citetalias{muller_etal_2010}]{muller_etal_2010} have recently presented numerical methods for multidimensional, general relativistic neutrino radiation hydrodynamics for the case where the conformal flatness condition (CFC) is imposed on the spatial metric \citep[e.g.,][]{wilson_etal_1996}.  
\citetalias{muller_etal_2010} employ the so-called ray-by-ray approach to multidimensional neutrino transport, where the radiation flux is assumed to be purely radial, and the radiation field is essentially obtained by solving independent spherically symmetric problems along each radial ray.  
(Lateral advection of neutrinos in optically thick regions and non-radial components of the radiation pressure gradient are still taken into account.)  
The radiation moment equations for the conformally flat spacetime become fully general relativistic when spherical symmetry is imposed.  
For the sake of comparing the conservative moment equations presented here with the equations solved by \citetalias{muller_etal_2010}, we adopt a spherically symmetric spacetime, and list---in full detail---general relativistic moment equations.  
The equations presented here can be obtained directly from Equations (\ref{eq:energyEquationCFC}), (\ref{eq:momentumEquationCFC}), and (\ref{eq:numberEquationCFC}).  

The invariant spacetime line element can be decomposed as
\begin{equation}
  ds^{2}=-\alpha^{2}dt^{2}+\gmdd{i}{j}\left(\,dx^{i}+\betau{i}dt\,\right)\left(\,dx^{j}+\betau{j}dt\,\right).  
  \label{eq:lineelementADM}
\end{equation}
We adopt spherical polar coordinates $(x^{1},x^{2},x^{3})=(r,\theta,\phi)$.  
With spherical symmetry imposed, non-spectral quantities are only functions of time $t$ and the radial coordinate $r$.  
The only nonzero component of the fluid three-velocity in the orthonormal tetrad basis is $\vbub{1}=\vbdb{1}\equiv v_{r}$.  
For the CFC spacetime, the spatial metric $\gmdd{i}{j}=\psi^{4}\,\gmbdd{i}{j}$ is diagonal, and the conformal metric is $\gmbdd{i}{j}=\mbox{diag}[\,1,\,r^{2},\,r^{2}\,\sin^{2}\theta\,]$, and $\psi$ is the conformal factor.  
Moreover, $\alpha$ is the lapse function and $\betau{i}$ is the shift vector (only the radial component of the shift vector $\betau{r}$ is nonzero when spherical symmetry is imposed).  
The lapse function, the shift vector, and the conformal factor can be obtained by solving a system of nonlinear elliptic equations \citep{wilson_etal_1996}.  
For the spherically symmetric spacetime, the determinant of the spatial metric is $\smdet=\psi^{6}\,r^{2}\,\sin\theta$, and we can set $\tetudb{\mu}{0}=\alpha^{-1}\,[\,1,\,-\betau{r},\,0,\,0\,]$, $\tetudb{0}{\imath}=[\,0,\,0,\,0\,]$, and $\tetudb{i}{\imath}=\psi^{-2}\,\mbox{diag}[\,1,\,r^{-1},\,(\,r\,\sin\theta\,)^{-1}\,]$.  

The monochromatic lab-frame radiation energy and momentum equations, expressed in terms of the comoving-frame angular moments $\mathcal{J}$, $\mathcal{H}$, $\mathcal{K}$, and $\mathcal{L}$, become
\begin{eqnarray}
  & &
  \f{1}{\alpha}\pderiv{}{t}
  \Big(
    W^{2}\Big[\mathcalh{J}+2\,v_{r}\,\mathcalh{H}+v_{r}^{2}\,\mathcalh{K}\Big]
  \Big) \nonumber \\
  & & \hspace{0.15in}
  +\f{1}{\alpha}\pderiv{}{r}
  \Big(
    \alpha W^{2}
    \Big\{
      \Big[
        \Big(\f{1}{\psi^{2}}-v_{r}\f{\betau{r}}{\alpha}\Big)+v_{r}\Big(\f{v_{r}}{\psi^{2}}-\f{\betau{r}}{\alpha}\Big)
      \Big]\mathcalh{H}
      +\Big[
        \f{v_{r}}{\psi^{2}}-\f{\betau{r}}{\alpha}
      \Big]\mathcalh{J}
      +v_{r}
      \Big[
        \f{1}{\psi^{2}}-v_{r}\f{\betau{r}}{\alpha}
      \Big]\mathcalh{K}
    \Big\}
  \Big) \nonumber \\
  & & \hspace{0.15in}
  +\f{1}{\psi^{2}}\pderiv{\ln\alpha}{r}W^{2}\Big[\Big(1+v_{r}^{2}\Big)\mathcalh{H}+v_{r}\Big(\mathcalh{J}+\mathcalh{K}\Big)\Big]
  +\Big(\tauderiv{\ln\psi^{2}}-\f{1}{\alpha}\pderiv{\betau{r}}{r}\Big)W^{2}\Big[\mathcalh{K}+2\,v_{r}\,\mathcalh{H}+v_{r}^{2}\,\mathcalh{J}\Big] \nonumber \\
  & & \hspace{0.30in}
  +\Big(\tauderiv{\ln\psi^{2}}-\f{1}{r}\f{\betau{r}}{\alpha}\Big)\Big[\mathcalh{J}-\mathcalh{K}\Big]
  \nonumber \\
  & & \hspace{0.15in}
  -\f{1}{\epsilon^{2}}\pderiv{}{\epsilon}
  \Big(
    \epsilon^{3}W^{2}
    \Big\{
      \Big[
        \f{1}{\psi^{2}}\pderiv{\ln\alpha}{r}
        +v_{r}\Big(\tauderiv{\ln\psi^{2}}-\f{1}{\alpha}\pderiv{\betau{r}}{r}\Big)
        +W^{2}\cderiv{v_{r}}{\tau}
      \Big]\Big(\mathcalh{H}+v_{r}\mathcalh{K}\Big) \nonumber \\
      & & \hspace{0.30in}
      +
      \Big[
        \f{v_{r}}{\psi^{2}}\pderiv{\ln\alpha}{r}
        +\Big(
          \tauderiv{\ln\psi^{2}}
          -\f{1}{\alpha}\pderiv{\betau{r}}{r}
        \Big)
        +W^{2}\Big(v_{r}\tauderiv{v_{r}}+\f{1}{\psi^{2}}\pderiv{v_{r}}{r}\Big)
      \Big]\Big(\mathcalh{K}+v_{r}\mathcalh{L}\Big) \nonumber \\
      & & \hspace{0.30in}
      +
      \Big[
        \cderiv{\ln\psi^{2}}{\tau}
        +\f{1}{r}\Big(\f{v_{r}}{\psi^{2}}-\f{\betau{r}}{\alpha}\Big)
      \Big]\Big(\Big[\mathcalh{J}-\mathcalh{K}\Big]+v_{r}\Big[\mathcalh{H}-\mathcalh{L}\Big]\Big)
    \Big\}
  \Big) \nonumber \\
  &&
  =W\Big(2\pi\int_{-1}^{1}\collisionh{f}\,d\mu+v_{r}\,2\pi\int_{-1}^{1}\collisionh{f}\mu\,d\mu\Big), 
  \label{eq:energyMomentCFCSphericalSymmetry}
\end{eqnarray}
and
\begin{eqnarray}
  & &
  \f{1}{\alpha}\pderiv{}{t}
  \Big(
    W^{2}\Big[\Big(1+v_{r}^{2}\Big)\mathcalh{H}+v_{r}\Big(\mathcalh{J}+\mathcalh{K}\Big)\Big]
  \Big) \nonumber \\
  & & \hspace{0.15in}
  +\f{1}{\alpha}\pderiv{}{r}
  \Big(
    \alpha W^{2}
    \Big\{
      \Big[\f{1}{\psi^{2}}-v_{r}\f{\betau{r}}{\alpha}\Big]\mathcalh{K}
      +\Big[
        \Big(\f{v_{r}}{\psi^{2}}-\f{\betau{r}}{\alpha}\Big)
        +v_{r}\Big(\f{1}{\psi^{2}}-v_{r}\f{\betau{r}}{\alpha}\Big)
      \Big]\mathcalh{H}
      +v_{r}\Big[\f{v_{r}}{\psi^{2}}-\f{\betau{r}}{\alpha}\Big]\mathcalh{J}
    \Big\}
  \Big) \nonumber \\
  & & \hspace{0.15in}
  + \f{1}{\psi^{2}}\pderiv{\ln\alpha}{r}
  W^{2}\Big[\mathcalh{J}+2\,v_{r}\,\mathcalh{H}+v_{r}^{2}\mathcalh{K}\Big]
  +\Big(\tauderiv{\ln\psi^{2}}-\f{1}{\alpha}\pderiv{\betau{r}}{r}\Big)
  W^{2}\Big[\Big(1+v_{r}^{2}\Big)\mathcalh{H}+v_{r}\Big(\mathcalh{J}+\mathcalh{K}\Big)\Big] \nonumber \\
  & & \hspace{0.30in}
  -\f{1}{\psi^{2}}\Big(\pderiv{\ln\psi^{2}}{r}+\f{1}{r}\Big)\Big[\mathcalh{J}-\mathcalh{K}\Big] \nonumber \\
  & & \hspace{0.15in}
  -\f{1}{\epsilon^{2}}\pderiv{}{\epsilon}
  \Big(
    \epsilon^{3}W^{2}
    \Big\{
      \Big[
        \f{1}{\psi^{2}}\pderiv{\ln\alpha}{r}
        +v_{r}\Big(\tauderiv{\ln\psi^{2}}-\f{1}{\alpha}\pderiv{\betau{r}}{r}\Big)
        +W^{2}\cderiv{v_{r}}{\tau}
      \Big]\Big(v_{r}\mathcalh{H}+\mathcalh{K}\Big) \nonumber \\
  & & \hspace{0.30in}
      +
      \Big[
        \f{v_{r}}{\psi^{2}}\pderiv{\ln\alpha}{r}
        +\Big(
          \tauderiv{\ln\psi^{2}}
          -\f{1}{\alpha}\pderiv{\betau{r}}{r}
        \Big)
        +W^{2}
        \Big(v_{r}\tauderiv{v_{r}}+\f{1}{\psi^{2}}\pderiv{v_{r}}{r}\Big)
      \Big]\Big(v_{r}\mathcalh{K}+\mathcalh{L}\Big) \nonumber \\
  & & \hspace{0.30in}
      +\Big[
        \cderiv{\ln\psi^{2}}{\tau}
        +\f{1}{r}\Big(\f{v_{r}}{\psi^{2}}-\f{\betau{r}}{\alpha}\Big)
      \Big]\Big(v_{r}\Big[\mathcalh{J}-\mathcalh{K}\Big]+\Big[\mathcalh{H}-\mathcalh{L}\Big]\Big)
    \Big\}
  \Big) \nonumber \\
  & &
  =W\Big(v_{r}\,2\pi\int_{-1}^{1}\collisionh{f}\,d\mu+2\pi\int_{-1}^{1}\collisionh{f}\mu\,d\mu\Big), 
  \label{eq:momentumMomentCFCSphericalSymmetry}
\end{eqnarray}
respectively.  
Equations (\ref{eq:energyMomentCFCSphericalSymmetry}) and (\ref{eq:momentumMomentCFCSphericalSymmetry}) are \emph{conservative} evolution equations for the monochromatic lab-frame radiation energy density and momentum density, respectively.  
The square root of the spatial metric determinant has been absorbed in quantities accented with a hat \citepalias[e.g., $\mathcalh{J}=\smdet\,\mathcal{J}$; cf.][]{muller_etal_2010}.  
The expression inside the time derivative in Equation (\ref{eq:energyMomentCFCSphericalSymmetry}) is the contribution from radiation to the ``matter sources" in the definition of the (conserved) ADM mass \citepalias[cf.][]{muller_etal_2010}.  
For flat spacetime (i.e., $\alpha=\psi=1$ and $\betau{r}=0$), Equations (\ref{eq:energyMomentCFCSphericalSymmetry}) and (\ref{eq:momentumMomentCFCSphericalSymmetry}) reduce to Equations (\ref{eq:energyMomentSRSphericalSymmetry}) and (\ref{eq:momentumMomentSRSphericalSymmetry}), respectively.  
In Equations (\ref{eq:energyMomentCFCSphericalSymmetry}) and (\ref{eq:momentumMomentCFCSphericalSymmetry}), we have defined the ``convective derivative" and the ``proper time derivative" along constant coordinate lines; 
\begin{equation}
  \cderiv{}{\tau}
  =\tauderiv{}+\f{v_{r}}{\psi^{2}}\pderiv{}{r}
  \,\mbox{ and }\,
  \tauderiv{}=\f{1}{\alpha}\pderiv{}{t}-\f{\betau{r}}{\alpha}\pderiv{}{r}, 
\end{equation}
respectively.  

The conservative, monochromatic lab-frame number equation, also expressed in terms of comoving-frame angular moments, can be obtained directly from Equation (\ref{eq:numberEquationCFC}), or by adding $W\epsilon^{-1}$ times Equation (\ref{eq:energyMomentCFCSphericalSymmetry}) and $-v_{r}W\epsilon^{-1}$ times Equation (\ref{eq:momentumMomentCFCSphericalSymmetry}).  
The result is
\begin{eqnarray}
  & &
  \f{1}{\alpha}\pderiv{}{t}
  \Big(
    W\Big[\mathcalhmcd{J}{N}+v_{r}\,\mathcalhmcd{H}{N}\Big]
  \Big)
  +\f{1}{\alpha}\pderiv{}{r}
  \Big(
    \alpha W
    \Big[
      \Big(\f{1}{\psi^{2}}-v_{r}\f{\betau{r}}{\alpha}\Big)\mathcalhmcd{H}{N}
      +\Big(\f{v_{r}}{\psi^{2}}-\f{\betau{r}}{\alpha}\Big)\mathcalhmcd{J}{N}
    \Big]
  \Big) \nonumber \\
  & & \hspace{0.15in}
  -\f{1}{\epsilon^{2}}\pderiv{}{\epsilon}
  \Big(
    \epsilon^{2}W
    \Big\{
      \Big[
        \f{1}{\psi^{2}}\pderiv{\ln\alpha}{r}
        +v_{r}\Big(\tauderiv{\ln\psi^{2}}-\f{1}{\alpha}\pderiv{\betau{r}}{r}\Big)
        +W^{2}\cderiv{v_{r}}{\tau}
      \Big]\mathcalh{H}
      +
      \Big[
        \f{v_{r}}{\psi^{2}}\pderiv{\ln\alpha}{r}
        +\Big(\tauderiv{\ln\psi^{2}}-\f{1}{\alpha}\pderiv{\betau{r}}{r}\Big) \nonumber \\
  & & \hspace{0.30in}
        +W^{2}\Big(v_{r}\tauderiv{v_{r}}+\f{1}{\psi^{2}}\pderiv{v_{r}}{r}\Big)
      \Big]\mathcalh{K}
      +
      \Big[
        \cderiv{\ln\psi^{2}}{\tau}
        +\f{1}{r}\Big(\f{v_{r}}{\psi^{2}}-\f{\betau{r}}{\alpha}\Big)
      \Big]\Big(\mathcalh{J}-\mathcalh{K}\Big)
    \Big\}
  \Big)
  =\f{2\pi}{\epsilon}
  \int_{-1}^{1}\collisionh{f}\,d\mu, 
  \label{eq:numberMomentCFCSphericalSymmetry}
\end{eqnarray}
where the comoving-frame number density and number flux are
\begin{equation}
  \big\{\,\mathcalmcd{J}{N},\,\mathcalmcd{H}{N}\,\big\}=2\pi\int_{-1}^{1}f\,\mu^{\{0,1\}}\,d\mu.  
  \label{eq:numberMomentsSphericalSymmetry}
\end{equation}
Note that $\{\,\mathcal{J},\,\mathcal{H}\,\}=\epsilon\,\{\,\mathcalmcd{J}{N},\,\mathcalmcd{H}{N}\,\}$ (cf. Equation (\ref{eq:energyMomentsSphericalSymmetry})).  
Equation (\ref{eq:numberMomentCFCSphericalSymmetry}) corresponds to the evolution equation for the neutrino number density given by \citetalias{muller_etal_2010} (their Equation (30); corrected here for a few misprints).  
For flat spacetime, Equation (\ref{eq:numberMomentCFCSphericalSymmetry}) reduces to the corresponding special relativistic equation (cf. Equation (\ref{eq:numberMomentSR})).  

\citetalias{muller_etal_2010} (cf. their Appendix B) have taken important initial steps in developing numerical methods for neutrino transport based on moments models where simultaneous conservation of energy and lepton number is considered \citep[see][for the Boltzmann case in spherical symmetry]{liebendorfer_etal_2004}.  
There may, however, be room for further improvements.  
Equations (\ref{eq:energyMomentCFCSphericalSymmetry}) and (\ref{eq:momentumMomentCFCSphericalSymmetry}) differ from the energy and momentum equations solved by \citetalias{muller_etal_2010} (their Equations (27) and (28), respectively).  
We obtain their energy equation by adding $W$ times Equation (\ref{eq:energyMomentCFCSphericalSymmetry}) and $-v_{r}\,W$ times Equation (\ref{eq:momentumMomentCFCSphericalSymmetry}).  
Similarly, we obtain their momentum equation by adding $-v_{r}\,W$ times Equation (\ref{eq:energyMomentCFCSphericalSymmetry}) and $W$ times Equation (\ref{eq:momentumMomentCFCSphericalSymmetry}).  
The relationship between the equation for the neutrino number density and the neutrino energy density in \citetalias{muller_etal_2010} (their Equations (30) and (27), respectively) is relatively simple: they differ only by a factor $\epsilon$.  
Note that there is a similarly simple relationship between Equation (\ref{eq:stressEnergyMomentComoving}) and the conservative number equation (Equation (\ref{eq:numberMoment})).  
This simple relationship is convenient when constructing a numerical method based on the radiation energy and momentum equations, that is simultaneously consistent with the conservative neutrino number equation (which ensures number conservation).  
However, the energy and momentum equations solved by \citetalias{muller_etal_2010} (obtainable directly from Equation (\ref{eq:stressEnergyMomentComoving})) are formulated in a non-conservative form.  
The ``source terms" depend on time and space derivatives of the fluid three-velocity, which do not vanish in the absence of gravity and neutrino-matter interactions.  
Conservation of energy is not guaranteed when the non-conservative formulation of the energy equation is used.  
(Additional complications may also arise in the presence of shocks.)  
As a possible improvement, a numerical scheme that simultaneously conserves energy and lepton number may be based on solving Equations (\ref{eq:energyMomentCFCSphericalSymmetry}) and (\ref{eq:momentumMomentCFCSphericalSymmetry}), which are conservative formulations of the radiation energy and momentum equations, respectively.   
However, such a scheme is potentially much more complicated than the algorithm outlined in Appendix B in \citetalias{muller_etal_2010}.  
When carrying out the the steps for obtaining the conservative number equation from the conservative energy and momentum equations, we observe that terms emanating from the time derivatives, space derivatives, and the geometry sources cancel with terms emanating from the energy derivatives.  
The remaining terms constitute the left-hand side of the number equation.  
Similar cancellations must occur in the discrete limit in order for the discretized energy and momentum to be consistent with neutrino number conservation (which is necessary to ensure lepton number conservation).


\begin{thebibliography}{}
  \bibitem[Anderson \& Spiegel (1972)]{andersonSpiegel_1972} Anderson, J.L. \& Spiegel, E.A. 1972, \apj, 171, 127
  \bibitem[Banyuls et al. (1997)]{banyuls_etal_1997} Banyuls, F., Font, J., Ib{\'a}{\~n}ez, J, Mart{\'i}, \& Miralles, J., \apj, 476, 221
  \bibitem[Baumgarte \& Shapiro (2010)]{baumgarteShapiro_2010} Baumgarte, T.W.L. \& Shapiro, S.L. 2010, Numerical Relativity (Cambridge: Cambridge University Press)
  \bibitem[Blondin et al. (2003)]{blondin_etal_2003} Blondin, J.M., Mezzacappa, A.,\& DeMarino, C. 2003, \apj, 584, 971
  \bibitem[Blondin \& Mezzacappa (2007)]{blondinMezzacappa_2007} Blondin, J.M. \& Mezzacappa, A. 2007, Nature, 445, 58
  \bibitem[Boffetta \& Ecke (2012)]{boffettaEcke_2012} Boffetta, G. \& Ecke, R.E. 2012, Annu. Rev. Fluid Mech., 44, 427
  \bibitem[Bruenn (1985)]{bruenn_1985} Bruenn, S.W. 1985, ApJS, 58, 771
  \bibitem[Bruenn et al. (2001)]{bruenn_etal_2001} Bruenn, S.W., De Nisco, K.R., \& Mezzacappa, A. 2001, \apj, 560, 326
  \bibitem[Bruenn et al. (2006)]{bruenn_etal_2006} Bruenn, S.W., Dirk, C.J., Mezzacappa, A., et al. 2006, J. Phys. Conf. Ser., 46, 393
  \bibitem[Bruenn et al. (2009)]{bruenn_etal_2009} Bruenn, S.W., Mezzacappa, A., Hix, W.R., et al. 2009, J. Phys. Conf. Ser., 180, 012018
  \bibitem[Bruenn et al. (2012)]{bruenn_etal_2012} Bruenn, S.W., Mezzacappa, A., Hix, W.R., et al. 2012, \apj, submitted (arXiv:1212.1747)
  \bibitem[Brunner \& Holloway (2001)]{brunnerHolloway_2001} Brunner, T.A. \& Holloway, J.P. 2001, JQSRT, 69, 543
  \bibitem[Buchler (1979)]{buchler_1979} Buchler, J.R. 1979, JQSRT, 22, 293
  \bibitem[Buchler (1983)]{buchler_1983} Buchler, J.R. 1983, JQSRT, 30, 395
  \bibitem[Buchler (1986)]{buchler_1986} Buchler, J.R. 1986, JQSRT, 36, 441
  \bibitem[Buras et al. (2006)]{buras_etal_2006a} Buras, R., Rampp, M., Janka, H.-Th., \& Kifonidis, K. 2006, A\&A, 447, 1049
  \bibitem[Burrows et al. (1995)]{burrows_etal_1995} Burrows, A. Hayes, J., \& Fryxell, B.A. 1995, \apj, 450, 830
  \bibitem[Burrows \& Thompson (2004)]{burrowsThompson_2004} Burrows, A. \& Thompson, T.A. 2004, 133, in Stellar Collapse, ed. C. Fryer (Kluwer Academic Publishers)
  \bibitem[Burrows et al. (2006)]{burrows_etal_2006} Burrows, A., Livne, E., Dessart, L., Ott, C.D., \& Murphy, J. 2006, \apj, 640, 878
  \bibitem[Burrows et al. (2007)]{burrows_etal_2007} Burrows, A., Livne, E., Dessart, L., Ott, C.D., \& Murphy, J. 2007, \apj, 655, 416
  \bibitem[Burrows et al. (2012)]{burrows_etal_2012} Burrows, A., Dolence, J.C., \& Murphy, J.W. 2012, \apj, 759, 5
  \bibitem[Cardall \& Mezzacappa (2003)]{cardallMezzacappa_2003} Cardall, C. \& Mezzacappa, A. 2003, Phys. Rev. D, 68, 023006
  \bibitem[Cardall et al. (2005)]{cardall_etal_2005} Cardall, C., Lentz, E.J., \& Mezzacappa, A. 2005, Phys. Rev. D, 72, 04307
  \bibitem[Cardall et al. (2012a)]{cardall_etal_2012a} Cardall, C., Budiardja, R., Endeve, E., \& Mezzacappa, A. 2012 (arXiv:1207.3392)
  \bibitem[Cardall et al. (2012b)]{cardall_etal_2012b} Cardall, C., Budiardja, R., Endeve, E., \& Mezzacappa, A. 2012 (arXiv:1207.3393)
  \bibitem[Cardall et al. (2012)]{cardall_etal_2012} Cardall, C., Endeve, E., \& Mezzacappa, A. 2012, Phys. Rev. D, submitted (arXiv:1209.2151)
  \bibitem[Castor (1972)]{castor_1972} Castor, J.I. 1972, ApJ, 178, 779
  \bibitem[Cernohorsky \& Bludman (1994)]{cernohorskyBludman_1994} Cernohorsky, J. \& Bludman, S.A. 1994, \apj, 433, 250
  \bibitem[Couch (2012)]{couch_etal_2012} Couch, S.M. 2012, \apj, submitted (arXiv:1212.0010v1)
  \bibitem[Ehlers (1971)]{ehlers_1971} Ehlers, J. 1971, in Proceedings of the International School of Physics ``Enrico Fermi" Course XLVII: General Relativity and Cosmology, 1, ed. B.K. Sachs (New York, NY: Academic Press)
  \bibitem[Font et al. (1994)]{font_etal_1994} Font, J.A., Ibanez, J.M., Marquina, A., \& Marti, M. 1994, A\&A, 282, 304
  \bibitem[Fryer \& Warren (2002)]{fryerWarren_2002} Fryer, C.L. \& Warren, M.S. 2002, \apj, 574, L65
  \bibitem[Fryer \& Young (2007)]{fryerYoung_2007} Fryer, C.L. \& Young, P.A. 2007, \apj, 659, 1438
  \bibitem[Hanke et al. (2012)]{hanke_etal_2012} Hanke, F., Marek, A., M{\"u}ller, B., \& Janka, H.-Th. 2012, \apj, 755, 138
  \bibitem[Hauck \& McClarren (2010)]{hauckMacclarren_2010} Hauck, C. \& McClarren, R. 2010, SIAM, J. Sci. Comput., 32, 2603
  \bibitem[Herant et al. (1994)]{herant_etal_1994} Herant, M., Benz, W., Hix, W.R., Fryer, C.L., \& Colgate, S.A. 1994, \apj, 435, 339
  \bibitem[Isenberg (2008)]{isenberg_2008} Isenberg, J.A. 2008, Int. J. Mod. Phys. D, 17, 265
  \bibitem[Ishihara et al. (2009)]{ishihara_etal_2009} Ishihara, T., Gotoh, T., \& Kaneda, Y. 2009, Annu. Rev. Fluid Mech., 41, 165
  \bibitem[Israel (1972)]{israel_1972} Israel, W. 1972, in General Relativity: Papers in Honour of J.L. Synge, 201, ed. L. O'Raifeartaigh (Oxford: Clarendon Press)
  \bibitem[Janka \& Muller (1996)]{jankaMuller_1996} Janka, H.-Th. \& M{\"u}ller, E. 1996, A\&A, 306, 167
  \bibitem[Janka (2012)]{janka_2012} Janka, H.-Th. 2012, Annu. Rev. Nucl. Part. Sci., 62, 407
  \bibitem[Kim et al. (2009)]{kim_etal_2009} Kim, J., Kim, H.I., \& Lee, H.M. 2009, MNRAS, 399, 229
  \bibitem[Kim et al. (2012)]{kim_etal_2012} Kim, J., Kim, H.I., Chaptuik, M.W., \& Lee, H.M. 2012, MNRAS, 424, 830
  \bibitem[Kaneko et al. (1984)]{kaneko_etal_1984} Kaneko, M., Morita, K., \& Maekawa, M. 1984, Ap\&SS, 107, 333
  \bibitem[Kitaura et al. (2006)]{kitaura_etal_2006} Kitaura, F.S., Janka, H.-Th., \& Hillebrandt, W. 2006, A\&A, 450, 345
  \bibitem[Kotake et al (2006)]{kotake_etal_2006} Kotake, K., Sato, Katsuhiko, W., \& Takahashi, K. 2006, Rep. Prog. Phys., 69, 971
  \bibitem[Kuroda et al. (2012)]{kuroda_etal_2012} Kuroda, T., Kotake, K., \& Takiwaki, T. 2012, \apj, 755, 11
  \bibitem[Landau \& Lifshitz (1959)]{landauLifshitz_1959} Landau, L. D. \& Lifshitz, E. M. 1959, Course of Theoretical Physics, Fluid Mechanics, Vol. 6 (Reading, MA: Addison-Wesley)
  \bibitem[Landau \& Lifshitz (1975)]{landauLifshitz_1975} Landau, L. D. \& Lifshitz, E. M. 1975, Course of Theoretical Physics, The Classical Theory of Fields, Vol. 2 (Oxford: Butterworth-Heineman)
  \bibitem[Lentz et al. (2012a)]{lentz_etal_2012a} Lentz, E.J., Mezzacappa, A., Messer, O.E.B., Liebend{\"o}rfer, M., Hix, W.R., \& Bruenn, S.W. 2012, \apj, 747, 73
  \bibitem[Lentz et al. (2012b)]{lentz_etal_2012b} Lentz, E.J., Mezzacappa, A., Messer, O.E.B., Hix, W.R., \& Bruenn, S.W. 2012, \apj, 760, 94
  \bibitem[Levermore (1984)]{levermore_1984} Levermore, C.D. 1984, JQSRT, 31, 149
  \bibitem[Levermore (1996)]{levermore_1996} Levermore, C.D. 1996, J. Stat. Phys., 83, 1021
  \bibitem[Liebend{\"o}rfer et al. (2001)]{liebendorfer_etal_2001} Liebend{\"o}rfer, Mezzacappa, A., Thielemann, F.-K., M., Messer, O.E.B., Hix, W.R., \& Bruenn, S.W. 2001, Phys. Rev. D, 63, 103004
  \bibitem[Liebend{\"o}rfer et al. (2004)]{liebendorfer_etal_2004} Liebend{\"o}rfer, M., Messer, O.E.B., Mezzacappa, A., Bruenn, S.W., Cardall, C.Y., \& Thielemann, F.-K. 2004, ApJS, 150, 263
  \bibitem[Liebend{\"o}rfer et al. (2009)]{liebendorfer_etal_2009} Liebend{\"o}rfer, M., Whitehouse, S.C., \& Fischer, T. 2009, \apj, 698, 1174
  \bibitem[Lindquist (1966)]{lindquist_1966} Lindquist, R.W. 1966, Ann. Phys., 37, 487
  \bibitem[Lowrie et al. (2001)]{lowrie_etal_2001} Lowrie, R.B., Mihalas, D., \& Morel, J.E. 2001, JQSRT, 69, 291
  \bibitem[Marek \& Janka (2009)]{marekJanka_2009} Marek, A. \& Janka, H.-Th. 2009, \apj, 694, 664
  \bibitem[Mezzacappa (2005)]{mezzacappa_2005} Mezzacappa A. 2005, Annu. Rev. Nucl. Part. Sci., 55, 467
  \bibitem[Mezzacappa \& Matzner (1989)]{mezzacappaMatzner_1989} Mezzacappa, A. \& Matzner R.A. 1989, ApJ, 343, 853
  \bibitem[Mihalas (1980)]{mihalas_1980} Mihalas, D. 1980, \apj, 237, 574
  \bibitem[Mihalas \& Mihalas (1999)]{mihalasMihalas_1999} Mihalas, D. \& Mihalas, B.W. 1999, Foundations of Radiation Hydrodynamics (New York: Dover)
  \bibitem[Minerbo (1978)]{minerbo_1978} Minerbo, G.N. 1978, JQSRT, 20, 541
  \bibitem[Misner et al. (1973)]{misner_etal_1973} Misner, C.W., Thorne, K.S., \& Wheeler, J.A. 1973 Gravitation (San Francisco, CA: W.H. Freeman)
  \bibitem[M{\"u}ller et al. (2010)]{muller_etal_2010} M{\"u}ller, B., Janka, H-Th., \& Dimmelmeier, H. 2010, ApJS, 189, 104
  \bibitem[M{\"u}ller et al. (2012a)]{muller_etal_2012a} M{\"u}ller, B., Janka, H.-Th., \& Marek, A. 2012, \apj, 756, 84
  \bibitem[M{\"u}ller et al. (2012b)]{muller_etal_2012b} M{\"u}ller, B., Janka, H.-Th., \& Heger, A. 2012, \apj, in press (arXiv:1205.7078v2)
  \bibitem[Munier \& Weaver (1986a)]{munierWeaver_1986a} Munier, A. \& Weaver R. 1986, Comp. Phys. Rep., 3, 125
  \bibitem[Munier \& Weaver (1986b)]{munierWeaver_1986b} Munier, A. \& Weaver R. 1986, Comp. Phys. Rep., 3, 165
  \bibitem[Nordhaus et al. (2010)]{nordhaus_etal_2010} Nordhaus, J., Burrows, A., Almgren, \& Bell, J. 2010, \apj, 720, 694
  \bibitem[Obergaulinger \& Janka (2011)]{obergaulingerJanka_2011} Obergaulinger, M \& Janka, H.-Th. 2011, A\&A, submitted (arXiv:1101.1198v1)
  \bibitem[Ott et al. (2008)]{ott_etal_2008} Ott, C.D., Burrows, A., Burrows, Dessart, L., \& Livne, E. 2008, \apj, 685, 1069
  \bibitem[Plewa \& M{\"u}ller (1999)]{plewaMuller_1999} Plewa, T. \& M{\"u}ller, E. 1999, A\&A, 342, 179
  \bibitem[Rampp \& Janka (2000)]{ramppJanka_2000} Rampp, M. \& Janka, H.-Th. 2000, \apj, 539, L33
  \bibitem[Rampp \& Janka (2002)]{ramppJanka_2002} Rampp, M. \& Janka, H.-Th. 2002, A\&A, 396, 361
  \bibitem[Riffert (1986)]{riffert_1986} Riffert, H. 1986, \apj, 310, 729
  \bibitem[Schutz (1985)]{schutz_1985} Schutz, B.F. 1985, A First Course in General Relativity (Cambridge)
  \bibitem[Shibata et al. (2011)]{shibata_etal_2011} Shibata, M., Kiuchi, K., Sekiguchi, Y., \& Suwa, Y. 2011, Prog. Theor. Phys., 125, 1255
  \bibitem[Smit et al. (2000)]{smit_etal_2000} Smit, J.M., van den Horn, L.J., \& Bludman, S.A. 2000, A\&A, 356, 559
  \bibitem[Sumiyoshi et al. (2005)]{sumiyoshi_etal_2005} Sumiyoshi, K., Yamada, S., Suzuki, H., Shen, H., Chiba, S., \& Toki, H. 2005, \apj, 629, 922
  \bibitem[Sumiyoshi \& Yamada (2012)]{sumiyoshiYamada_2012} Sumiyoshi, K. \& Yamada, S. 2012, ApJS, 199, 17
  \bibitem[Suwa et al. (2010)]{suwa_etal_2010} Suwa, Y., Kotake, K., Takiwaki, T., Whitehouse, S.C., Liebend{\"o}rfer, M., \& Sato, K., PASJ, 62, L49
  \bibitem[Swesty \& Myra (2009)]{swestyMyra_2009} Swesty, D.F. \& Myra, E.S. 2009, ApJS, 181, 1
  \bibitem[Takiwaki et al. (2009)]{takiwaki_etal_2009} Takiwaki, T., Kotake, K, \& Sato, K. 2009, \apj, 691, 1360
  \bibitem[Takiwaki et al. (2012)]{takiwaki_etal_2012} Takiwaki, T., Kotake, K., \& Suwa, Y. 2012, \apj, 749, 98
  \bibitem[Thompson et al. (2003)]{thompson_etal_2003} Thompson, T.A., Burrows, A., \& Pinto, P.A. 2003, \apj, 592, 434
  \bibitem[Thorne (1981)]{thorne_1981} Thorne, K.S. 1981, MNRAS, 194, 439
  \bibitem[Wilson et al. (1996)]{wilson_etal_1996} Wilson, J.R., Mathews, G.J., \& Marronetti, P. 1996, Phys. Rev. D, 54, 1317
  \bibitem[Woosley \& Janka (2005)]{woosleyJanka_2005} Woosley, S.E. \& Janka, H.-Th. 2005, Nature Physics, 1, 147
  \bibitem[Zhang et al. (2012)]{zhang_etal_2012} Zhang, W., Howell, L., Almgren, A., Burrows, A., Dolence, J., \& Bell, J. 2012, ApJS, in press (arXiv:1207.3845v2)
\end{thebibliography}
\end{document}